\documentclass[11pt]{article}
\usepackage{geometry}                
\pdfoutput=1

\usepackage{setspace}
\usepackage{graphicx}
\usepackage{amssymb}
\usepackage{bm}
\usepackage{epstopdf}
\usepackage{dblaccnt}
\newcommand{\ra}[1]{\renewcommand{\arraystretch}{#1}}

\usepackage[square,authoryear]{natbib}
\usepackage[titletoc,title]{appendix}
\usepackage[font=small,labelfont=bf]{caption}
\usepackage{placeins}

\title{Isostasy with Love: I Elastic equilibrium}
\author{Mikael Beuthe\\
\it Royal Observatory of Belgium, Brussels, Belgium\\
mikael.beuthe@observatory.be
 }      
\date{}                                             

\begin{document}

\maketitle

\begin{abstract}

Isostasy explains why observed gravity anomalies are generally much weaker than what is expected from topography alone, and why planetary crusts can support high topography without breaking up.
On Earth, it is used to subtract from gravity anomalies the contribution of nearly compensated surface topography.
On icy moons and dwarf planets, it constrains the compensation depth which is identified with the thickness of the rigid layer above a soft layer or a global subsurface ocean.
Classical isostasy, however, is not self-consistent, neglects internal stresses and geoid contributions to topographical support, and yields ambiguous predictions of geoid anomalies.
Isostasy should instead be defined either by minimizing deviatoric elastic stresses within the silicate crust or icy shell, or by studying the dynamic response of the body in the long-time limit.
In this paper, I implement the first option by formulating Airy isostatic equilibrium as the linear response of an elastic shell to a combination of surface and internal loads.
Isostatic ratios are defined in terms of deviatoric Love numbers which quantify deviations with respect to a fluid state.
The Love number approach separates the physics of isostasy from the technicalities of elastic-gravitational spherical deformations, and provides flexibility in the choice of the interior structure.
Since elastic isostasy is invariant under a global rescaling of the shell shear modulus, it can be defined in the fluid shell limit, which is simpler and reveals the deep connection with the asymptotic state of dynamic isostasy.
If the shell is homogeneous, minimum stress isostasy is dual to a variant of elastic isostasy called zero deflection isostasy, which is less physical but simpler to compute.
Each isostatic model is combined with general boundary conditions applied at the surface and bottom of the shell, resulting in one-parameter isostatic families.
At long wavelength, the thin shell limit is a good approximation, in which case the influence of boundary conditions disappears as all isostatic families members yield the same isostatic ratios.
At short wavelength, topography is supported by shallow stresses so that Airy isostasy becomes similar to either pure top loading or pure bottom loading.
The isostatic ratios of incompressible bodies with three homogeneous layers are given in analytical form in the text and in complementary software.

\end{abstract}


\vspace{\stretch{1}}
\newpage

{\small
\tableofcontents
\scriptsize
\listoffigures
\listoftables
}

\newpage

\section{Introduction}

The geophysical concept of isostasy was originally motivated by two questions \citep{airy1855}.
First, why are gravity anomalies close to Earth's mountains typically much weaker than the gravitational attraction of the mountains themselves?
Second, how can a rigid crust above a fluid-like interior support high relief without breaking up?
According to ``Jeffreys' theorem'' \citep{melosh2011}, topography creates non-hydrostatic differential stresses (or deviatoric stresses) in the subsurface, reaching at least one-third to one-half of the load pressure $\rho{}gH$, and having a non-negligible amplitude down to a depth comparable to the load width or half-wavelength \citep{jeffreys1943,jeffreys1959}.
A key idea of isostatic equilibrium is that deviatoric stresses vanish below a certain depth (or \textit{compensation depth}), where hydrostatic pressure becomes the only force available \citep{melosh2011}.
For silicate planets and moons, this hypothesis fits well with the decreasing strength of rock at depth, up to a point (lithosphere-asthenosphere boundary) where rocks start to flow viscously over long time scales.
The notion of compensation depth is even more relevant to icy moons, where an ice-to-water phase transition at depth may result in the creation of a global subsurface ocean.
In both cases, the deviatoric stresses supporting topography are restricted to the zone (silicate crust or icy shell) above the compensation depth
and are larger than they would be if the compensation depth was infinite.
In order to avoid crustal breakage, long-wavelength loads of large amplitude must thus be supported by counteracting underground loads, either in the form of crustal density variations (Pratt isostasy), or in the form of crustal bottom topography (Airy isostasy).

Additional assumptions are necessary in order to determine the amount of compensation \citep{lambeck1980}.
Classical models of isostasy often assume \textit{local compensation} in the sense that crustal columns are free to move vertically; in such a model, shear stresses on vertical surfaces play no part in supporting topography.
This assumption conveniently results in isostatic balance (assuming that topographic stresses are above the yield stress) but is only appropriate for long-wavelength anomalies \citep{dahlen1982}.
It is also difficult to justify a priori, especially when the effect of spherical geometry becomes significant, as crustal columns are generally expected to exert shear stresses on each other.
Regional isostasy, also called flexural or Vening Meinesz isostasy, rejects local compensation in attributing part of the support to lithospheric flexure, but this model is not strictly isostatic because the elastic thickness of the lithosphere appears as an additional parameter.
Whereas the assumption of local compensation is optional, it is absolutely necessary to choose a constraint (or \textit{isostatic prescription}) determining the distribution of compensating masses.
In classical approaches to isostasy, this constraint has been often formulated as `equality of pressure at the compensation depth'.
In practice, it is implemented either as the equality of mass in conical crust-mantle columns, or as the equality of lithostatic/hydrostatic pressure along radial lines.
The debate about which implementation is best has been going on and off for a long time \citep{lambert1930,vening1946,heiskanen1958,hemingway2017}.

Most authors agree that equality of mass is a rather arbitrary choice, conveniently approximating the `equality of pressure at the compensation depth', but without any good physical justification.
By contrast, equality of lithostatic/hydrostatic pressure seems at first to make good physical sense, until its fundamental flaws become manifest.
First, the assumption of a lithostatic (that is, hydrostatic) crust is incompatible with the existence of non-zero deviatoric stresses supporting the topography, as pithily expressed by Jeffreys: \textit{`supposing the Earth to be a perfect liquid (...) an elevated region $100\,$km in width would be drastically altered in appearance in a few minutes'} \citep[][Section 6.01]{jeffreys1959}.
The lithostatic/hydrostatic assumption suppresses vertical shear support, modifies the vertical support due to tangential stresses in spherical geometry (since tangential stresses are supposed to be equal to radial stresses), and predicts a pressure at the bottom of the crust deviating from the actual pressure.
Quoting \citet{lambert1930}:
\textit{`The existence of hydrostatic pressure implies that the matter in any truncated spherical cone forming a part of the crust derives no support from its connection with adjoining portions of the crust, an assumption that does not seem probable.'}
Second, this prescription does not take into account the effect of geoid deviations on topographical support.
Actually, the rigorous implementation of equality of pressure at a fixed depth (thus on a spherical surface) results in a completely different model, predicting geoid anomalies which are clearly way off \citep{garland1965,dahlen1982}.
While the equality of lithostatic/hydrostatic pressure is easy to write down mathematically, it does not correspond to equality of pressure on a pre-defined physical surface, although it is always possible to find a surface of equal pressure  after the fact. 
One should realize that, whatever the model (isostatic or not),  the hydrostatic layer below the compensation depth contains surfaces of equal pressure which are identical to equipotential surfaces.

Moreover, equal pressure isostasy cannot be considered as a special case of a more complete isostatic model, which would reduce to equal pressure isostasy in some limit.
In that respect, it is instructive to study the various isostatic models in the thin shell limit.
Although the difference between equal mass and equal pressure prescriptions is sometimes attributed to the nonnegligible shell thickness \citep{hemingway2017},
long-wavelength geoid anomalies also vary widely between thin shell isostatic models \citep{dahlen1982}.
For example, equal mass isostasy (with conical columns) and equal (lithostatic) pressure isostasy predict geoid anomalies respectively proportional to
\begin{eqnarray}
\delta V_{eq.\,mass} &\sim& 1- (1-d/R)^n \, \sim \, n \, d/R \, ,
\\
\delta V_{eq.\,press.} &\sim& 1 - (1-d/R)^{n+2} \, \sim \, (n+2) \, d/R \, ,
\end{eqnarray}
where $(d,R,n)$ are respectively the crustal thickness, surface radius, and harmonic degree (the dependence of gravity on depth was neglected here).
At harmonic degree two, geoid anomalies differ by a factor of two between these approaches (Jeffreys noted similar discrepancies much earlier:
\textit{`with different distributions of the stress in the upper layer we obtain values of the external potential associated with a given external form differing as much as in the ratio 1 to 2'} \citep{jeffreys1932}).
We will show in this paper that all well-grounded isostatic approaches have the same thin shell limit, but that it is not true of equal mass isostasy and equal (lithostatic) pressure isostasy.

If the classical prescriptions of equal mass and equal pressure are incorrect, what is then the right approach?
A first path forward is Jeffrey's idea of minimizing elastic deviatoric stresses within the crust:
\textit{`We shall say that isostasy means (1) below a certain depth the stress is hydrostatic, (2) above the depth the maximum stress-difference is as small as possible, consistently with the surface load'} \citep{jeffreys1959}.
Jeffreys considered both the maximum stress difference \citep{jeffreys1932} and the second stress invariant \citep{jeffreys1943}.
His idea did not catch on at the time, but was revived first by \citet{dahlen1982} to predict geoid anomalies on Earth (in the thin shell limit and with local compensation), and then by \citet{beuthe2016} for Enceladus (without the thin shell and local assumptions).
In the same spirit, \citet{kaula1963} determined internal loads within the Earth's mantle under a constraint of minimum stress.
Minimum stress isostasy consists in analyzing the final elastic state and is closely related to the energy minimization of a static mechanical system: one solves for compensating density anomalies without considering the loading history.
It is not known whether the isostatic configuration results from faulting, plastic deformations or viscous evolution, but the final state is assumed to be held in place by elastic stresses.
This assumption is less restrictive than it seems because the elastic solution is a good approximation of the absolute minimum stress solution (including non-elastic solutions), as argued by Jeffreys with a principle of minimum energy (Chapter 6 of \citet{jeffreys1959}; Section 3.3.1 of  \citet{melosh2011}).
As an alternative to minimum stress isostasy, dynamic models (either viscoelastic or purely viscous) include the loading history and generate the final isostatic state as the end product of a long time evolution.
Dynamic models will be examined in a companion paper (\citet{beuthe2020b}, hereafter called Paper~II), and will shown to be closely related to elastic isostasy.
Both elastic and dynamic models have the physically desirable property of being nonlocal (regional compensation), although locality can be achieved with transversely isotropic elasticity if one wishes a model more similar to classical isostasy.

\begin{figure}
\centering
    \includegraphics[width=\textwidth]{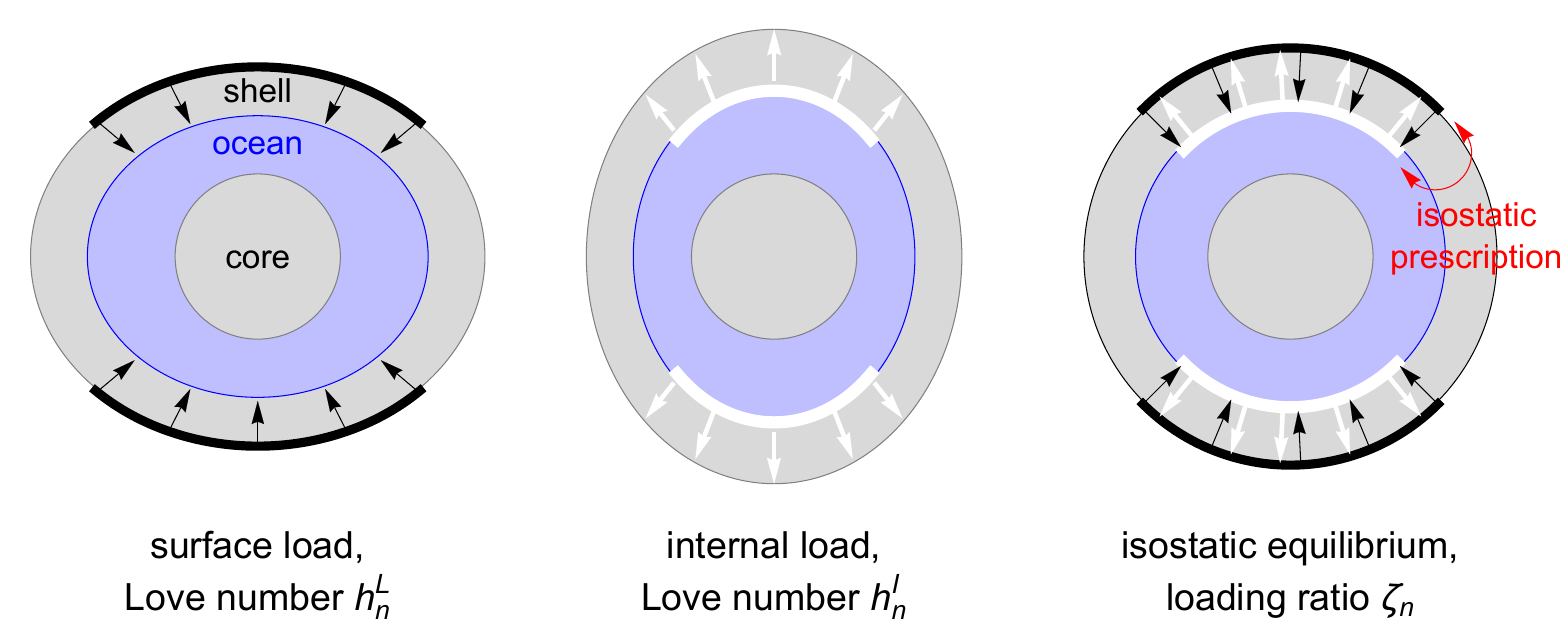}
   \caption[Isostasy with Love as a loading problem]{
   Isostasy with Love: isostatic equilibrium results from the linear combination of the (visco)elastic-gravitational deformations due to a surface load and a counteracting internal load at the shell-ocean boundary.
   The choice of an isostatic prescription imposes a relation between the surface load and the internal load.
   Deformations are quantified by Love numbers.
 }
   \label{FigLoadingSketch}
\end{figure}

Isostatic models going beyond classical `naive' isostasy typically suffer from being too complex, because they combine two problems in one: defining a precise isostatic model and solving for gravitational-elastic (or viscoelastic) deformations of the body.
If deformations are small, using Love numbers is a practical way to keep these tasks separate (Fig.~\ref{FigLoadingSketch}).
Love numbers have the advantages of modularity, generality, and flexibility:
\begin{itemize}
\item Modularity means decoupling the isostatic analysis from the complicated, but well-known, computation of gravitational-(visco)elastic linear deformations (described by Love numbers).
\item Generality means that completely different isostatic approaches can be implemented in the same framework.
In particular, Love numbers describe as well elastic and viscoelastic deformations, with the latter including viscous deformations as a limiting case.
Another example is that the isostatic compensation can be either regional (isotropic material) or local (transversely isotropic material).
\item Flexibility means that Love numbers can be computed with different assumptions about the internal structure of the body (density stratification, depth-dependent rheology of the crust, deformable core, compressibility, etc).
Moreover, the second stress invariant (or the elastic shear energy) of a homogeneous shell can be directly computed in terms of partial derivatives of Love numbers, so that the minimum stress configuration of such a shell can be found without computing stresses.
\end{itemize}
The main limitation of Love numbers is that deformations should be small so that the gravitational-elastic equations can be linearized.
Load Love numbers are part of the standard geophysical toolbox since 1960 \citep{munk1960,longman1963,kaula1963,farrell1972}; their computation can be done with different methods and software, and will be treated in this paper as a black box yielding the required numbers  (general principles are outlined in Appendix~\ref{AppendixLoveNumbers}, and analytical formulas for the Love numbers of a 3-layer model are given in complementary software, see \citet{beuthe2020a}).
While this paper emphasizes analytical results, there is absolutely no problem in evaluating numerically the Love numbers before using them in the various isostatic ratios.

The scope of this paper is limited to Airy isostasy, which is mathematically simpler than Pratt isostasy because Airy loads are applied on shell boundaries, whereas Pratt loads can be distributed in many ways within the shell.
Loading at boundaries makes it possible to express Airy isostatic ratios in terms of newly defined \textit{deviatoric Love numbers} which represent the deviation of the full Love numbers from their fluid limit.
It will turn out that elastic isostatic ratios are invariant under a global rescaling of the shear modulus.
This property will be used to compute elastic isostasy in the fluid limit, which is not only much simpler analytically but makes obvious their connection with dynamic isostasy (see Paper~II).
In keeping with tradition, the compensating topography is supposed to be located at the compensation depth, meaning that the bottom of the crust coincides with the lithosphere-asthenosphere boundary.
This situation is perfectly realized in icy satellites where the ice-to-water phase transition causes both a density contrast and a mechanical discontinuity.
Note that the Love number approach can be extended to Pratt isostasy by solving numerically the gravitational-elastic equations for distributed internal loads \citep{kaula1963}.

If the shell is thick, there is more freedom in choosing boundary conditions.
Stress must indeed be minimized over a subset of all possible configurations which can be defined in different ways:
by imposing the surface shape, by imposing the bottom shape, or by keeping constant a combination of both.
Thus, minimum stress isostasy (MSI) becomes a one-parameter isostatic family.
Besides MSI, I discuss an alternative model of elastic isostasy based on a no-deformation constraint \citep{banerdt1982}.
This model of `zero deflection isostasy' (ZDI) is simpler to formulate than MSI and is mathematically equivalent to stationary dynamic isostasy (see Paper~II), and therefore interesting in its own right.
Like MSI, it belongs to a one-parameter isostatic family characterized by where the no-deformation constraint is applied.
I will show here that MSI and ZDI families are dual if the body is incompressible and has a homogeneous shell.
Thanks to this duality, MSI can be computed exactly from ZDI without evaluating the second stress invariant or partial derivatives of Love numbers.

The rest of the paper is organized as follows.
Section~\ref{SectionSurfaceInternalLoading} shows how to express shape and gravity perturbations in terms of deviatoric Love numbers.
Section~\ref{SectionIsostaticRatios} defines isostatic ratios in terms of Love numbers, without applying yet a specific isostatic prescription.
Special attention is paid to relations between isostatic ratios in 3-layer models, degree-one isostasy, and the property of $\mu$-invariance which justifies the study of elastic isostasy in the fluid limit.
Section~\ref{SectionIsostasyWithoutLove} describes in that framework previous approaches not based on Love numbers, namely classical or `naive' isostasy and Dahlen's thin shell isostasy (viscous isostasy is examined in Paper~II).
Section~\ref{ElasticIsostasyMSI} expounds elastic isostasy, including zero deflection isostasy and minimum stress isostasy.
Mathematica notebooks and Fortran code implementing the analytical formulas for an incompressible body with three homogeneous layers are freely available online \citep{beuthe2020a}.

\section{Surface and internal loads}
\label{SectionSurfaceInternalLoading}

\subsection{Isostasy as a linear perturbation}

In their papers dealing with isostasy (cited in the Introduction), Jeffreys and Dahlen explicitly solved the equations of static equilibrium (involving stress and gravity) and Poisson's equation for the gravitational potential before applying a condition of minimum stress.
Their approach forces them to choose a specific interior model to start with.
As in \citet{beuthe2016}, I will instead use the fact that the solution of the linearized equations of equilibrium and Poisson's equation can be represented, at least in part, by way of Love numbers.
Love numbers quantify the linear response function of the body to a small forcing, which corresponds in this case to the surface and internal loads involved in Airy isostasy.
The `Isostasy with Love' approach consists in combining the responses to the two loads so as to satisfy some physical criterion defining isostasy (Fig.~\ref{FigLoadingSketch}).

Is linear perturbation theory applicable to isostasy?
Jeffreys' theorem tells us that the minimum stress difference required to support a surface load of height $H_s$ is about $\sigma_{min}\sim\rho_s{}g_sH_s/3$ \citep{melosh2011}.
If $H_s\sim1\rm\,km$, the corresponding strain $\epsilon_{min}\sim\sigma_{\min}/2\mu_s$ is of the order of $10^{-4}$ on Earth, and even less on icy satellites where $\rho_s{}g_s/\mu_s$ is smaller.
Thus it is generally a good approximation to work with infinitesimal strains and with a linear stress-strain relation (Hooke's law).
In addition, treating the load as a surface density is a good approximation for gravity anomalies as long as $H_s/R_s\ll1$.
The only real difficulty arises when trying to disentangle the isostatic shape from the hydrostatic figure of equilibrium, especially at harmonic degrees 2 and 4.
The simplest procedure consists in computing separately the isostatic and hydrostatic contributions before adding them, each one being considered as a perturbation of the spherically symmetric initial state \citep{iess2014}.
For small bodies rotating rapidly such as Enceladus, the hydrostatic deformation must be computed to second order in the flattening \citep{mckinnon2015,beuthe2016}.
The error on isostatic gravity coefficients can be estimated to be 10 to 20\% at harmonic degrees 2 and 3, which is less than the present-day accuracy of the data \citep{beuthe2016}.
The interpretation of future gravity data will however require a self-consistent computation of isostatic and hydrostatic deformations to second order in the flattening.

\subsection{Unperturbed model}
\label{SectionUnperturbedModel}

\begin{figure}
\centering
    \includegraphics[width=0.4\textwidth]{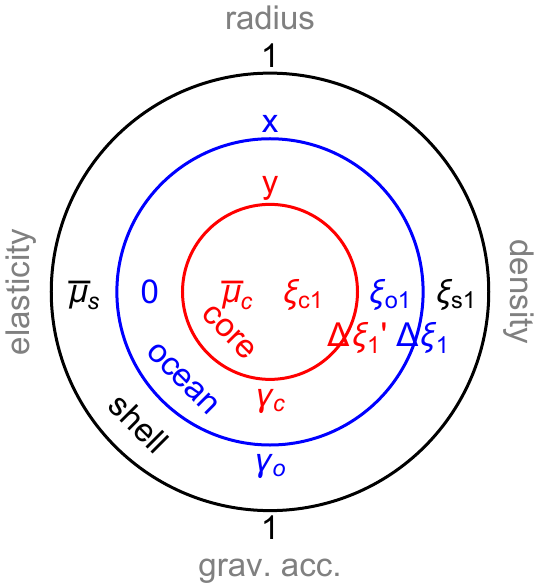}
   \caption[3-layer model with homogeneous layers: nondimensional parameters]{
   3-layer model with homogeneous layers: nondimensional parameters (see Table~\ref{TableParam}).
}
   \label{FigInternalStructure}
\end{figure}

The initial state, or unperturbed model, is spherically symmetric and possesses a shell (or crust), floating on a fluid layer (ocean or asthenosphere).
In its simplest spherical version, Airy isostasy can be modelled with a 2-layer body made of a homogeneous shell surrounding a homogeneous ocean (or liquid core) with a density contrast at the shell-ocean interface.
The problem with this model is that the density stratification, and thus the dependence of gravity on depth, is not realistic.
For example, the density of the shell and ocean of the Galilean moon Europa is three times smaller than the bulk density.
This flaw is easily corrected by adding a solid core within the ocean (or an inner core within the liquid core).
For simplicity, the terms core/ocean/shell will be used indifferently for icy satellites and silicate planets, with the understanding that the ocean and shell represent the asthenosphere and crust of silicate planets.
The density (of all layers) and the rheology (of solid layers) may depend on radius, although incompressible models with homogeneous layers (see Table~\ref{TableParam} and Fig.~\ref{FigInternalStructure}) will receive special attention because they can be solved analytically.
In Airy isostasy, the shell is loaded by the surface topography (surface load denoted `L') and bottom topography (bottom load denoted `I' for internal), the latter taking the form of undulations of the shell-ocean or crust-asthenosphere boundary.
There must be a density contrast at this boundary otherwise compensation cannot occur.
As usual, the shape of an interface denotes the deviations from the initial spherical state, while the topography of an interface denotes the shape of the interface minus the local geoid.

\begin{table}[h]\centering
\ra{1.3}
\small
\caption[Internal structure parameters for a 3-layer model with homogeneous layers]{
Internal structure parameters for a 3-layer model with homogeneous layers.
}
\vspace{1.5mm}
\begin{tabular}{@{}lll@{}}
\hline
\vspace{0.3mm}
Parameter &  Symbol & Nondimensional version \\
\hline
Surface (or shell) radius & $R_s$ & 1 \\
Ocean (or asthenosphere) radius & $R_o$ & $x=R_o/R_s$ \\
Core radius & $R_c$ & $y=R_c/R_s$ \\
Shell thickness & $d_s$ & $\varepsilon=d_s/R_s=1-x$ \\
Bulk density & $\rho_b$ & 1 \\
Density of layer $(j=s,o,c)$ & $\rho_j$ & $\xi_{jn}=\frac{3}{2n+1} \,  (\rho_j/\rho_b)$ \\
Density contrast (ocean-shell) & $\rho_o-\rho_s$ & $\Delta\xi_n=\xi_{on}-\xi_{sn}$ \\
Density contrast (core-ocean) & $\rho_c-\rho_o$ & $\Delta\xi'_n=\xi_{cn}-\xi_{on}$ \\
Gravitational acceleration at radius $R_j$ ($j=s,o,c$) & $g_j$ & $\gamma_j=g_j/g_s$ \\
Elastic shear modulus (shell) & $\mu_{\rm s}$ & $\bar\mu_{\rm s}=\mu_{\rm s}/(\rho_b g_s R_s)$ \\
Elastic shear modulus (core) & $\mu_{\rm c}$ & $\bar\mu_{\rm c}=\mu_{\rm c}/(\rho_b g_s R_s)$ \\
\hline
\end{tabular}
\label{TableParam}
\end{table}%

Nondimensional parameters are used throughout this work (Table~\ref{TableParam}).
Nondimensionalization has the advantage of reducing the number of parameters to a minimum, which makes it easier to obtain analytical formulas in simple models and reveals characteristic properties of the system.
Regarding the density, it is convenient to define the degree-$n$ density ratio of the layer $j$ by
\begin{equation}
\xi_{jn} = \frac{3}{2n+1} \,  \frac{\rho_j}{\rho_b} \, ,
\end{equation}
because this factor systematically appears when computing gravitational perturbations.
The degree-one density ratio is simply equal to the relative density: $\xi_{j1}=\rho_j/\rho_b$.
If the shell is of uniform density, the nondimensional gravitational acceleration at the shell-ocean boundary is given by
\begin{equation}
\gamma_o = \frac{g_o}{g_s} \, = \, \frac{1}{x^2} \left( 1+ (x^3-1) \, \xi_{s1} \right) .
\label{gammao} \\
\end{equation}
If the ocean is also of uniform density, the nondimensional gravitational acceleration at the core-ocean boundary is given by
\begin{equation}
\gamma_c = \frac{g_c}{g_s}  \, = \, \frac{1}{y^2} \left( 1- \xi_{s1} - x^3\Delta\xi_1 + y^3 \, \xi_{o1} \right) .
\label{gammac}
\end{equation}
If each layer is of uniform density, the nondimensional density contrast at the core-ocean boundary is equal to $\Delta \xi'_1$ with
\begin{equation}
\Delta \xi'_n =
\frac{3}{2n+1} \, \frac{1 - \xi_{s1} - x^3 \Delta\xi_1}{y^3} \, .
\label{CoreOceanDensContrast}
\end{equation}

If the core is made of a homogeneous material (silicates or iron),  it deforms much less than the icy shell not only because of the much higher shear modulus of the core material, but mostly because the global ocean plays the role of a decoupling layer (see Section 7.2 of \citet{beuthe2015b} for a quantitative analysis).
If this is correct, it is an excellent approximation to assume that the core does not deform at all or, equivalently, that it is infinitely rigid.
Once the core density has been determined by mass conservation (Eq.~(\ref{CoreOceanDensContrast})), the solution becomes independent of the core radius.
Mathematically, this is equivalent to the assumption of a point-like core.

The assumption of a very rigid core could fail for small icy satellites if the silicate core is unconsolidated and porous, with pores filled by interstitial ice \citep{roberts2015} or water in hydrothermal circulation caused by tidal heating \citep{travis2015,choblet2017,liao2020}.
More generally, it is possible that a soft core has had enough time to relax viscoelastically under the isostatic loads, especially if the latter are maintained as a dynamic equilibrium between viscous flow and melting (or freezing) at the bottom of the shell.
Such behaviour can be modelled with a fluid-like core.

\subsection{Shape and gravity perturbation}

\begin{figure}
\centering
   \includegraphics[width=0.5\textwidth]{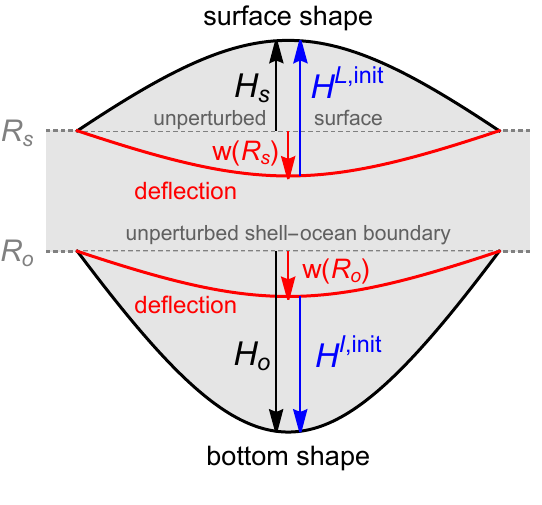}
   \caption[Surface and internal loading of the shell: shape, elevation, and deflection]{
   Surface and internal loading of the shell: shape $(H_{s},H_{o})$, initial elevation $(H^{L,init},H^{I,init})$, and deflection $(w(R_s),w(R_o))$ of the surface and shell-ocean boundary.
   The shape of the core (not shown) is equal to its deformation.
}
   \label{FigDeformation}
\end{figure}

Consider a surface density perturbation distributed on a sphere of radius $R_j$ ($j=s$ or $o$).
Since the unperturbed state is spherically symmetric, the induced deformations and gravitational perturbations can be computed separately at each harmonic degree and are independent of the harmonic order.
The surface density perturbation is thus expanded in spherical harmonics, with $\sigma_{n}$ denoting the degree-$n$ coefficient of the expansion (the harmonic order is kept implicit).
The gravitational potential is given by \citep[e.g.][]{kaula1968,grefflefftz1997}
\begin{eqnarray}
U_n(r\geq R_j) &=& 4\pi{}GR_j \Big( \frac{R_j}{r}\Big)^{n+1} \, \frac{\sigma_{n}}{2n+1} \, ,
\label{PotExt} \\
U_n(r<R_j) &=& 4\pi{}GR_j \Big( \frac{r}{R_j}\Big)^{n} \,  \frac{\sigma_{n}}{2n+1} \, .
\label{PotInt}
\end{eqnarray}
The load potentials are conventionally defined by their value at the radii where they are emplaced.
Thus, the surface load potential $U_n^L$ and the bottom (or internal) load potential $U_n^I$ read
\begin{eqnarray}
U_{n}^L &=& \frac{4\pi{}G \,R_s}{2n+1} \, \sigma_{n}^L \, ,
\\
U_{n}^I &=& \frac{4\pi{}G \,R_o}{2n+1}  \, \sigma_{n}^I \, .
\end{eqnarray}
In Airy isostasy, the surface density results from the initial elevation, where `initial' means that the elevation is measured with respect to the sphere before deformation: $\sigma_n^L=\rho_sH^{L,init}_n$ and $\sigma_n^I=(\rho_o-\rho_s)H_n^{I,init}$.
The initial elevation is related to the forcing load by
\begin{eqnarray}
H_n^{L,init} &=& \frac{1}{\xi_{sn}} \, \frac{U_n^L}{g_s} \, ,
\\
H_n^{I,init} &=& \frac{1}{ x \Delta\xi_n } \, \frac{U_n^I}{g_s} \, .
\end{eqnarray}
The degree-$n$ radial deformation, or \textit{deflection}, is characterized by a transfer function called the radial Love number $h$, specified by a superscript denoting the type of forcing ($L$ for surface load and $I$ for internal or bottom load) and a subscript denoting the location where the response is evaluated ($j=s$ for surface, $j=o$ for shell-ocean boundary, and $j=c$ for core-ocean boundary).
The deflection of the interface of radius $R_j$ reads
\begin{equation}
w_{n}(R_j) = \frac{1}{g_s} \left( h_j^L \, U_{n}^L+ h_j^I  \, U_{n}^I \right) .
\label{wn}
\end{equation}
The shape of the boundaries is obtained by combining the initial elevation and the deformation (see Fig.~\ref{FigDeformation}):
\begin{eqnarray}
H_{sn} &=& \left( \frac{1}{\xi_{sn}} + h_s^L \right) \frac{U_{n}^L}{g_s} + h_s^I  \, \frac{U_{n}^I}{g_s} \, ,
\label{ShapeSurf}\\
H_{on} &=& h_o^L \, \frac{U_{n}^L}{g_s} + \left( \frac{1}{ x \Delta\xi_n } + h_o^I  \right) \frac{U_n^I}{g_s} \, ,
\label{ShapeBot} \\
H_{cn} &=& h_c^L \, \frac{U_{n}^L}{g_s}  + h_c^I \, \frac{U_n^I}{g_s} \, .
\label{ShapeCore}
\end{eqnarray}
The shape $H_{jn}$ is measured with respect to the unperturbed sphere of radius $R_j$.

The total gravitational perturbation $\Gamma$ is due to the load potential (direct effect) and to the gravitational potential induced by the deformation (indirect effect or self-gravity).
The former depends on the relative position of the forcing and where the potential is computed through Eqs.~(\ref{PotExt})-(\ref{PotInt}).
The latter is characterized by its transfer function, called the gravitational Love number $k$, with the same superscripts $(L,I)$ and subscripts $(s,o,c)$ specifying radial Love numbers.
At the surface and at the shell-ocean boundary, the total gravitational perturbation reads
\begin{eqnarray}
\Gamma_{sn} &=& \left( 1 + k_s^L \right) U_{n}^L + \left( x^{n+1} + k_s^I \right) U_{n}^I \, ,
\label{GravSurf} \\
\Gamma_{on} &=& \left( x^n + k_o^L \right) U_{n}^L + \left( 1 + k_o^I \right) U_{n}^I \, ,
\label{GravBot} \\
\Gamma_{cn} &=& \left( y^n + k_c^L \right) U_{n}^L + \left( (y/x)^n + k_c^I \right) U_{n}^I \, .
\label{GravCore}
\end{eqnarray}
In each expression between brackets, the first term represents the direct effect of the potential, which is normalized so that it is equal to one where the load is emplaced.
The direct effect decreases by the factor $(R_j/r)^{n+1}$ above the load and by the factor $(r/R_j)^n$ below the load.

Computing Love numbers requires solving the equations of equilibrium and Poisson's equation for the gravitational potential, which can be formulated as a system of six differential equations of the first order constraining six scalar variables (Appendix~\ref{AppendixVariablesyi}).
The same equations are used whatever the forcing: free oscillations, tidal forcing, surface loads, internal loads, pressure loads, etc.
Different forcings are characterized by different sets of boundary conditions; those applicable to Airy isostasy are given in Appendix~\ref{AppendixBoundaryConditions}.
In general, the elastic-gravitational equations must be integrated numerically but analytical solutions exist for a body with homogeneous and incompressible layers if the material is fully isotropic (Appendix~\ref{AppendixMethodsIsotropic}).
But the transversely isotropic Love numbers required for isostatic models with local compensation must always be evaluated numerically (Appendix~\ref{AppendixMethodsLocal}).

Different sets of Love numbers (corresponding to different boundary conditions) are not independent \citep{molodensky1977,saito1978}.
In  Appendix~\ref{AppendixSaitoMolodensky}, I show that the Saito-Molodensky relation between surface and internal load Love numbers reads
\begin{equation}
k_s^I - h_s^I = x \, \Big( k_o^L - \gamma_o \, h_o^L \Big) \, .
\label{SMrelation}
\end{equation}
This equation is true whatever the interior structure as long as it is spherically symmetric.
The deviatoric Love numbers defined in Section~\ref{FluidDeviatoricLoveNumbers} satisfy the same relation: $\hat k_s^I - \hat h_s^I = x \, ( \hat k_o^L - \gamma_o \, \hat h_o^L )$.

\subsection{Fluid and deviatoric Love numbers}
\label{FluidDeviatoricLoveNumbers}

\begin{table}[h]\centering
\ra{1.3}
\small
\caption[Fluid and deviatoric load Love numbers]{
Fluid and deviatoric load Love numbers (implicitly at harmonic degree $n$).
The functions $y_1(r)$ and $y_5(r)$ represent the degree-$n$ linear response functions for the radial displacement and the gravitational perturbation (see Appendix~\ref{AppendixLoveNumbers}).
}
\vspace{1.5mm}
\begin{tabular}{@{}llllll@{}}
\hline
\vspace{0.3mm}
Symbol  & Type of forcing & Response at & Relation to $y_i(r)$ & Fluid & Deviatoric-fluid \\
              &                          &  which interface?  &                 &  limit & decomposition \\
\hline
\multicolumn{6}{l}{Radial Love numbers} \\
$h_s^L$ & surface load & surface & $g_s y_1^L(R_s)$ & $-1/\xi_{sn}$ & $\hat h_s^L-1/\xi_{sn}$ \\
$h_o^L$ & surface load & shell-ocean &  $g_s y_1^L(R_o)$ & \hspace{1.5mm} $0$ & $\hat h_o^L$ \\  
$h_c^L$ & surface load & core-ocean &  $g_s y_1^L(R_c)$ & \hspace{1.5mm} $0$ & $\hat h_c^L$ \\  
$h_s^I$ & internal load & surface & $g_s y_1^I(R_s)$ & \hspace{1.5mm} $0$& $\hat h_s^I$ \\ 
$h_o^I$ & internal load & shell-ocean &  $g_s y_1^I(R_o)$ & $-1/(x \Delta\xi_n)$ & $\hat h_o^I-1/(x \Delta\xi_n)$\\ 
$h_c^I$ & internal load & core-ocean &  $g_s y_1^I(R_c)$ & \hspace{1.5mm} $0$& $\hat h_c^I$\\ 
\multicolumn{6}{l}{Gravitational Love numbers} \\
$k_s^L$ & surface load & surface & $y_5^L(R_s)-1$ & $-1$ & $\hat k_s^L-1$ \\ 
$k_o^L$ & surface load & shell-ocean & $y_5^L(R_o)-x^n$ & $-x^n$ & $\hat k_o^L-x^n$ \\ 
$k_c^L$ & surface load & core-ocean & $y_5^L(R_c)-y^n$ & $-y^n$ & $\hat k_c^L-y^n$ \\ 
$k_s^I$ & internal load & surface & $y_5^I(R_s)-x^{n+1}$ & $-x^{n+1}$ & $\hat k_s^I-x^{n+1}$ \\ 
$k_o^I$ & internal load & shell-ocean & $y_5^I(R_o)-1$ & $-1$ & $\hat k_o^I-1$ \\ 
$k_c^I$ & internal load & core-ocean & $y_5^I(R_c)-(y/x)^n$ & $-(y/x)^n$ & $\hat k_c^I-(y/x)^n$ \\ 
\hline
\end{tabular}
\label{TableLove}
\end{table}%

In the limit of a fluid shell, loads at the top and bottom of the shell are not supported by elastic stresses.
Instead, they `float' and `sink' into the shell, resulting in the full compensation of their gravitational signal (see Appendix~\ref{AppendixFluidCrust}):
\begin{equation}
\lim_{\mu_0\rightarrow0}\Gamma_{jn} = 0 \, ,
\end{equation}
where $\mu_0$ is a reference shear modulus for the whole shell.
More precisely, the depth-dependent shear modulus of the shell is written as $\mu(r) = \mu_0 f(r)$ where $\mu_0$ is an arbitrary reference value (for example the shear modulus at the surface) and $f(r)$ is a nondimensional radial function.

In the fluid shell limit, Love numbers tend to well-defined values called \textit{fluid-crust Love numbers} \citep{beuthe2015b}.
For tidal forcing, fluid-crust Love numbers differ from \textit{fluid Love numbers} (i.e.\ the Love numbers of a completely fluid body), because the elastic core tidally deforms under the ocean and the fluid shell.
For loads applied on the shell, there is no such distinction because the fully compensated loads do not perturb the ocean or the core.
For this reason, fluid-crust Love numbers are identified here with fluid Love numbers and denoted by
\begin{equation}
\left( h_j^{J\circ} , k_j^{J\circ}  \right) = \lim_{\mu_0\rightarrow0} \left( h_j^J , k_j^J \right) , 
\label{FCLove}
\end{equation}
where $j=(s,o,c)$ and $J=(L,I)$.
On an interface carrying a load, the fluid radial Love numbers have the value required to cancel the initial elevation in Eqs.~(\ref{ShapeSurf})-(\ref{ShapeBot}):
\begin{eqnarray}
h_s^{L\circ} &=& -1/\xi_{sn} \, ,
 \\
h_o^{I\circ} &=& -1/(x \Delta\xi_n) \, ,
\label{hoI0}
\end{eqnarray}
whereas they are zero elsewhere (proof given in Appendix~\ref{AppendixFluidCrust}).
Thus the shape of all interfaces is zero, in agreement with the fact that a fluid shell cannot support loads.
Similarly, the fluid gravitational Love numbers cancel the direct gravitational effect of the load.
Table~\ref{TableLove} lists the fluid Love numbers at the interfaces of a 3-layer model.

Each Love number can be written as the sum of the fluid Love number and a remainder called in this paper \textit{deviatoric Love number}, which vanishes in the fluid limit:
\begin{eqnarray}
h_j^J =  h_j^{J \circ} + \hat h_j^J \, ,
 \\
k_j^J =  k_j^{J \circ} + \hat k_j^J \, ,
\label{LoveDecomp}
\end{eqnarray}
where $j=(s,o,c)$ and $J=(L,I)$.
Since the fluid gravitational Love numbers are opposite in sign to the direct gravitational effect, the gravitational deviatoric Love numbers measure the total gravitational perturbation.

At the shell boundaries, the shape (Eqs.~(\ref{ShapeSurf})-(\ref{ShapeBot})) and the total gravitational perturbation (Eqs.~(\ref{GravSurf})-(\ref{GravBot})) have a simple form in terms of deviatoric Love numbers:
\begin{eqnarray}
H_{jn} &=& \frac{1}{g_s} \left( \hat h_j^L \, U_{n}^L + \hat h_j^I \, U_n^I \right) ,
\label{ShapeDev} \\
\Gamma_{jn} &=& \hat k_j^L \, U_{n}^L + \hat k_j^I \, U_{n}^I \, .
\label{GravDev}
\end{eqnarray}
These formulas make it obvious that the shape and total gravitational perturbation tend to zero on all interfaces in the fluid limit.

As we will see later, elastic isostasy can be studied in the fluid limit.
With that purpose in mind, I expand 
deviatoric Love numbers around their zero fluid value:
\begin{equation}
\left( \hat h_j^J , \hat k_j^J \right) = \left( \mu_0 \, \dot h_j^J , \mu_0 \, \dot k_j^J \right) + {\cal O}(\mu_0^2) \, ,
\label{LoveExpansion}
\end{equation}
where $(\dot h_j^J,\dot k_j^J)$ denote the partial derivatives of Love numbers (full or deviatoric, it does not matter) with respect to the reference shear modulus $\mu_0$ and evaluated at $\mu_0=0$:
\begin{equation}
\left( \dot h^J_j ,  \dot k^J_j \right) = \Big( \partial_{\mu_0} h^J_j , \partial_{\mu_0} k^J_j \Big)\Big|_{\mu_0=0} \, .
\label{LovePartial}
\end{equation}

\section{Isostatic ratios with Love}
\label{SectionIsostaticRatios}

\subsection{Definitions}
\label{IsostaticRatiosDefinitions}

\textit{Isostatic ratios} are nondimensional quantities characterizing the output of a model of Airy isostasy.
When analyzing data, the only isostatic ratio that matters is the one relating the gravitational and shape perturbations observed at the surface.
When building isostatic models, it is convenient to define other isostatic ratios which are not as easily measurable (though they could in principle be observed if detailed radar or seismic data were available).
Their multiplicity does not pose a problem because they can easily be transformed into each other (see Eq.~(\ref{LoadingRatioSn}) and Section~\ref{IsostaticRelations}).

The first of these is the \textit{shape ratio}, defined in the harmonic domain as the ratio of the shape of the shell-ocean boundary to the surface shape:
\begin{equation}
S_n = \frac{H_{on}}{H_{sn}} \, .
\label{ShapeRatio}
\end{equation}
A quantity related to the shape ratio but depending also on the gravitational perturbation is the \textit{topographic ratio}, which is defined in the harmonic domain as the ratio of the topography (shape minus local geoid) of the shell-ocean boundary to the surface topography:
\begin{equation}
T_n = \frac{H_{on}-\Gamma_{on}/g_o}{H_{sn}-\Gamma_{sn}/g_s} \, .
\label{TopoRatio}
\end{equation}
Contrary to the compensation factor and (to a lesser extent) the shape ratio, the topographic ratio cannot be directly observed but, if it can be computed analytically, the result is much simpler than for the shape ratio or compensation factor.
For example, the topographic ratio in zero deflection isostasy does not depend on the internal structure below the ocean top layer.

The \textit{potential-shape admittance} $Z_n$ is defined in the harmonic domain as the ratio of the total gravitational potential perturbation to the surface shape:
\begin{equation}
Z_n = \frac{\Gamma_{sn}}{H_{sn}} \, .
\label{Admittance}
\end{equation}
If different harmonic orders are present, the admittance is more generally defined as the ratio of cross-powers $\langle\Gamma_{sn}H_{sn}\rangle/\langle{}H_{sn}H_{sn}\rangle$ where $\langle{}A_{sn}B_{sn}\rangle=\sum_mA_{snm}B_{snm}$ in a real spherical harmonic basis, but this definition reduces to Eq.~(\ref{Admittance}) in predictions of Airy isostasy because surface and bottom loads are in phase.
In the literature, the admittance often denotes the transfer function between the gravity anomaly and either the shape or the topography \citep{turcotte2014,wieczorek2015}.
This should not cause any confusion here, because we will work instead with the \textit{compensation factor} $F_n$.
It is the nondimensional equivalent of the admittance, varying between $F_n=0$ (full compensation) and $F_n=1$ (no compensation, see below):
\begin{equation}
F_n = \frac{1}{g_s \xi_{sn}} \, Z_n \, .
\label{CompensFactor}
\end{equation}
Finally, the \textit{gravitational ratio} is defined in the harmonic domain as the ratio of the total gravitational perturbation at the shell-ocean boundary to the same quantity at the surface:
\begin{equation}
G_n = \frac{\Gamma_{on}}{\Gamma_{sn}} \, .
\label{GraviRatio}
\end{equation}
This ratio, though not common, is useful in classical isostasy (see Section~\ref{SectionUnperturbedOcean}).

\subsection{Shell of finite strength}

The isostatic ratios defined in Section~\ref{IsostaticRatiosDefinitions} will now be expressed in terms of the deviatoric Love numbers and of
the \textit{loading ratio} $\zeta_n$, which is the yet-to-be-determined ratio of internal load to surface load:
\begin{equation}
\zeta_n = \frac{U_n^I}{U_n^L} \, .
\label{LoadingRatio}
\end{equation}
Using Eqs.~(\ref{ShapeDev})-(\ref{GravDev}), I can write the shape and topographic ratios as
\begin{eqnarray}
S_n &=& \frac{ \hat h_o^L + \zeta_n \, \hat h_o^I}{\hat h_s^L + \zeta_n \, \hat h_s^I} \, ,
\label{ShapeRatioLove} \\
T_n &=& \frac{1}{\gamma_o} \, \frac{ \hat t^{\, L}_o + \zeta_n \, \hat t^{\, I}_o}{ \hat t^{\, L}_s + \zeta_n \, \hat t^{\,I} _s} \, ,
\label{TopoRatioLove}
\end{eqnarray}
where
\begin{equation}
\hat  t^{\, J}_j = \gamma_j \, \hat  h^{J}_j - \hat  k^{J}_j \, .
\label{tHat}
\end{equation}
Doing the same for the compensation factor and the gravitational ratio, I write
\begin{eqnarray}
F_n &=& \frac{1}{\xi_{sn}} \, \frac{ \hat k_s^L + \zeta_n \, \hat k_s^I }{ \hat h_s^L + \zeta_n \, \hat h_s^I } \, ,
\label{CompensFactorLove} \\
G_n &=& \frac{ \hat k_o^L + \zeta_n \, \hat k_o^I }{ \hat k_s^L + \zeta_n \, \hat k_s^I } \, .
\label{GraviRatioLove}
\end{eqnarray}
An uncompensated surface load means no internal load ($\zeta_n=0$) and no deformation ($h_s^L=k_s^L=0$ so that $\hat h_s^L=1/\xi_{sn}$ and $\hat k_s^L=1$), in which case $F_n=1$.

The shape ratio and compensation factor diverge if the surface shape vanishes, which occurs for a particular combination of surface and internal loads:
\begin{equation}
\zeta^{sing}_{n} = - \hat h_s^L/\hat h_s^I \, .
\label{zetaSingular}
\end{equation}
This problem is not specific to isostasy, as it occurs whenever there is a combination of surface and internal loads.
One should be aware that singularities can occur and take the necessary steps to exclude them by restricting the range of admissible boundary conditions.

Love numbers depend on the unperturbed internal structure and can be considered as known, whereas the loading ratio must be determined from an additional constraint, or \textit{isostatic prescription}.
In classical isostasy, the isostatic prescription sometimes directly specifies the shape ratio.
In that case, the loading ratio can be obtained by inverting Eq.~(\ref{ShapeRatioLove}):
\begin{equation}
\zeta_n = - \frac{\hat h_o^L - S_n \, \hat h_s^L}{\hat h_o^I - S_n \, \hat h_s^I} \, .
\label{LoadingRatioSn}
\end{equation}
The topographic ratio, the gravitational ratio, and the compensation factor can be directly related to the shape ratio by substituting Eq.~(\ref{LoadingRatioSn}) into Eqs.~(\ref{TopoRatioLove})-(\ref{GraviRatioLove}).

\subsection{Isostatic relations}
\label{IsostaticRelations}

With some more assumptions, gravitational perturbations can be computed in terms of the deformations of the shell boundaries, meaning that gravitational Love numbers can be dispensed with.
The assumptions are twofold: the body should be incompressible and stratified in (possibly many) layers of homogeneous density, so that density perturbations only occur at the interfaces between layers;
the shell should be of uniform density, so that the inside of the shell does not perturb the gravity field.
The procedure is detailed in Appendix~\ref{AppendixRelationLoveIncompressible}.
The result can be written in terms of deviatoric Love numbers ($J=L$ or $I$):
\begin{equation}
\left(
\begin{array}{c}
\hat k_s^J \\
\hat k_o^J
\end{array}
\right)
=
\left(
\begin{array}{ll}
a & b  \\
c & b \, x^{-n-1}
\end{array}
\right)
\left(
\begin{array}{c}
\hat h_s^J \\
\hat  h_o^J
\end{array}
\right) ,
\label{khRelation}
\end{equation}
where the coefficients $(a,b,c)$, as well as an auxiliary coefficient $d$ used below, are given in Table~\ref{TableCoeff}.
These coefficients will turn out to be useful to compute minimum stress isostasy from zero deflection isostasy.
The corresponding equation for non-deviatoric Love numbers is given by Eq.~(\ref{khRelationNonDev}).
All gravitational perturbations originating below the shell are included in the parameter $K_{n}$, which is known analytically if the body is made of three incompressible homogeneous layers (see Table~\ref{TableCoeff} and Appendix~\ref{AppendixRelationLoveIncompressible}).
Since the number of layers is arbitrarily large, Eq.~(\ref{khRelation}) remains valid if density varies continuously with depth within the core and ocean, but it becomes laborious to compute $K_n$ analytically.
The parameter $K_n$ can instead be computed from the gravitational and radial Love numbers (which have been computed numerically) by inverting Eq.~(\ref{khRelation}).
The first relation yields
\begin{equation}
K_n = \frac{\hat k_s^J - \xi_{sn} \, \hat h_s^J - x^{n+2} \, \Delta\xi_n \, \hat h_o^J}{\xi_{sn} \, \hat h_s^J + x^{1-n} \, \Delta\xi_n \, \hat h_o^J} \, ,
\label{KnLove}
\end{equation}
which can be computed with $J=L$ or $J=I$ since $(a,b,c)$ do not depend on $J$ in Eq.~(\ref{khRelation}).

The Saito-Molodensky relation (Eq.~(\ref{SMrelation})), combined with Eq.~(\ref{khRelation}), gives a constraint on radial Love numbers:
\begin{equation}
\left( a -1\right) h_s^I + b\, h_o^I = x \, \Big( c \, h_s^L - d \, h_o^L \Big) \, .
\label{SMrelationRadial}
\end{equation}
A similar relation holds for deviatoric Love numbers.
 
\begin{table}[htp]
\ra{1.3}
\small
\caption[Coefficients relating gravitational and radial perturbations of shell boundaries]{
Coefficients relating gravitational and radial perturbations of shell boundaries in incompressible models with a shell of uniform density.
The parameter $K_{n}$ includes all contributions from below the shell (Eqs.~(\ref{defAelastic})-(\ref{Kc1}));
$f_n$ is a degree-dependent number (Eq.~(\ref{fndef})) while $\gamma_c$ and $\Delta\xi'_n$ are given by Eqs.~(\ref{gammac})-(\ref{CoreOceanDensContrast}).
$K_{n}=0$ if the ocean is homogeneous and the core is infinitely either rigid/non-deformable ($\bar \mu_{\rm c}\rightarrow\infty$), or point-like ($y\rightarrow0$).
}
\begin{center}
\begin{tabular}{@{}ll@{}}
\hline
\multicolumn{2}{l}{Homogeneous shell:}\\
$a$ & $\xi_{sn} \left(1+K_{n}\right)$ \\
$b$ &  $\Delta\xi_n \, x^{n+2}\left( 1 + K_{n} \, x^{-2n-1} \right)$ \\
$c$ & $\xi_{sn} \, x^{n} \left( 1+K_{n}\,x^{-2n-1} \right)$ \\
$d$ & $\gamma_o - b \, x^{-n-1}$ \\
\multicolumn{2}{l}{If homogeneous shell, ocean, and core:}\\
$K_{n}$ & $\Delta\xi'_n \, y^{2n+2} / \left(\gamma_c + f_n \, \bar \mu_{\rm c}/ (y\,\Delta\xi'_1) - y \, \Delta \xi'_n \right)$ \\
\hline
\end{tabular}
\end{center}
\label{TableCoeff}
\end{table}%

Substituting Eq.~(\ref{khRelation}) into Eqs.~(\ref{TopoRatioLove})-(\ref{CompensFactorLove}), I can relate the various isostatic ratios:
\begin{eqnarray}
\gamma_o T_n \, = \,  \frac{ - d \, S_n + c }{ b \, S_n  -(1-a)}
\hspace{7mm}
&\Leftrightarrow&
\hspace{3mm}
S_n \, \stackrel{\mathrm{*}}{=} \, \frac{ (1-a) \left(\gamma_o T_n \right) +  c }{ b \left(\gamma_o T_n \right) + d }  \, ,
\label{TnSnRelation} \\
G_n \,=\, \frac{ b\, x^{-n-1} \, S_n + c}{ b \, S_n + a}
\hspace{6mm}
&\Leftrightarrow&
\hspace{3mm}
S_n \,=\, \frac{-a \, G_n + c}{b \, G_n - b \, x^{-n-1}} \, ,
\label{GnSnRelation} \\
\xi_{sn} F_n \,=\, b \, S_n + a
\hspace{17mm}
&\Leftrightarrow&
\hspace{3mm}
S_n \,=\, \frac{\xi_{sn} F_n - a}{b} \, ,
\label{FnSnRelation} \\
 \gamma_o T_n = \frac{- d \left(\xi_{sn} F_n \right) + a d + b c }{b \left(\xi_{sn} F_n \right) - b }
&\Leftrightarrow&
\xi_{sn}F_n \stackrel{\mathrm{*}}{=} \frac{b \left( \gamma_o T_n\right) + a d + b c}{ b \left( \gamma_o T_n\right) + d}  \, .
\label{TnFnRelation}
\end{eqnarray}
Starred equalities are not valid at degree one because the transformations giving $T_1$ in terms of $S_1$ and $F_1$ cannot be inverted (see Section~\ref{DegreeOneCompensation}).

\subsection{Degree-one compensation}
\label{DegreeOneCompensation}

The incompressible model with a shell of uniform density (Section~\ref{IsostaticRelations}) is helpful to understand what happens at degree one.
Since degree-one deformations include rigid translations, an additional boundary condition is required in order to fix the frame.
In particular, the centre-of-mass frame (typically used to describe the planetary shape) is specified by the condition of zero gravitational perturbation at the surface, $\Gamma_{s1}=0$ \citep{grefflefftz1997}, which means that there is full compensation:
\begin{equation}
F_1 = 0 \, .
\label{F1}
\end{equation}
Contrary to the compensation factor (or the shape ratio), the topographic ratio does not depend on the frame because it is defined as a height difference: shape minus geoid.
The dependence on $S_1$ and $F_1$ in Eqs.~(\ref{TnSnRelation}) and (\ref{TnFnRelation}) is thus only apparent.
This results from the property
\begin{equation}
\left. \frac{d}{b} \right|_{n=1}
=\left.  \frac{c}{1-a} \right|_{n=1}
= \left. \frac{ad+bc}{b} \right|_{n=1} \, ,
\label{property}
\end{equation}
which can be solved to yield
\begin{equation}
K_{1} = \frac{1-\xi_{s1}-x^3\Delta\xi_1}{\xi_{o1}} \, .
\label{K1}
\end{equation}
For example, this property holds for the model with three homogeneous layers (Eq.~(\ref{Kc1}))).
Eqs.~(\ref{TnSnRelation}) and (\ref{TnFnRelation}) thus yield the same frame-independent result whatever the values of $S_1$ and $F_1$:
\begin{equation}
\gamma_o \, T_1
= - \left. \frac{d}{b} \right|_{n=1}
= - \frac{\xi_{s1}}{\Delta\xi_1} \, \frac{1}{x^2} \, .
\label{T1}
\end{equation}
The degree-one shape ratio, computed with Eq.~(\ref{FnSnRelation}), does not depend on the loading ratio:
\begin{equation}
S_1 = \left. - \frac{a}{b} \right|_{n=1}
=  - \frac{\xi_{s1}}{\Delta\xi_1} \, \frac{1+\Delta\xi_1 \, (1-x^3)}{1-\xi_{s1} \, (1-x^3)} \, ,
\label{S1}
\end{equation}
where the second equality results from substituting Eq.~(\ref{K1}) into $a$ and $b$.

These results are consistent with the use of degree-one Love numbers.
If applied separately on surface loads and internal loads, the centre-of-mass condition can be written $\hat k_s^L=\hat k_s^I=0$.
Eq.~(\ref{khRelation}) can then be solved for $(\hat h_o^J,\hat k_o^J)$ in terms of $\hat h_s^J$, and the results are substituted into Eqs.~(\ref{ShapeRatioLove})-(\ref{TopoRatioLove}).
The frame-independence of the topographic ratio is guaranteed by the frame-independence of the four combinations $\gamma_j \hat h_j^J - \hat k_j^J$.
Alternatively, $S_1$ and $T_1$ can be derived without introducing surface and internal loads, by solving Eqs.~(\ref{GammaHRelation1}) and (\ref{GammaHRelation3}) with the constraints $\Gamma_{s1}=0$ (centre-of-mass frame) and $\Gamma_{c1}=g_cH_{c1}$ (equipotential core surface) \citep{trinh2019}.
Since the isostatic ratios are the same whatever the relative amount of surface and internal loads, there is no need for an isostatic prescription at degree one.
Isostasy truly starts at degree two.

\subsection{$\mu$-invariance and fluid limit}
\label{SectionFluidCrust}

While it is convenient to treat isostasy as an elastic loading problem, we will see that the isostatic prescriptions of elastic isostasy have the interesting property of $\mu$-\textit{invariance}, meaning that the isostatic ratios are invariant under a global rescaling of the shear modulus of the shell.
If the shell is homogeneous, this invariance means that isostasy does not depend on the uniform shear modulus of the shell.
Thus, it does not matter whether the shell is soft or very rigid: the partition between surface and internal loads is different in the two cases, but it has no effect on the shape ratio and the compensation factor.
If in addition the shell is incompressible, it is sufficient to establish the $\mu$-invariance of the shape ratio, because Eqs.~(\ref{GnSnRelation})-(\ref{TnFnRelation}) imply that the other ratios $(G_n,F_n,T_n)$ are invariant as well.
This conclusion can be generalized to incompressible bodies with stratified ocean and core.

If $\mu$-invariance holds, the various isostatic ratios can be evaluated in the fluid limit according to L'H\^{o}pital's rule, by expanding Love numbers around $\mu_0=0$ (Eq.~(\ref{LoveExpansion})).
This procedure has two advantages.
It makes it easier to derive analytical formulas, and it clarifies the connection between the elastic, viscous, and viscoelastic approaches (see Paper~II).
The fluid limit of the loading ratio is denoted
\begin{equation}
\zeta_n^\circ \equiv \lim_{\mu_0\rightarrow0} \zeta_n \, .
\label{LoadingRatioFluidLimit}
\end{equation}
Under the assumption of $\mu$-invariance, the shape ratio, topographic ratio, and compensation factor are equal to
\begin{eqnarray}
S_n &=& \frac{\dot h_o^L + \zeta_n^\circ \, \dot h_o^I}{\dot h_s^L + \zeta_n^\circ \, \dot h_s^I} \, ,
\label{ShapeRatio0} \\
T_n &=& \frac{1}{\gamma_o} \, \frac{ \dot t^{\, L}_o + \zeta_n^\circ \, \dot t^{\, I}_o }{ \dot t^{\, L}_s + \zeta_n^\circ \, \dot t^{\, I}_s } \, ,
\label{TopoRatio0} \\
F_n &=& \frac{1}{\xi_{sn}} \, \frac{ \dot k_s^L + \zeta_n^\circ \, \dot k_s^I }{ \dot h_s^L + \zeta_n^\circ \, \dot h_s^I } \, ,
\label{CompensFactor0}
\end{eqnarray}
where the partial derivatives $(\dot h_j^J,\dot k_j^J)$ are defined by Eq.~(\ref{LovePartial}) and $\dot t^{\, J}_j=\gamma_j\,\dot h_j^J-\dot k_j^J$.
These expressions remain valid if $\zeta_n^\circ$ diverges.
The shape ratio and compensation factor diverge if the surface shape is zero (see Eq.~(\ref{zetaSingular})), which occurs if the fluid loading ratio takes the value
\begin{equation}
\zeta_{n}^{\circ} \rightarrow - \dot h_s^L/\dot h_s^I \, .
\label{zetaSingular0}
\end{equation}
A similar divergence exists for the topographic ratio.

\subsection{Thin shell limit}
\label{SectionThinShellLimit}

Isostatic compensation often occurs at shallow depths compared to the body radius.
In that case, the thin shell expansion in the parameter $\varepsilon=1-x\rightarrow0$ provides an interesting point of comparison for all models.
The accuracy of the expansion, however, decreases at short wavelength because the expansion parameter is actually
\begin{equation}
n \, \varepsilon \cong 2\pi \, d_s/\lambda \ll 1 \, .
\end{equation}
In the thin shell limit, Love numbers evaluated at the shell boundaries tend to
\begin{eqnarray}
\lim_{\varepsilon\rightarrow0} h^J_j &=& - 1/\xi_{on} \, ,
\label{hThinShell} \\
\lim_{\varepsilon\rightarrow0} k^J_j &=& -1
\hspace{3mm}
\Rightarrow
\hspace{3mm}
 \lim_{\varepsilon\rightarrow0} \hat k^J_j \,\, = \,\, 0 \, .
\label{kThinShell} 
\end{eqnarray}
These limits can be computed from the membrane formulas for $(h^L_s,k^L_s)$ (Eqs.~(78) and (81) of \citet{beuthe2015b}).
The other Love numbers have the same limit because there is no difference between top and bottom of shell, and between surface and internal loads, given that the shell has zero thickness.
Expanding the isostatic ratios (Eqs.~(\ref{ShapeRatioLove})-(\ref{CompensFactorLove})) to zeroth order in $\varepsilon$, I get
\begin{eqnarray}
S_n &\cong& T_n \,\, \cong \,\, - \frac{\xi_{s1}}{\Delta\xi_1} + {\cal O}(\varepsilon) \, ,
\\
F_n &\cong& {\cal O}(\varepsilon) \, .
\end{eqnarray}
Thus, the thin shell expansion should be done to first order in $\varepsilon$ otherwise the compensation factor vanishes.
This last property is explained by the fact that isostatic geoid anomalies measure the dipole moment of the density distribution \citep{turcotte2014} and are thus second-order quantities in the perturbating layer thickness \citep{dahlen1982}, whereas shape anomalies are of first order.
This is also the reason why the compensation factor is much more sensitive  to the details of the isostatic model than the shape and topographic ratios.

\section{Isostasy without Love}
\label{SectionIsostasyWithoutLove}

The following isostatic prescriptions were originally obtained without using Love numbers, hence their denomination `Isostasy without Love'.

\subsection{Naive isostasy}
\label{SectionNaive}

\subsubsection{Equal mass}

The \textit{equal-mass} assumption states that conical columns have equal mass \citep{lambert1930,vening1946,heiskanen1958}:
\begin{equation}
\rho_s \, H_{sn} + x^2 \left( \rho_o-\rho_s \right) H_{on} = 0 \, .
\end{equation}
The shape ratio directly follows:
\begin{equation}
S_n = - \frac{\xi_{s1}}{\Delta\xi_1} \, \frac{1}{x^2} \, .
\label{SnEM}
\end{equation}
If the shell and ocean are homogeneous and incompressible and if the core is infinitely rigid, substituting Eq.~(\ref{SnEM}) into Eq.~(\ref{FnSnRelation}) with $K_{n}=0$ yields the classical `equal-mass' formula for the compensation factor ($n\geq2$):
\begin{equation}
F_n = 1 - x^n
\cong n \, \varepsilon \, ,
\label{fnEM}
\end{equation}
where the approximation holds in the thin shell limit.

\subsubsection{Equal weight}

A variant of the equal mass model consists in imposing that conical columns have \textit{equal weight}, where the weight is computed from the topography \citep{cadek2019}:
\begin{equation}
\rho_s \left( g_s \, H_{sn} - \Gamma_{sn} \right) + x^2 \left( \rho_o-\rho_s \right) \left( g_o \, H_{on} - \Gamma_{on} \right) = 0 \, .
\end{equation}
The topographic ratio directly follows:
\begin{equation}
T_n = - \frac{\xi_{s1}}{\Delta\xi_1} \, \frac{1}{\gamma_o} \, \frac{1}{x^2} \, .
\label{TnEW}
\end{equation}
This formula is fortuitously the same as the degree-one topographic ratio (Eq.~(\ref{T1})).

If the shell and ocean are homogeneous and incompressible and if the core is infinitely rigid, Eq.~(\ref{TnFnRelation}) yields for $n\geq2$:
\begin{equation}
F_n = 1 - x^n \, \frac{1-\xi_{sn}-\Delta\xi_n \, x^{n+2}}{\gamma_o - \xi_{sn} \, x^n - \Delta\xi_n \, x}
\cong \frac{n+2-3\, \xi_{o1}}{1-\xi_{on}} \, \varepsilon \, ,
\label{fnEW}
\end{equation}
where the approximation holds in the thin shell limit.

\subsubsection{Equal (lithostatic) pressure}

The assumption of \textit{equal (lithostatic) pressure} means that `the pressure on deeper surfaces is not affected by the presence of the combined topography and compensation' \citep{vening1946} (see also \citet{lambert1930,heiskanen1958,hemingway2017}).
The pressure means here the lithostatic (i.e.\ hydrostatic) pressure of the shell taking into account the deviation from the spherical shape at both shell boundaries:
\begin{equation}
\rho_s \, g_s \, H_{sn} + \left( \rho_o-\rho_s \right) g_o \, H_{on} = 0 \, .
\label{condEP}
\end{equation}
The shape ratio directly follows:
\begin{equation}
S_n = - \frac{\xi_{s1}}{\Delta \xi_1} \, \frac{1}{\gamma_o} \, .
\label{SnEP}
\end{equation}
If the shell and ocean are homogeneous and incompressible and if the core is infinitely rigid, substituting Eq.~(\ref{SnEP}) into Eq.~(\ref{FnSnRelation}) with $K_{n}=0$ yields the standard `equal (lithostatic) pressure' formula for the compensation factor:
\begin{equation}
F_n = 1 - \frac{x^{n+2}}{\gamma_o}
\cong \left( n + 4 - 3 \, \xi_{s1} \right) \varepsilon \, ,
\label{fnEP}
\end{equation}
where the approximation holds in the thin shell limit.
A variant of this model consists in replacing shape by topography in Eq.~(\ref{condEP}) \citep{cadek2019}.

\subsubsection{Equal (spherical) pressure, or unperturbed ocean}
\label{SectionUnperturbedOcean}

Beside the equality of lithostatic pressure, the assumption of equal pressure has also been understood as meaning that the ocean is unperturbed \citep{garland1965,dahlen1982}: isobaric surfaces, which are also equipotential surfaces in a fluid, are spherical.
Thus the total gravitational perturbation (and the gravitational ratio) vanishes at the bottom of the shell (Eqs.~(\ref{GravDev}) and (\ref{GraviRatio})):
\begin{equation}
\Gamma_{on} = 0
\,\, \Leftrightarrow \,\,
G_n = 0  \, .
\label{zetaUO}
\end{equation}
If the shell is incompressible and of homogenous density, the shape and compensation ratios follow from Eqs.~(\ref{GnSnRelation})-(\ref{FnSnRelation}):
\begin{eqnarray}
S_n &=& - \frac{\xi_{s1}}{\Delta \xi_1} \, x^{n-1} \, ,
\label{SnUO}\\
F_n &=& 1 - x^{2n+1}
\cong \left( 2n+1 \right) \varepsilon \, ,
\label{fnUO}
\end{eqnarray}
where the approximation holds in the thin shell limit.
In this model, all perturbations vanish below the shell so that the core remains spherical.

\subsection{Thin shell (minimum stress) isostasy}
\label{DahlenIsostasy}

\citet{dahlen1982} was motivated to find a more solid theoretical foundation for the relation between the isostatic geoid anomaly $N=\Gamma_s/g_s$ and the underlying dipole density distribution, which can be written as
\begin{equation}
\Gamma_{sn} = 2\pi G \, \nu \int_d z \, \delta\rho_n \, dz \, ,
\end{equation}
where $\nu=1$ in the textbook derivation (Eq.~(5.144) of \citet{turcotte2014}).
Dahlen works with a highly idealized model based on the following assumptions: the shell is thin, the anomalies are of long wavelength, the density of the ocean and shell are the same, the core is point-like, and the isostatic equilibrium is purely local.
The last condition means that there is no shear stress between vertical columns (i.e.\ $\sigma_{r\theta}=\sigma_{r\phi}=0$).
Using the scalar representation of tensors on the sphere \citep{backus1967}, Dahlen finds a family of solutions depending on a parameter (denoted here $\nu$) fixed by the choice of an isostatic prescription.
Thus, various isostatic prescriptions result in different $\nu$ values.
Dahlen proves that the conditions of equal (spherical) pressure, equal mass, and minimum deviatoric stress imply respectively that $\nu=2$, $\nu=2n/(2n+1)$, and $\nu=\nu^{\rm min}$ with
\begin{equation}
\nu^{\rm min} = \frac{2}{1-\xi_{on}} \, \left( \frac{(2n-1)(n+2)^2}{(2n+1)(2n^2+2n-1)} - \xi_{on} \right) .
\end{equation}
I will now translate these results into Airy isostatic ratios.
Given that the observed surface shape is the same whatever the isostatic prescription (denoted $A$ or $B$), the ratio of the compensation factors for two different prescriptions is equal to the ratio of the corresponding gravitational perturbations: $F_n^A/F_n^B=\Gamma_{sn}^A/\Gamma_{sn}^B=\nu^A/\nu^B$.
Combining the values of $\nu$ given above with the thin shell limit of Eq.~(\ref{fnEM}) (or Eq.~(\ref{fnUO})), I get the compensation factor for Dahlen's minimum stress isostasy:
\begin{equation}
F_n^{\rm TSI} \cong \varepsilon \, \frac{2n+1}{2} \, \nu^{\rm min} \, .
\label{CompensFactorTSI}
\end{equation}
The approximate equality ($\cong$) reminds us of the fact that this result only holds in the thin shell limit.
As I will show later, this approximation plays a more important role in Dahlen's isostasy than the assumptions of local compensation and minimum stress.
For that reason, Dahlen's approach is labelled here `TSI', for `Thin Shell Isostasy'.

The corresponding shape ratio and topographic ratio are obtained from Eqs.~(\ref{FnSnRelation})-(\ref{TnFnRelation}) with $K_c=0$ (the core is point-like):
\begin{eqnarray}
S_n^{\rm TSI} &=& - \frac{\xi_{s1}}{\Delta\xi_1} \left( 1 + \varepsilon \, \frac{n-1}{2n+1-3\xi_{o1}} \left( 3 \xi_{o1} - \frac{ \left(2n+1\right) \left(n+2\right)}{2n^2+2n-1} \right) \right) ,
\label{ShapeRatioTSI} \\
T_n^{\rm TSI} &\cong& - \frac{\xi_{s1}}{\Delta\xi_1} \, \frac{1}{\gamma_o} \left( 1 + \varepsilon \, \frac{3n \left( n+1 \right)}{2n^2+2n-1} \right) .
\label{TopoRatioTSI}
\end{eqnarray}
Since the factors multiplying $\varepsilon$ are of order unity, the shape and topographic ratios are nearly constant functions of harmonic degree in this approximation.
Eq.~(\ref{TopoRatioTSI}) was previously obtained by \citet{cadek2019} using an Eulerian approach.

While Dahlen's approach is not based on Love numbers, it is a consistent treatment of elastic isostasy.
For the same interior model (homogenous shell and ocean, infinitely rigid core), we will see that the isostatic ratios for a thick shell with isotropic elasticity (that is, regional compensation) reduce to Eqs.~(\ref{CompensFactorTSI})-(\ref{TopoRatioTSI}) in the thin shell limit.
Remarkably, the assumption that isostatic compensation is regional or local does not matter in that limit.
This can be understood by noting that vertical shear stresses $(\sigma_{r\theta}, \sigma_{r\phi})$ are negligible in a thin shell, which is approximately in a state of plane stress.
The minimum stress condition is not essential either, because the zero deflection prescription has the same thin shell limit (see Section~\ref{ZeroDeflectionIsostasy}).

\subsection{Comparison of isostatic prescriptions}
\label{SectionComparison}

\begin{figure}
\centering
    \includegraphics[width=\textwidth]{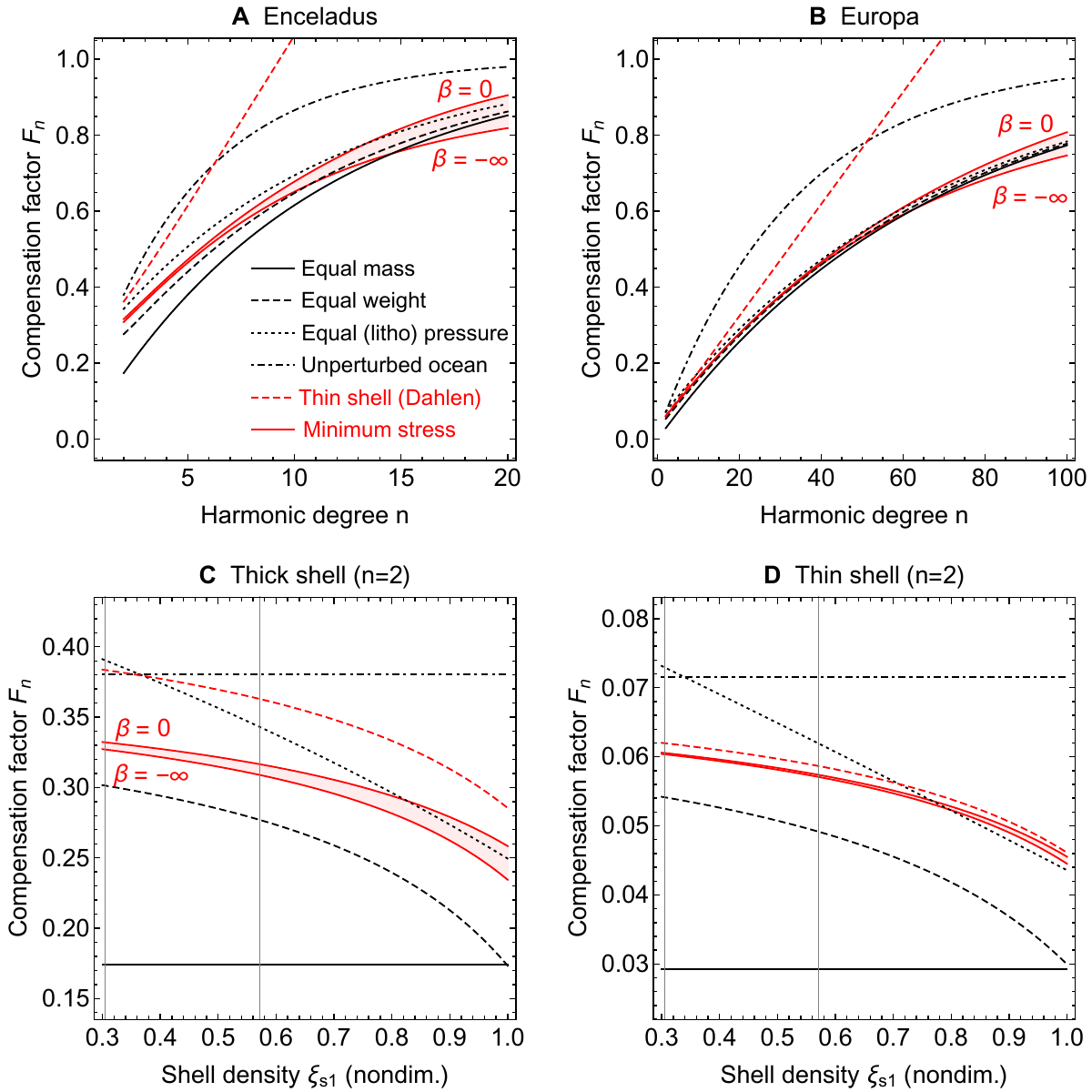}
   \caption[Compensation factor for different isostatic prescriptions]{
   Compensation factor for different isostatic prescriptions as a function of harmonic degree (A,B) or nondimensional shell density (C,D).
   In panel (A), interior parameters correspond to a simplified 3-layer model of Enceladus with an infinitely rigid core \citep{beuthe2016} ($\rho_s=920\rm\,kg/m^3$; $\rho_o=1020\rm\,kg/m^3$; $\rho_b=1610\rm\,kg/m^3$; $d=23\rm\,km$; $R_s=252.1\rm\,km$).
   In panel (B), interior parameters are the same except that the bulk density and the surface radius are those of Europa ($\rho_b=3013\rm\,kg/m^3$; $R_s=1560.8\rm\,km$).
   Panels (C,D) are drawn for hypothetical bodies with the ratio $d/R_s$ of Enceladus (panel C, thick shell) or Europa (panel D, thin shell); in both cases the ocean density is 10\% higher than the shell density.
   Vertical lines show the nondimensional shell densities of Enceladus and Europa.
   The shaded area shows the predictions of minimum stress isostasy for the whole range of the isostatic family parameter $\beta$ (Section~\ref{CrustFiniteStrengthMSI}).
 }
   \label{FigComparisonfn}
\end{figure}

The compensation factors for the isostatic prescriptions `without Love' are easily compared in the thin shell limit (see Eqs.~(\ref{fnEM}), (\ref{fnEW}), (\ref{fnEP}), (\ref{fnUO}), and (\ref{CompensFactorTSI})).
They can differ at low degree by more than a factor of two \citep{dahlen1982}.
At degree 2, the compensation factor for unperturbed ocean isostasy is $2.5$ times larger than if it is computed with equal mass isostasy.
Results for other prescriptions depend on density.
For example, if the density of the crust and of the ocean are approximately 2/3 of the bulk density (not far from Enceladus's case), the compensation factors for equal weight isostasy and for equal (lithostatic) pressure isostasy are respectively $5/3$ and 2 times the result for equal mass isostasy.

Fig.~\ref{FigComparisonfn} shows the compensation factor for the different isostatic prescriptions, as a function of harmonic degree (panels A, B) or shell density (panels C, D) for Enceladus-like and Europa-like bodies.
Equal mass isostasy generally gives the lowest predictions, whereas unperturbed-ocean isostasy gives the highest ones.
Equal weight and equal pressure (lithostatic) isostasy are in-between, as is Dahlen's thin shell isostasy if the harmonic degree is not too high.
These prescriptions differ widely at low harmonic degree, but converge at high harmonic degree.

For reference, the plots show the minimum stress isostatic family (shaded area).
At high degree, minimum stress isostasy tends to 1 as the other prescriptions, but the speed of convergence depends on the isostatic family parameter $\beta$ (Section~\ref{CrustFiniteStrengthMSI}, Eq.~(\ref{MinStressCons1})).
Panels C and D show that equal-pressure (lithostatic) isostasy predicts a degree-two compensation factor that is too high by 8 to 11\% for Enceladus (the exact value depending on the choice of the isostatic family parameter) and by 20\% for Europa (the discrepancy increases as the body is more stratified in density).
Similar errors occur when inferring the shell thickness from the compensation factor: the error is 0\% if the core is made of ice (e.g.\ Tethys), 10\% if there is a hydrated silicate core (e.g.\ Enceladus), and 20\% if the core is metallic (e.g.\ Europa).

\section{Elastic isostasy}
\label{ElasticIsostasyMSI}

Elastic isostasy means choosing an isostatic prescription constraining the stress, strain or displacement of the shell.
Before discussing minimum stress isostasy, I will briefly describe \textit{zero deflection isostasy} (ZDI for short), which is a model of elastic isostasy defined by a constraint on the radial displacement of the shell boundaries.
This model gives a shortcut to computing minimum stress isostasy, and plays a key role in the relation between elastic and viscous isostasy.
I will then compute stress invariants in terms of Love numbers, and use this formulation to build up minimum stress isostasy.

\subsection{Zero deflection isostasy}
\label{ZeroDeflectionIsostasy}

The motivation for zero deflection isostasy (ZDI) is twofold:
\begin{itemize}
\item Minimum stress isostasy (MSI) can be computed in terms of ZDI (which is much easier) if the shell is homogeneous and incompressible. In particular, it makes it possible to find MSI isostatic ratios in analytical form.
\item ZDI is mathematically equivalent to stationary viscous isostasy (see Paper~II). With this equivalence, it is possible to establish a precise relation between MSI and viscous isostasy.
\end{itemize}
Because of its importance, the ZDI model is investigated in detail in Appendix~\ref{AppendixZDI}.

\begin{figure}
\centering
    \includegraphics[width=0.5\textwidth]{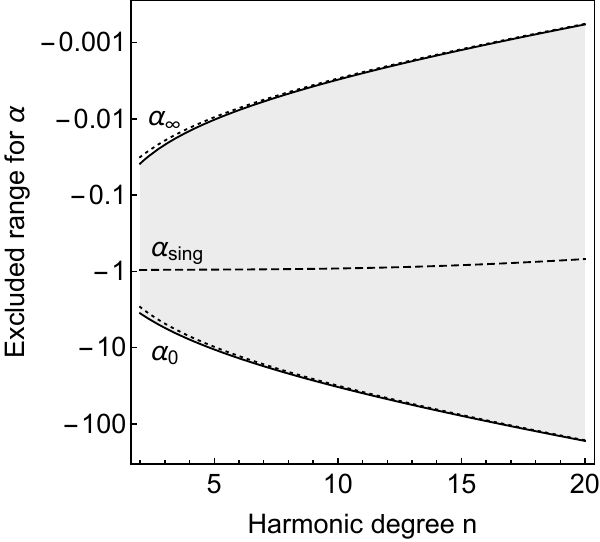}
   \caption[Zero deflection isostasy: excluded range of the isostatic family parameter]{
   Zero deflection isostasy: excluded range of the isostatic family parameter $\alpha$ as a function of the harmonic degree (shaded area).
   Solid curves show the bounds $\alpha_0$ and $\alpha_\infty$ while dotted curves show their asymptotic limits.
   The value $\alpha_{\rm sing}$ for which the shape ratio and compensation factor are singular (dashed curve) belongs to the excluded range.
   The interior model is the one used for Enceladus in Fig.~\ref{FigComparisonfn}.
   See Appendix~\ref{AppendixZDIalpha} for more details.
 }
   \label{FigAlpha01}
\end{figure}

When isostasy is formulated as an elastic loading problem, it seems natural to require that the shell be deformed as little as possible.
For example, one could minimize the volume-integrated second invariant of the deviatoric strain, but this procedure is not fundamentally different from stress minimization. 
Another choice is to set to zero the radial displacement at the surface (or surface deflection, see Eq.~(\ref{wn})) as done by \citet{banerdt1982}.
Constraining the surface instead of the bottom of the shell is an arbitrary choice.
For this reason, I define zero deflection isostasy as a one-parameter family by imposing that the deformations of the top and bottom boundaries are in a constant ratio:
\begin{equation}
w_{n}(R_s)  + \alpha \, w_{n}(R_o) = 0
\,\, \Leftrightarrow \,\,
\zeta_n^{\rm ZDI}
= - \frac{h_s^L + \alpha \, h_o^L}{h_s^I + \alpha \, h_o^I} \, ,
\label{zetaZDI}
\end{equation}
where $\alpha$ is a real number which may depend on the harmonic degree.
The cases $\alpha=0$ and $\alpha=\pm\infty$ correspond to zero deflection at the surface and at the bottom of the shell, respectively.
The isostatic ratios follow from substituting $\zeta_n^{\rm ZDI}$ into Eqs.~(\ref{ShapeRatioLove})-(\ref{CompensFactorLove}).
Alternatively, the loading ratio can be evaluated numerically before substitution.
The ZDI shape ratio reads
\begin{equation}
S_n^{\rm ZDI} = \frac{ X_{os} + \alpha X_{oo} }{ X_{ss} +  \alpha X_{so} } \, ,
\label{SnZDI}
\end{equation}
where
\begin{equation}
X_{jk} = \hat h_j^L \, h_k^I - \hat h_j^I \, h_k^L \, .
\label{Xjkdef}
\end{equation}
If the body is incompressible and has a shell of uniform density, the other isostatic ratios are related to $S_n$ by Eqs.~(\ref{TnSnRelation})-(\ref{FnSnRelation}).
Some values of $\alpha$ lead to non-isostatic models, for example when the surface shape vanishes (Eqs.~(\ref{zetaSingular}) and (\ref{alphaSing})).
The duality between ZDI and MSI constrains the admissible range for $\alpha$ (Eq.~(\ref{ExcludedRange})).
Fig.~\ref{FigAlpha01} shows that the excluded range tends to cover the whole negative axis as the harmonic degree becomes large.

The shape ratio is $\mu$-invariant if the terms $X_{jk}$ scale by the same factor under a global rescaling of the shear modulus of the shell, that is if
\begin{equation}
\partial_{\mu_0} \left( X_{jk}/X_{ss} \right) = 0
\, ,
\label{BilinearInvariant}
\end{equation}
where $j, k=s$ or $o$ and $\mu_0$ is the reference shear modulus of Section~\ref{FluidDeviatoricLoveNumbers}.
These conditions can be checked analytically for an incompressible body with three homogeneous layers (see Eqs.~(\ref{IdentityAB})-(\ref{IdentityBB}), Eqs.~(\ref{Xjs})-(\ref{Xjo}), and complementary software), and numerically for a model with depth-dependent shell rheology.
Let us then conjecture that Eq.~(\ref{BilinearInvariant}) is verified whatever the interior model, meaning that $\mu$-invariance is a general property of zero deflection isostasy.
If $\mu$-invariance holds, the shape ratio and the compensation factor can be computed in the fluid limit (Eq.~(\ref{LoadingRatioFluidLimit})).
Substituting the fluid Love numbers of Table~\ref{TableLove} into Eq.~(\ref{zetaZDI}), I obtain the ZDI fluid loading ratio:
\begin{equation}
\zeta_n^{ \circ \rm ZDI} = - \frac{\Delta\xi_1}{\xi_{s1}} \, \frac{x}{\alpha} \, .
\label{zetaZDI0}
\end{equation}
I assume that $\alpha$ is a non-zero finite number; the cases $\alpha=0$ and $\alpha\rightarrow\pm\infty$ can be treated separately but the results are consistent with taking the limits $\alpha\rightarrow0$ and $\alpha\rightarrow\infty$ of Eq.~(\ref{zetaZDI0}).
The isostatic ratios follow from substituting $\zeta_n^{ \circ \rm ZDI}$ into Eqs.~(\ref{ShapeRatio0})-(\ref{CompensFactor0}).

The ZDI topographic ratio is independent of the internal structure below the shell (save for the density contrast between the shell and the top of the ocean).
I checked this property analytically for the incompressible body with three homogeneous layers (see Section~\ref{ZDI3layers}), and numerically for models with ocean and core stratified in two layers.
Similarly to $\mu$-invariance, I will conjecture that this property is generally valid.
For an incompressible body with three homogeneous layers, the ZDI topographic ratio is given by a simple analytical formula (Eq.~(\ref{TopoRatioZDI})), from which the shape ratio and compensation factor can easily be generated (Appendix~\ref{AppendixIsoRatiosIncompressible}).
If the core is rigid, the ZDI isostatic ratios have the same thin shell limit as Dahlen's thin shell isostasy (Eq.~(\ref{TopoRatioZDIthinshell})).

\subsection{Stress and energy}

\subsubsection{Stress and strain invariants}

The zero deflection prescription is easy to understand but lacks a robust physical basis  in the context of elastic equilibrium, because setting to zero the displacement looks more like an initial constraint than the outcome of a physical process.
If isostasy is the final state obtained after the shell broke in places where the stress is maximum, it makes sense to try to minimize the stress within the shell.
One possibility consists in minimizing the maximum value of the deviatoric stress within the shell \citep{jeffreys1932}.
In practice, it is easier to minimize the average over the shell volume of the second invariant of the deviatoric stress \citep{jeffreys1943,dahlen1982}.
Besides facilitating computations, it has the advantage of being related to the strain energy (see below).
The second invariant of the deviatoric stress is given by
\begin{equation}
\sigma_{\rm II} = \frac{1}{2} \, \sigma'_{ij} \, \sigma'_{ij} = 2 \mu^2 \, \epsilon'_{ij} \, \epsilon'_{ij} \, ,
\label{sigma2}
\end{equation}
where there is an implicit sum over repeated indices; the prime denotes that stress and strain are deviatoric.

A closely related approach consists in minimizing the elastic energy.
The elastic strain energy density $\sigma_{ij}\epsilon_{ij}/2$ can be split into a part ${\cal E}_\mu$ depending on deviatoric stresses and strains, called shear energy or distortional strain energy, and a part ${\cal E}_\kappa$ depending on the volume change, called compressional energy (\citet{jaeger2007}, Section 5.8; \citet{dahlen1999}, Eq.~(8.29)):
\begin{eqnarray}
{\cal E}_\mu &=& \frac{1}{2} \, \sigma'_{ij} \, \epsilon'_{ij} = \mu \, \epsilon'_{ij} \, \epsilon'_{ij} \, ,
\label{Emu} \\
{\cal E}_\kappa &=& \frac{1}{2} \left( \frac{1}{3} \, \sigma \, \epsilon \right) = \frac{1}{2} \, \kappa \, \epsilon^2 \, ,
\label{Ekappa}
\end{eqnarray}
where $\sigma$ and $\epsilon$ are the traces of $\sigma_{ij}$ and $\epsilon_{ij}$, respectively.

I choose to neglect the compressional energy (which vanishes anyway in the incompressible limit).
With this choice, the energy approach is closely related to the stress approach because $\sigma_{\rm II}=2\mu{\cal E}_\mu$.
If the shell is homogeneous, it is equivalent to minimize the volume integral of $\sigma_{\rm II}$ or of ${\cal E}_\mu$.

The energy associated with the perturbation due to a load of harmonic degree $n$ can be computed in the same way as the energy associated with free oscillations or with tidal deformations.
Since \citet{alterman1959}, it is common to express the strain, through the displacement, in vector spherical harmonics (spheroidal-toroidal decomposition, see Appendix~12 of \citet{dahlen1999}).
Since isostatic loads do not cause toroidal deformations, the strain depends on two spheroidal scalars associated with radial and tangential displacements (Appendix~\ref{AppendixVariablesyi}).
The second invariant of the deviatoric stress (or the energy density) can then be expressed in terms of spheroidal scalars and scalar differential operators, before being integrated over the volume of the shell (tidal energy is computed in such a way in \citet{beuthe2013}).
This procedure yields the same result as the computation of the elastic energy of spheroidal free oscillations from a variational principle \citep{pekeris1958,takeuchi1972,dahlen1999}, except for the presence of the forcing potential (as for tidal energy, see \citet{tobie2005}).
For example, Eq.~(8.128) of \citet{dahlen1999}, multiplied by the square of the degree-$n$ forcing potential $U_n$, gives the shear energy stored in the shell:
\begin{equation}
E_{shear} =  \frac{1}{2} \left( \int_d \, \mu \, {\cal H}_\mu \, dr \right) \big( U_n \big)^2 \, ,
\label{defEshear}
\end{equation}
where
\begin{equation}
{\cal H}_\mu = \frac{4}{3} \Big( r \partial_r y_1 - y_1 - \frac{\delta_n}{2} y_3 \Big)^2 - \delta_n \Big( r \partial_r y_3 - y_3 + y_1 \Big)^2 + \delta_n (\delta_n+2) \big( y_3 \big)^2 \, ,
\label{defHmu}
\end{equation}
in which $\delta_n=-n(n+1)$ is the eigenvalue of the spherical Laplacian.
The spheroidal scalars are denoted $(y_1,y_3)$ as in \citet{takeuchi1972} because the notation $(U,V)$ used by \citet{dahlen1999} would create confusion with the loading potential.
Spherical harmonics are real and normalized to one.
If different harmonic orders contribute to degree $n$, the terms of mixed order cancel in the angular integration and the squared potential should be interpreted as a sum of squares: $\big( U_n \big)^2=\sum_m\big( U_{nm} \big)^2$.

\subsubsection{Loading basis}

In Airy isostasy, the full solution is a superposition of the two solutions associated with the surface load and the internal load.
The substitution $y_i\,U_n\rightarrow{}y_i^L\,U_n^L+y_i^I\,U_n^I$ in Eq.~(\ref{defEshear}) yields the shear energy stored in the shell:
\begin{equation}
E_{shear} =  \sum_{J,K}  E_\mu^{JK} \, \bar U^J_{n} \, \bar U^{K}_{n} \, ,
\end{equation}
where $\bar U^J_{n}=U^J_{n}/(g_s R_s)$ are the nondimensional load potentials ($J=L,I$; same for $K$).
The energy coefficients are given by
\begin{equation}
E_\mu^{JK} =  \frac{(g_s R_s)^2}{2} \int_d \, \mu \, {\cal H}^{JK}_\mu \, dr \, ,
\label{defEJKmu}
\end{equation}
where
\begin{equation}
{\cal H}^{JK}_\mu = f^{JK}_A + f^{JK}_B + f^{JK}_C \, ,
\label{defHJKmu}
\end{equation}
in which
\begin{eqnarray}
f^{JK}_A &=& \frac{4}{3} \Big( r \partial_r y^J_1 - y^J_1 - \frac{\delta_n}{2} y^J_3 \Big) \Big( r \partial_r y^{K}_1 - y^{K}_1 - \frac{\delta_n}{2} y^{K}_3 \Big) ,
\\
f^{JK}_B &=& - \delta_n \Big( r \partial_r y^J_3 - y^J_3 + y^J_1 \Big) \Big( r \partial_r y^{K}_3 - y^{K}_3 + y^{K}_1 \Big) \, ,
 \\
f^{JK}_C &=& \delta_n (\delta_n+2) \, y^J_3 \, y^{K}_3 \, .
\label{weights}
\end{eqnarray}
The functions $f^{JK}_{A,B,C}$ are actually the contributions of the three basic spatial patterns \citep{beuthe2013}, which are of interest if one needs to map the spatial variations of the energy within the shell.
The function $f^{JK}_B$ can always be written without derivatives as $- \delta_n{}r^2y^J_4 y^K_4/\mu^2$, where $y_4$ is the scalar for the radial-tangential shear stress (Appendix~\ref{AppendixVariablesyi}).
If the material is incompressible, $f^{JK}_A$ reduces to $12 (y^J_1 + (\delta_n/2) y^J_3)(y^{K}_1 + (\delta_n/2) y^{K}_3)$.
Thus, the derivatives of the functions $y_i^J$ are not required to compute the energy if the shell is incompressible.

Eq.~(\ref{defEJKmu}) can be generalized to group the minimum stress and minimum energy approaches:
\begin{equation}
E^{JK}_p =  \frac{p}{2}\, (g_s R_s)^2 \int_d \, \mu^p \, {\cal H}^{JK}_\mu \, dr \, ,
\label{defEJKmup}
\end{equation}
where $p=1$ for the shear energy and $p=2$ for the second invariant of the deviatoric stress.

If the shear modulus of the shell is uniform, the factor $\mu^p$ can be taken out of the integral.
In that case, it is possible to express $E^{JK}_\mu$ in terms of the partial derivatives of Love numbers with respect to $\mu$ (see Appendix \ref{AppendixVariationLove}) using the following identities:
\begin{eqnarray}
\int_d \, {\cal H}^{LJ}_\mu \, dr &=& \chi \, \partial_\mu \Big( h_s^J - k_s^J \Big) \, , 
\label{LoveDeriv1} \\
 \int_d \, {\cal H}^{IJ}_\mu \, dr &=& \chi \, x \, \partial_\mu \Big( \gamma_o \, h_o^J - k_o^J \Big) \, ,
\label{LoveDeriv2}
\end{eqnarray}
where $J=L$ or $I$ and $\chi=(2n+1)R_s/(4 \pi G)$.
The identity $\int_d \, {\cal H}^{LI}_\mu \, dr=\int_d \, {\cal H}^{IL}_\mu \, dr$ is consistent with the Saito-Molodensky relation (Eq.~(\ref{SMrelation})).
If necessary, similar identities for the compressional energy can be obtained by substituting $\kappa$ to $\mu$.
These identities remain valid if the material is transversely isotropic, which is useful when assuming local compensation.
The partial derivatives with respect to $\mu$ can be computed explicitly if Love numbers are known in analytical form, otherwise they should be evaluated numerically. 
Identities (\ref{LoveDeriv1})-(\ref{LoveDeriv2}) are the static analogue of the eigenfrequency perturbation formulas of seismic theory; ${\cal H}^{JK}_\mu$ is called a Fr\'echet kernel \citep{dahlen1999}.

\subsubsection{Shape basis}

In the loading basis, the shear energy stored in the shell ($p=1$) or the second stress invariant integrated over the volume of the shell ($p=2$) reads
\begin{equation}
E_{shear,p} =  
E^{LL}_p \left( \bar U^L_n \right)^2 + 2 \, E^{LI}_p \, \bar U^L_n \, \bar  U^I_n + E^{II}_p\left( \bar  U^I_n \right)^2 .
\label{ShearEnergyLoad}
\end{equation}
Using Eq.~(\ref{ShapeDev}), I write it as a quadratic form in the surface and bottom shapes (using nondimensional shapes $\bar H_{jn}=H_{jn}/R_s$):
\begin{equation}
E_{shear,p} = E_{oo,p} \left(\bar H_{sn} \right)^2 - 2 \, E_{so,p} \, \bar H_{sn} \, \bar H_{on} + E_{ss,p} \left( \bar H_{on} \right)^2 ,
\label{ShearEnergyShape}
\end{equation}
where
\begin{equation}
E_{jk,p} = \frac{1}{N^2 } \left( \hat h_j^I \,\hat h_k^I \, E^{LL}_p  -  \left(  \hat h_j^L \, \hat h_k^I + \hat h_j^I \, \hat h_k^L \right) E^{LI}_p + \hat h_j^L \, \hat h_k^L \, E^{II}_p \right) ,
\label{Ejkp}
\end{equation}
and
\begin{equation}
N = \hat h_s^L \, \hat h_o^I - \hat h_s^I \, \hat h_o^L \, .
\label{defN}
\end{equation}
If the shell is homogeneous and incompressible, the coefficients $E_{jk,p}$ are proportional to $\mu^{p-2}$.
For the incompressible body with a homogeneous shell, this property follows from the necessary conditions for ZDI $\mu$-invariance (see Appendix~\ref{AppendixFluidCrustLoadingRatio} and complementary software).
If the shapes of the shell boundaries are given, the shear energy is thus inversely proportional to $\mu$ whereas the volume-integrated stress invariant is independent of $\mu$.

\begin{figure}
    \includegraphics[width=\textwidth]{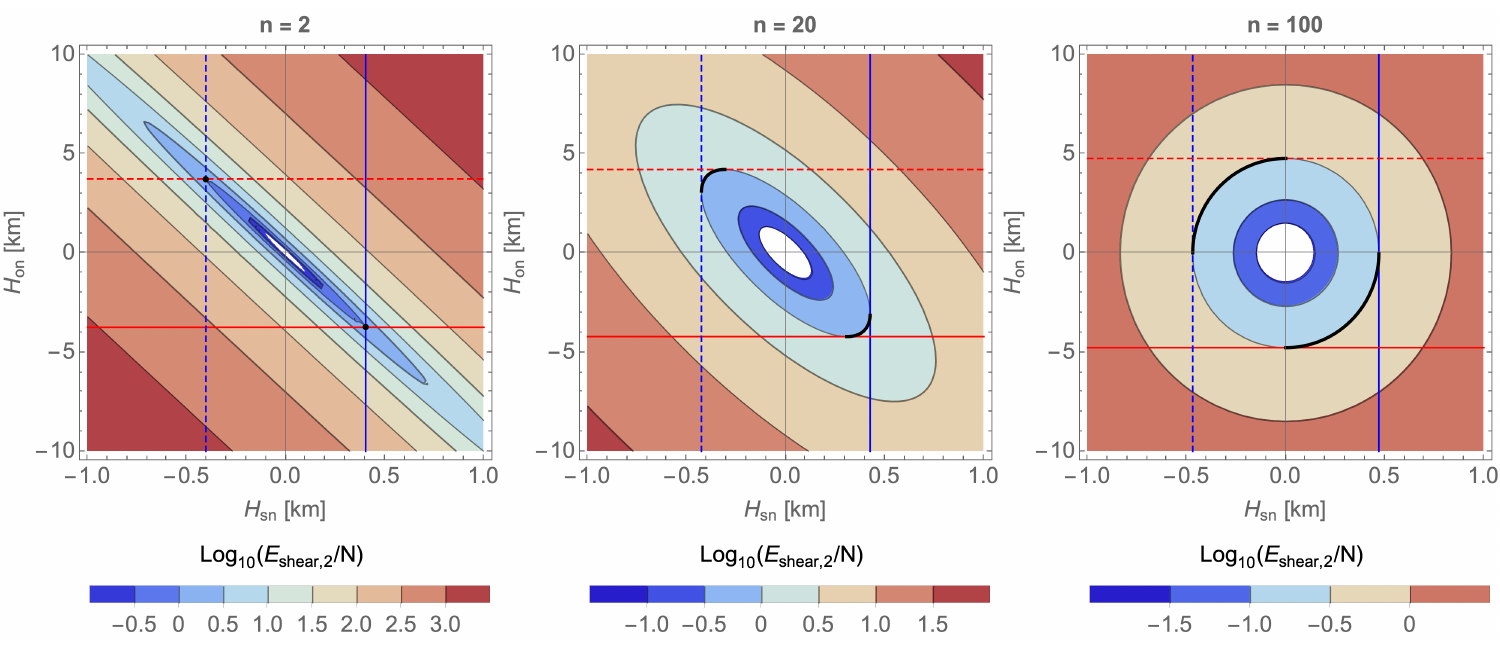}
   \caption[Volume-integrated stress invariant as a function of the surface and bottom shapes]{
   Volume-integrated stress invariant as a function of the surface and bottom shapes.
    The three panels correspond to harmonic degrees 2, 20, and 100.
   The contour scale is logarithmic and the volume-integrated stress invariant is nondimensionalized by $N=10^{-9}(Mg_s)^2/R_s$ where $M$ is the mass of the body.
   These examples are computed for an incompressible 3-layer body with homogeneous shell/ocean and infinitely rigid core.
   Interior parameters are those of Enceladus used in Fig.~\ref{FigComparisonfn}.
   Straight blue (resp.\ red) lines illustrate the constraint of constant surface (resp.\ bottom) shape; the surface shape is either positive (solid lines) or negative (dashed lines).
   Thicker parts (in black) of the contours show the other solutions having the same volume-averaged stress invariant and satisfying the condition $\beta\leq0$ (Eq.~(\ref{MinStressCons1})).
 }
   \label{FigEnergyABC}
\end{figure}

\subsection{Minimum stress isostasy}

\subsubsection{Shell of finite strength}
\label{CrustFiniteStrengthMSI}

The total shear energy can be seen as a quadratic form in the loads (Eq.~(\ref{ShearEnergyLoad})) or in the shapes (Eq.~(\ref{ShearEnergyShape})).
Its absolute minimum occurs if the loads vanish ($U^L_n=U^I_n=0$) or if the shapes vanish ($H_{cn}=H_{on}=0$), which is the unperturbed state.

An additional constraint must be imposed in order to find a non-trivial minimum.
\citet{beuthe2016} supposed the constancy of the surface load, which is a good approximation if the shell deforms very little (as in Enceladus's case), but this constraint does not represent an observable quantity.
A better choice consists in imposing the surface shape (which can be observed from space) and determining the bottom shape minimizing the energy \citep{trinh2019}.
Another possibility consists in imposing the bottom shape (which can in principle be detected by radar or from seismic data) and determining the surface load minimizing the energy.
In parallel to what was done for zero deflection isostasy, I consider the more general case where a linear combination of the surface shape and the bottom shape is held constant:
\begin{equation}
H_{sn} + \beta H_{on} = H_{\beta n} \, ,
\label{MinStressCons1}
\end{equation}
where $\beta\leq0$ and $H_{\beta n}$ is a constant.
The surface shape is held constant if $\beta=0$ whereas the bottom shape is held constant if $\beta\rightarrow-\infty$.
At this stage, it is not obvious how to give physical meaning to non-zero finite values of $\beta$, but we will see that they are needed to relate minimum stress isostasy to other isostatic models.
Positive values of $\beta$ are excluded because they can lead to manifestly non-isostatic models, in which the shapes of the shell boundaries have the same sign, or in which the shape of either one of the boundaries vanishes (see Fig.~\ref{FigEnergyABC}).

In the loading basis, the constraint (\ref{MinStressCons1}) reads
\begin{equation}
\left( \hat h_s^L + \beta \, \hat h_o^L \right) U_{n}^L+ \left( \hat h_s^I + \beta \, \hat h_o^I \right) U_n^I = g_s H_{\beta n} \, .
\label{MinStressCons2}
\end{equation}
Minimizing the energy or the stress in the loading basis (Eq.~(\ref{ShearEnergyLoad})) under the constraint (\ref{MinStressCons2}) yields the loading ratio:
\begin{equation}
\zeta_n^{\rm MSI}
= \frac{ \left( \hat h_s^I \, E^{LL}_p - \hat h_s^L \, E^{LI}_p \right)  + \beta \left( \hat h_o^I \, E^{LL}_p - \hat h_o^L \, E^{LI}_p \right) }{ \left( \hat h_s^L \,  E^{II}_p - \hat h_s^I \, E^{LI}_p \right) + \beta \left(  \hat h_o^L \, E^{II}_p - \hat h_o^I \, E^{LI}_p \right)  } \, ,
\label{zetaMING}
\end{equation}
where MSI stands for Minimum Stress Isostasy.
The shape ratio can be computed by substituting the loading ratio into Eq.~(\ref{ShapeRatioLove}):
\begin{equation}
S^{\rm MSI}_n = \frac{ E_{so,p} + \beta \, E_{oo,p} }{ E_{ss,p} + \beta \, E_{so,p} } \, .
\label{SnMING}
\end{equation}
The same result follows from minimizing the energy or the stress in the shape basis (Eq.~(\ref{ShearEnergyShape})) under the constraint (\ref{MinStressCons1}).
If the shell is homogeneous and incompressible, the $\mu$-invariance of the shape ratio (as well as the other isostatic ratios) follows from the necessary conditions for ZDI $\mu$-invariance (see Appendix~\ref{AppendixFluidCrustLoadingRatio}).
In all generality, the compensation factor is computed by substituting Eq.~(\ref{zetaMING}) into Eq.~(\ref{CompensFactorLove}), but it is quicker to use Eq.~(\ref{FnSnRelation}) for an incompressible body with three homogeneous layers.

\subsubsection{Fluid limit}
\label{MinimumStressFluidLimit}

Recall that $\mu$-invariance implies that isostatic ratios can be evaluated in the fluid limit.
In addition, let us assume that the body is incompressible and has a homogeneous shell.
The MSI fluid loading ratio can then be written as
\begin{equation}
\zeta_n^{\circ \rm MSI} =
\frac{ \left( \dot h_s^I \, H^{LL}_0 - \dot h_s^L \, H^{LI}_0 \right)  + \beta \left( \dot h_o^I \, H^{LL}_0 - \dot h_o^L \, H^{LI}_0 \right) }
{ \left( \dot h_s^L \, H^{II}_0 - \dot h_s^I \, H^{LI}_0  \right) + \beta \left(  \dot h_o^L \, H^{II}_0 -  \dot h_o^I \, H^{LI}_0 \right)  } \, ,
\label{zetaMING0Appendix}
\end{equation}
where $H^{JK}_0$ results from the fluid limit of Eq.~(\ref{defEJKmup}),
\begin{equation}
H^{JK}_0 = \frac{1}{\chi} \left[ \int_d \, {\cal H}^{JK}_\mu \, dr \right]_{\mu=0} \, ,
\label{defEJK0}
\end{equation}
in which $\chi=(2n+1)R_s/(4 \pi G)$.
The expressions within brackets can be transformed into derivatives of Love numbers with Eqs.~(\ref{LoveDeriv1})-(\ref{LoveDeriv2}).
Gravitational Love numbers are then eliminated with Eq.~(\ref{khRelation}).
The resulting expressions are all proportional to $X_{so}^0 = \dot h_s^L \, \dot h_o^I - \dot h_s^I \, \dot h_o^L$:
\begin{eqnarray}
\dot h_s^I \, H^{LL}_0 - \dot h_s^L \, H^{LI}_0 
&=&  \dot h_s^I \left( \dot h_s^L - \dot k_s^L \right) - \dot h_s^L \left( \dot h_s^I - \dot k_s^I \right) 
\,\, = \,\,  b \,  X_{so}^0 \, ,
\\
\dot h_o^I \, H^{LL}_0 -  \dot h_o^L  \,  H^{LI}_0
 &=& \dot h_o^I \left(  \dot h_s^L - \dot k_s^L \right) - \dot h_o^L \left( \dot h_s^I - \dot k_s^I \right)
 \,\, = \,\,  \left( 1-a \right) X_{so}^0 \, ,
 \\
 \dot h_s^L \, H^{II}_0 -  \dot h_s^I \, H^{LI}_0
 &=& \dot h_s^L \, x \left( \gamma_o \, \dot h_o^I - \dot k_o^I \right) - \dot h_s^I \, x \left( \gamma_o \, \dot h_o^L - \dot k_o^L \right) 
\,\, = \,\,  d\, x \, X_{so}^0 \, ,
\\
\dot h_o^L  \, H^{II}_0 - \dot h_o^I \, H^{LI}_0
&=& \dot h_o^L \, x \left( \gamma_o \, \dot h_o^I - \dot k_o^I \right) - \dot h_o^I \, x \left( \gamma_o \, \dot h_o^L - \dot k_o^L \right)
\,\, = \,\,  c \, x \, X_{so}^0 \, .
\label{zetaMINGbrackets}
\end{eqnarray}
If the ocean and the core are homogeneous, the coefficients $(a,b,c,d)$ are given analytically in Table \ref{TableCoeff}; otherwise, they can be computed numerically with Eq.~(\ref{khRelation}) from the values of the gravitational and radial Love numbers.
In the fluid limit, the MSI loading ratio is thus given by
\begin{equation}
\zeta_n^{\circ \rm MSI} = \frac{1}{x} \, \frac{ b + \beta \left(1-a\right)}{ d+\beta \, c} \, .
\label{zetaMING0}
\end{equation}

\subsubsection{MSI-ZDI duality}
\label{MSIZDIDuality}

\begin{figure}
   \centering
    \includegraphics[width=\textwidth]{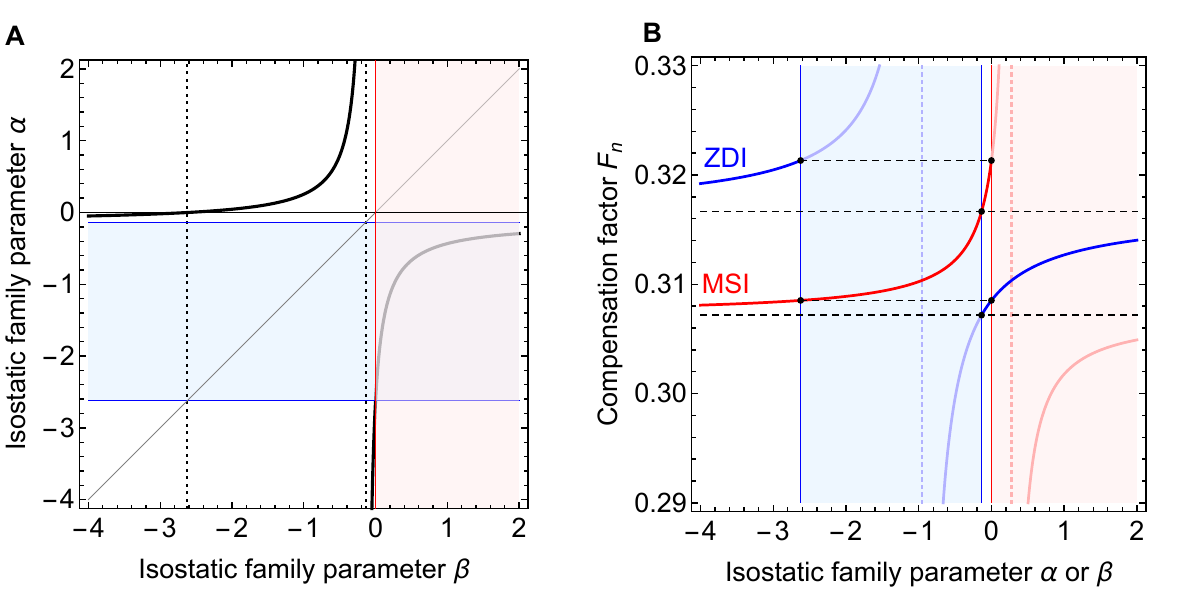}
   \caption[Relation between minimum stress isostasy and zero deflection isostasy]{
   (A) Relation between the isostatic family parameters $\beta$ (MSI) and $\alpha$ (ZDI), at degree 2 (solid black curve).
    (B) Compensation factor at degree 2 as a function of the isostatic family parameter $\alpha$ (solid blue curve) or $\beta$ (solid red curve).
   The interior model is an incompressible body with homogeneous shell and ocean and an infinitely rigid core
   with $(\xi_{s1},\xi_{o1},x)=(0.7,0.9,0.9)$.
   The excluded ranges $\alpha_0<\alpha<\alpha_\infty$ and $\beta>0$ are shaded in blue and red, respectively.
  In panel (A), vertical dotted lines show where $\alpha$ vanishes ($\beta=\alpha_0$) or diverges ($\beta=\alpha_\infty$),
     while horizontal dotted lines show where $\beta$ vanishes ($\alpha=\alpha_0$) or diverges ($\alpha=\alpha_\infty$).
        The graph is symmetric with respect to the diagonal.
  In panel (B), vertical dotted lines show where $F_n^{\rm ZDI}$ diverges ($\alpha=\alpha_{sing}$) and where $F_n^{\rm MSI}$ diverges ($\beta=\beta_{sing}$).
  Horizontal dashed lines relate values of $\alpha$ and $\beta$ yielding the same compensation factor.
   }
   \label{Figalphabeta}
\end{figure}

At a given harmonic degree, minimum stress isostasy gives the same result as zero deflection isostasy if their loading ratios are equal.
Equating Eqs.~(\ref{zetaZDI0}) and (\ref{zetaMING0}) and noting that $b/c=x^2\Delta\xi_1/\xi_{s1}$, I get a mapping between the isostatic family parameters $\alpha$ and $\beta$:
\begin{equation}
\alpha = - \frac{b}{c} \, \frac{c \, \beta + d}{\left(1-a\right) \beta + b}
\hspace{3mm} \Leftrightarrow \hspace{3mm}
\beta = - \frac{b}{c} \, \frac{c \, \alpha + d}{\left(1-a\right) \alpha + b} \, .
\label{alphabetaRelation}
\end{equation}
This function, being its own inverse, has a symmetric graph with respect to the line $\alpha=\beta$ (Fig.~\ref{Figalphabeta}).
Although the mapping between $\alpha$ and $\beta$ was derived in the fluid limit, it is valid whatever the value of the shear modulus of the shell because the parameters $(a,b,c,d)$ and $(\alpha,\beta)$ are all independent of $\mu$.
The two most interesting cases are:
\begin{eqnarray}
\beta = 0 \hspace{5mm} &\Leftrightarrow& \hspace{5mm} \alpha = \alpha_0 = - \frac{d}{c} \, ,
\label{beta0} \\
\beta = \pm \infty \hspace{5mm} &\Leftrightarrow& \hspace{5mm} \alpha = \alpha_\infty = - \frac{b}{1-a} \, .
\label{betainf}
\end{eqnarray}
By symmetry, $\alpha=0$ corresponds to $\beta=\alpha_0$ and $\alpha=\pm\infty$ corresponds to  $\beta=\alpha_\infty$ (see Fig.~\ref{FigAlpha01} for the dependence of $\alpha_0$ and $\alpha_\infty$ on harmonic degree).
Fig.~\ref{Figalphabeta} shows the relation between $\alpha$ and $\beta$ at degree two for a particular choice of interior parameters, as well as the MSI and ZDI degree-two compensation factors as a function of their isostatic family parameter.
At high harmonic degree, the relation between $\alpha$ and $\beta$ becomes
\begin{equation}
\alpha \sim - \frac{\Delta\xi_1}{\xi_{s1}} \, \frac{\gamma_o \,  x^2}{\beta} \, ,
\end{equation}
so that the MSI model with $\beta=0$ (resp.\ $\beta=-\infty$) becomes indistinguishable from the ZDI model with $\alpha=\infty$ (resp.\ $\alpha=0$).

The shape ratio and compensation factor diverge if Eq.~(\ref{zetaMING0}) coincides with the singular loading ratio (Eq.~(\ref{zetaSingular0})), which happens if $\beta$ takes the value
\begin{equation}
\beta_{sing}= \frac{b \, \dot h_s^I + d \, x \, \dot h_s^L}{(a-1) \, \dot h_s^I - c \, x \, \dot h_s^L}
= - \frac{b}{c} \, \frac{c \, \alpha_{sing} + d}{\left(1-a\right) \alpha_{sing} + b} \, .
\label{betaSing}
\end{equation}
The second equality can be checked by substituting $\alpha_{sing}$ from Eq.~(\ref{alphaSing}). 
As expected, the singular values of the isostatic family parameters $\alpha$ and $\beta$ are related by Eq.~(\ref{alphabetaRelation}).
I checked numerically that $\beta_{sing}$ is strictly positive for physically reasonable values of the interior parameters (i.e.\ $0<\xi_{s1}<\xi_{o1}\leq1$; note that $\beta_{sing}=0$ if $\xi_{s1}=\xi_{o1}$).
Thus, $\beta_{sing}$ does not pose a problem for the minimum stress isostatic family, for which $\beta\leq0$.

Using the mapping between $\alpha$ and $\beta$, the isostatic ratios in minimum stress/energy isostasy can be computed from those for zero deflection isostasy, for which an analytical solution exists if the body is made of three incompressible homogeneous layers:
\begin{equation}
\Big( S_n^{\rm MSI} , T_n^{\rm MSI} , F_n^{\rm MSI} \Big) = \left. \Big( S_n^{\rm ZDI}, T_n^{\rm ZDI} , F_n^{\rm ZDI} \Big) \right|_{\alpha=\alpha_{\rm MSI}} \, ,
\label{IsoRatioMSI}
\end{equation}
where
\begin{equation}
\alpha_{\rm MSI} =  - \frac{b}{c} \, \frac{ \beta \, c + d }{  \beta \left(1-a\right) + b  } \, .
\label{alphaMSI}
\end{equation}
Isostatic ratios for the model with three homogeneous layers are given in Appendix~\ref{AppendixIsoRatiosIncompressible}.

\subsection{Discussion}
\label{MSIproperties}

\begin{figure}
   \centering
    \includegraphics[width=0.98\textwidth]{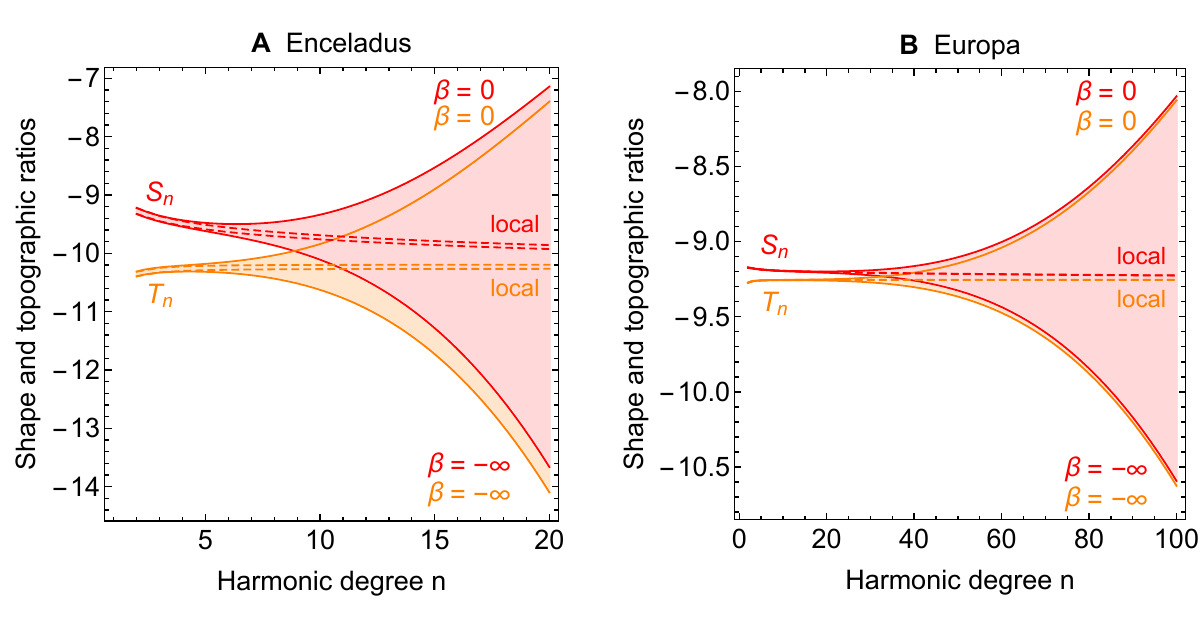}
   \caption[Minimum stress isostasy: shape and topographic ratios]{
   Minimum stress isostasy: shape and topographic ratios as a function of harmonic degree: (A) Enceladus; (B) Europa.
   Interior models are those used in Fig.~\ref{FigComparisonfn}.
   At high harmonic degree, both ratios tend to zero if $\beta=0$ whereas they diverge if $|\beta|$ is large.
   Dashed lines show the results under the assumption of local compensation (Appendix~\ref{AppendixMethodsLocal}).
   }
   \label{FigShapeMSI}
\end{figure}

Although the compensation factor is the only ratio which is actually observed, it is instructive to analyze how the shape (or topographic) ratio changes with harmonic degree (Fig.~\ref{FigShapeMSI}).
At low harmonic degree, the shape ratio is nearly the same whatever the value of the isostatic family parameter.
This can be understood with the thin shell approximation.
In the thin shell limit, isostatic ratios are the same in minimum stress/energy isostasy and in zero deflection isostasy, because the latter are independent of $\alpha$ up to first order in $\varepsilon=1-x$ (Eq.~(\ref{TopoRatioZDIthinshell})).
Thus, they also coincide with the isostatic ratios of Dahlen's model (Eq.~(\ref{TopoRatioTSI})) which were derived in the thin shell approximation.
As a corollary, the thin shell approximation erases the difference between keeping the surface shape or the bottom shape constant.
Moreover, the extension of the MSI topographic ratio to $n=1$ is independent of $\beta$ (see Eq.~(\ref{Tau1})) and predicts the correct degree-one topographic ratio (Eq.~(\ref{T1})), though the shape ratio and compensation factor cannot be similarly extended.

\begin{figure}
   \centering
    \includegraphics[width=\textwidth]{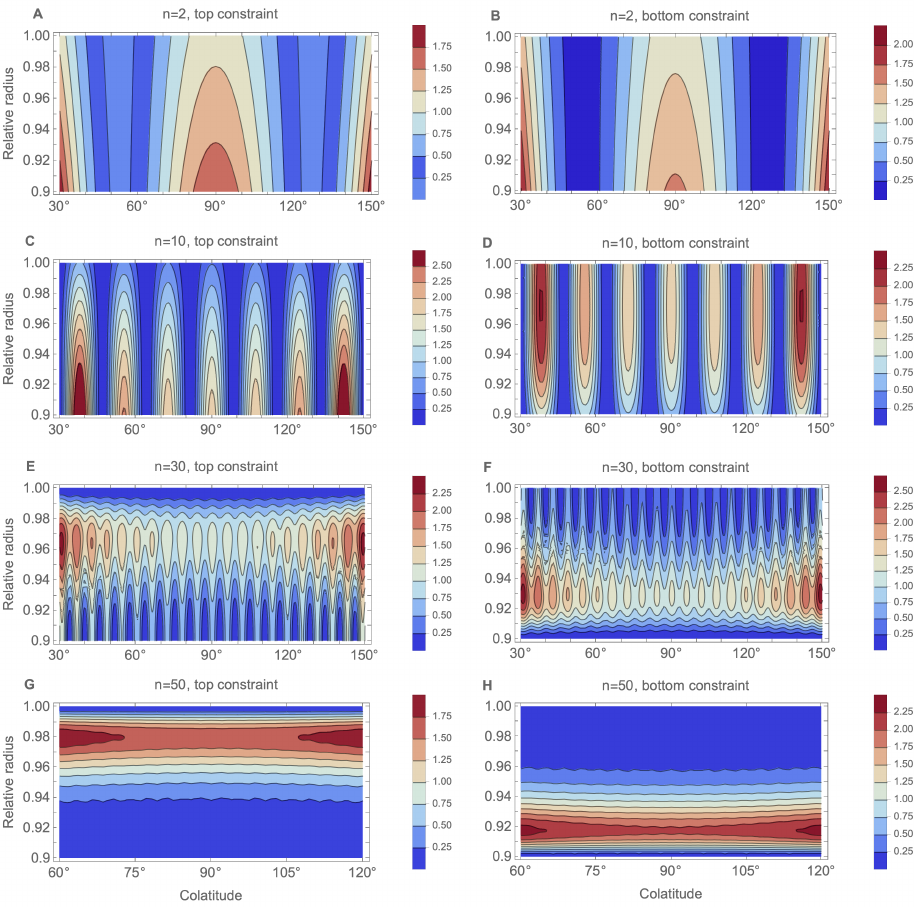}
   \caption[Second invariant of deviatoric stress within Enceladus's shell]{
   Second invariant of deviatoric stress within Enceladus's shell, as a function of radius and colatitude.
   Left (resp.\ right) panels show minimum stress isostasy with surface (resp.\ bottom) shape held fixed.
   The shape is zonal with a given harmonic degree (from top to bottom: $n=2, 10, 30, 50$).
   The stress invariant is normalized by its mean.
   The shell thickness is 10\% of the surface radius and has a shear modulus of $3.5\rm\,GPa$ (other parameters as in Fig.~\ref{FigComparisonfn}).
      Polar areas are cut in order to reduce the stress range.
   At high degree, deviatoric stresses are maximum at a distance $R_s/n$ from the constraint: isostasy becomes similar to either pure top loading (left) or pure bottom loading (right).
   The angular dependence is computed as in \citet{beuthe2013}.}
   \label{FigStressMax}
\end{figure}

At high harmonic degree, the shape ratio tends to zero if $\beta=0$ whereas it diverges to $-\infty$ if $|\beta|$ is large.
This behaviour is explained by the fact that, at high harmonic degree, the maximum stress occurs at a depth of about $R_s/n\approx\lambda/2\pi$ where $\lambda$ is the load wavelength \citep{jeffreys1943,jeffreys1959,melosh2011}.
It is thus not possible to decrease the stresses due to a top load of small extent by adjusting a bottom load, and vice versa: MSI isostasy tends to a pure top (resp.\ bottom) loading configuration if the surface (resp.\ bottom) load is held constant.
Fig.~\ref{FigStressMax} illustrates this behaviour for zonal harmonic loading on Enceladus.
At the longest wavelength ($n=2$), deviatoric stresses are maximum at the bottom of the shell, whatever the boundary condition.
At intermediate wavelengths (here $n=10$), the stress remains widely distributed through the thickness of the shell.
At short wavelengths (here $n=30$ and 50), the maximum is at a depth of $R_s/n$ if the surface shape is held constant, as in a homogeneous body loaded on its surface (\citet{jeffreys1943}; Fig.~3.3 of \citet{melosh2011}).
If the bottom shape is held constant, the maximum is at a distance of $R_s/n$ from the bottom.

Small-wavelength loads receive significant support from vertical shear stresses, as can be seen by computing the shape and topographic ratios under the assumption of local compensation (see Appendix~\ref{AppendixMethodsLocal}).
In that case, the isostatic ratios tend at high degree to a constant value which depends little on the choice of boundary conditions (dashed curves in Fig.~\ref{FigShapeMSI}).

\begin{figure}
\centering
    \includegraphics[width=\textwidth]{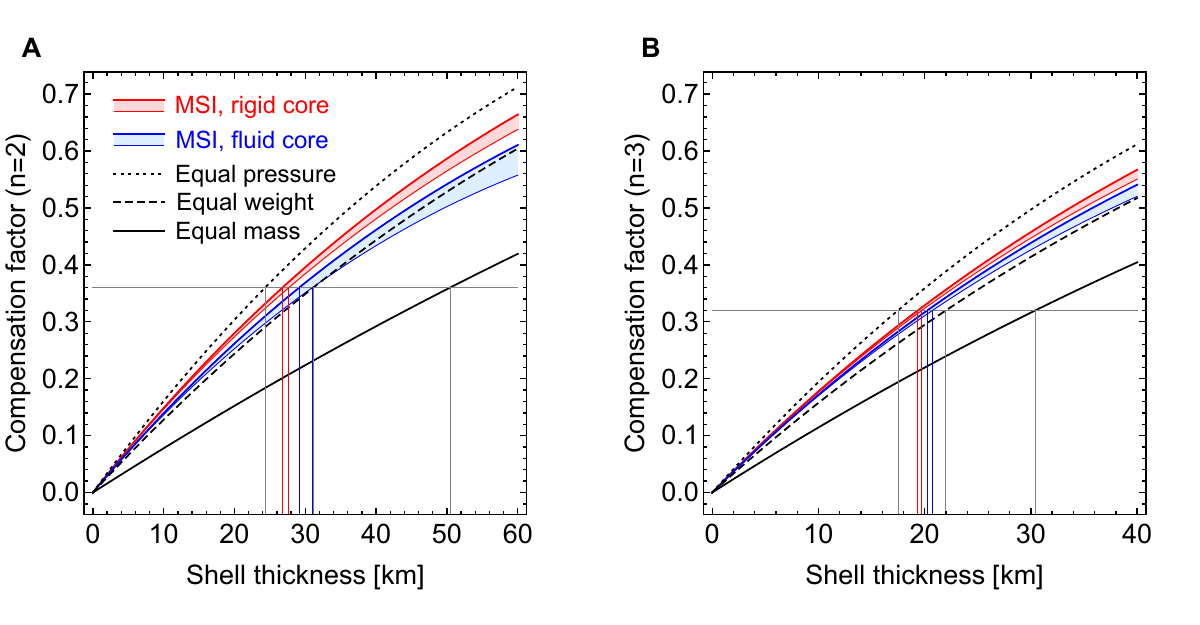}
   \caption[Compensation factor for Enceladus as a function of shell thickness]{
   Compensation factor for Enceladus as a function of shell thickness at (A) degree 2, (B) degree 3.
   Colored curves are the predictions of minimum stress isostasy (MSI) with an infinitely rigid or fluid-like core (thickest curve corresponds to $\beta=0$, thinnest curve to $\beta=-\infty$).
   Shaded areas show MSI predictions for the whole range of the isostatic family parameter $\beta$.
   Black curves are the predictions of classical isostasy with equal mass, equal weight, or equal (lithostatic) pressure.
   The interior model is the same as in Fig.~\ref{FigComparisonfn}; the core radius is equal to $192\,$km.
   Horizontal lines show the compensation factors $(F_2,F_3)=(0.36,0.32)$ inferred from observations (e.g.\ \citet{beuthe2016}).
   Vertical lines show inferred values of shell thickness.
 }
   \label{FigCompensFactorThickness}
\end{figure}

In Section~\ref{SectionUnperturbedModel}, I mentioned the possibility that the core could have had the time to relax viscoelastically while the shell reached isostatic equilibrium.
In such a case, the core should be modelled as fluid-like.
Fig.~\ref{FigCompensFactorThickness} compares the predictions of minimum stress isostasy for the compensation factor of Enceladus at degrees 2 and 3.
It shows that assuming a relaxed core results in a 10\% increase of the inferred shell thickness from degree-two observations, but the effect is smaller at degree~3.
Models of classical isostasy are shown for comparison; their predictions are dispersed over a larger range.
The effect of fluid-like core is even larger for Europa, but on the other hand the associated degree-two compensation factor is small and difficult to measure (see  Fig.~\ref{FigComparisonfn}).

\FloatBarrier

\section{Summary}

There are nowadays three main approaches to Airy isostasy: classical (or naive) isostasy based on a simplified view of mechanical equilibrium, elastic isostasy based on a principle of minimum stress (or energy), and dynamic isostasy based on the viscous (or viscoelastic) time evolution of topography.
Classical isostasy was originally proposed in a very restricted physical setting: the wavelength of the topography should be neither too large so that spherical geometry effects do not need to be considered, nor too small so that local compensation is a good approximation (local compensation certainly breaks down if the wavelength is much smaller than the depth of compensation).
The extension of classical isostasy out of this range raised issues which have never been satisfactorily resolved, as witnessed by the century-old debate of equal mass versus equal pressure prescriptions.
As argued in the Introduction, classical isostasy has gone too far in simplifying the mechanics of isostasy, by eliminating any consideration of internal stress and self-gravity.
The conclusive proof that its predictions are wrong at very long wavelengths (comparable to the body radius) is that it does not have the correct thin shell limit, as predicted by the more complete model of elastic isostasy (and by dynamic isostasy as well, see Paper~II).
While equal pressure isostasy is generally a better approximation than equal mass isostasy, it is not especially better than equal weight isostasy.
One should thus give up debating the various models of classical isostasy.

In this paper, elastic isostasy is formulated as a loading problem under surface and internal loads, in which the complicated response of the body is represented by Love numbers.
More specifically, Airy isostatic ratios depend on \textit{deviatoric Love numbers}, which quantify deviations with respect to a fluid body, and on the loading ratio, which quantifies the relative amount of internal and surface loading.
A physical constraint on the loading ratio, or \textit{isostatic prescription}, determines the particular form taken by elastic isostasy.
In an incompressible model stratified into homogeneous layers, the different isostatic ratios are linked by simple transformations (\textit{isostatic relations}), so that it is enough to know one isostatic ratio to determine them all.
At degree one, isostatic ratios are fully known without isostatic prescription, because full compensation always occurs in the centre-of-mass frame (this property can be understood  as well in terms of degree-one Love numbers).
Isostatic ratios in elastic isostasy have the nice property of being invariant under a global rescaling of the shear modulus of the shell (\textit{$\mu$-invariance}).
This property was verified here for 3-layer incompressible models, but a general proof would be welcome.
Thanks to $\mu$-invariance, isostatic ratios can be computed in the limit of a fluid shell.
Besides greatly simplifying analytical computations, the fluid limit will be used in Paper~II to reveal the connection between elastic isostasy and dynamic isostasy, in which the shell evolves in the long-time limit to a quasi-fluid state.
With Love numbers, it becomes easy to quantify the effect on isostasy of the structure below the shell.
For example, the state of Enceladus's core (relaxed or not) has a 10\% effect on isostatic anomalies (Fig.~\ref{FigCompensFactorThickness}).
For Titan-like icy satellites, the effect of ocean stratification could be an interesting avenue to explore, although the nearly complete compensation on these large bodies makes it difficult to measure their long-wavelength gravity anomalies.
If needed, it is possible to consider isostatic models with local compensation by assuming that the elasticity of the shell is transversely isotropic (Fig.~\ref{FigShapeMSI}).

The most meaningful prescriptions of elastic isostasy rely either on the principle of minimum stress or on the principle of minimum energy.
There are nevertheless good reasons to introduce another isostatic prescription called zero deflection isostasy (ZDI).
In its most general form, it is a one-parameter isostatic family based on specifying the ratio of the deflections of the shell boundaries.
It has the advantages of having a simple formulation and a close relation to minimum stress isostasy, and of being equivalent to a family of viscous/viscoelastic isostatic models (see Paper~II).
In a model with three incompressible homogeneous layers, the ZDI topographic ratio is given by a simple analytical formula, which can be used to generate the ZDI shape ratio and compensation factor (see Table~\ref{TableFormulas}).
Although these formulas are not as compact as those of classical isostasy, they similarly depend on four nondimensional parameters associated with shell thickness, shell density, shell-ocean (or asthenosphere) density contrast, and harmonic degree.
If the core is elastic, the shape ratio and compensation factor also depend on the size and shear modulus of the core, but the topographic ratio does not.

The version of minimum stress isostasy (MSI) proposed in this paper is based on finding the minimum of the second invariant of the deviatoric stress averaged over the volume of the shell, which is equivalent to the minimum of the deviatoric strain energy of the shell if the shell is  elastically homogeneous.
It differs from the  minimum stress model of \citet{dahlen1982} by relying neither on local compensation nor on the thin shell assumption.
The MSI isostatic prescription must be completed by a constraint on the space of physical configurations that must be considered when searching for a minimum (otherwise the unperturbed state would always be the winner).
The two obvious choices, consisting in specifying either the surface shape or the shape of the bottom of the shell, are generalized here to a one-parameter isostatic family specified by the ratio of the bottom shape to the surface shape.
In the general case of a shell with depth-dependent shear modulus and density, the MSI loading ratio must be evaluated numerically after computing the Love numbers and the stress (or energy) coefficients (see Table~\ref{TableFormulas}).
If the shell is elastically uniform, the second invariant of the deviatoric stress (or equivalently the deviatoric strain energy) can be computed in terms of partial derivatives of Love numbers.
If in addition the shell is incompressible and has a uniform density, the Love numbers can be factored out of the MSI fluid loading ratio, which becomes a simple function of the isostatic family parameter.
It is thus possible, if the shell is incompressible and homogeneous, to establish a duality between MSI and ZDI by equating their loading parameters, with the implication that all MSI isostatic ratios can be computed from ZDI formulas (see Table~\ref{TableFormulas}).
The existence of this duality depends on the $\mu$-invariance of MSI, which is guaranteed by the $\mu$-invariance of ZDI.
Analytical formulas for MSI and ZDI are included in the complementary software \citep{beuthe2020a}.

The introduction of one-parameter isostatic families (both for MSI and ZDI) shifts the focus on the right choice of boundary conditions.
Isostatic ratios differ if boundary conditions are applied at the surface or at the bottom of the shell.
This difference, which is most notable in the shape and topographic ratios (Fig.~\ref{FigShapeMSI}),  increases with harmonic degree and with shell thickness, and is mainly due to shear support between vertical columns (regional compensation).
Fortunately, predictions for the compensation factor are not very sensitive to the choice of boundary conditions (Fig.~\ref{FigComparisonfn}).
In the thin shell limit, vertical shear stresses are negligible so that predictions of elastic isostasy do not depend on boundary conditions and agree with the thin shell/local compensation/minimum stress model of \citet{dahlen1982}, at least if the same simple interior model is adopted.

The new focus on boundary conditions reintroduces the physical picture.
For example, the purpose of minimum stress isostasy with constant surface shape is to explain topography observed at the surface.
At long wavelength, isostasy attributes its support to topography at the shell-ocean boundary.
At short wavelength (or high harmonic degree), surface topography is supported by shallow stresses (assuming regional compensation).
Thus, it cannot be explained by supporting topography at the shell-ocean boundary and, as a corollary, the shape ratio tends to zero at high degree.
The reverse situation is true if the aim is to explain topography that has been observed (with radar or seismic measurements) at the shell-ocean boundary: at short wavelength, it cannot be supported by surface topography and the shape ratio diverges.
One can say that isostatic models with regional compensation naturally lose their isostatic properties at short wavelength, whereas isostatic models with local compensation remain isostatic at all degrees, but at the cost of physical realism.

\begin{table}[h]\centering
\ra{1.3}
\small
\caption[Isostasy with Love: key formulas]{Isostasy with Love: key formulas}
\begin{tabular}{@{}ll@{}}
\hline
\vspace{0.3mm}
\textit{Arbitrary interior model} &  Eqs. \\
Isostatic ratios (shell of finite strength) & (\ref{ShapeRatioLove})-(\ref{CompensFactorLove}) \\
Isostatic ratios (fluid limit) & (\ref{ShapeRatio0})-(\ref{CompensFactor0}) \\
ZDI loading ratio (shell of finite strength) & (\ref{zetaZDI}) \\
ZDI loading ratio (fluid limit) & (\ref{zetaZDI0}) \\
Stress (or energy) coefficients & (\ref{defEJKmup}) \\
MSI loading ratio & (\ref{zetaMING}) \\
MSI shape ratio & (\ref{SnMING}) \\
\hline
\textit{Incompressible shell with uniform density} &  Eqs. \\
Gravity-deformation coefficients $(a,b,c,d,K_n)$ & Table~\ref{TableCoeff} and Eq.~(\ref{KnLove}) \\
Isostatic relations & (\ref{TnSnRelation})-(\ref{TnFnRelation}) \\
\hline
\textit{Shell with uniform elasticity} &  Eqs. \\
Stress/energy coefficients from derivatives of Love numbers & (\ref{LoveDeriv1})-(\ref{LoveDeriv2}) \\
\hline
\textit{Incompressible homogeneous shell} &  Eqs. \\
ZDI $\alpha$-parameter range & (\ref{ExcludedRange}) \\
MSI loading ratio (fluid limit) & (\ref{zetaMING0}) \\
MSI isostatic ratios by duality & (\ref{IsoRatioMSI})-(\ref{alphaMSI}) \\
\hline
\textit{Three incompressible homogeneous layers} &  Eqs. \\
ZDI topographic ratio${}^{(1)}$ & (\ref{TopoRatioZDI})-(\ref{GeometricalFactorT}) \\
ZDI shape ratio and compensation factor${}^{(1,2)}$ &(\ref{ShapeRatioZDI})-(\ref{CompensFactorZDI})  \\
MSI isostatic ratios${}^{(1,2)}$ & (\ref{TopoRatioMSI})-(\ref{FnMSILFT}) \\
\hline
\multicolumn{2}{l}{\scriptsize ${}^{(1)}$ Conjectured to be valid if stratified core and ocean.}\\
\multicolumn{2}{l}{\scriptsize ${}^{(2)}$ If stratified core or ocean, the coefficients $(a,b,c,d)$ should be computed with $K_n$ given by Eq.~(\ref{KnLove}).}
\end{tabular}
\label{TableFormulas}
\end{table}%

\small
\section*{Acknowledgments}
I am grateful to Antony Trinh for his insights, in particular about degree-one compensation and the constant shape constraint in minimum stress isostasy.
I thank Isamu Matsuyama and David Al-Attar for taking the time to read and comment the manuscript.
All data used in the paper are publicly available.
Mathematica and Fortran codes are available on https://zenodo.org (see \citet{beuthe2020a} in the reference list).
This work is financially supported by the Belgian PRODEX program managed by the European Space Agency in collaboration with the Belgian Federal Science Policy Office.

\normalsize

\FloatBarrier

\normalsize
\begin{appendices}

\section{Surface and internal load Love numbers}
\label{AppendixLoveNumbers}
\renewcommand{\theequation}{A.\arabic{equation}} 
\setcounter{equation}{0}  

\subsection{Variables $y_i$ }
\label{AppendixVariablesyi}

Love numbers are computational byproducts of the linear solution for stress and strain inside a spherically symmetric self-gravitating body submitted to a small forcing.
In this paper, the forcing is expressed in the form of a potential with degree-$n$ harmonic coefficient $U^J_n$ and is either due to a surface load ($J=L$) or to an internal load applied at the shell-ocean boundary ($J=I$).
At each harmonic degree, the linearized equations of motion and Poisson's equation can be written as a set of six differential equations of first order, the solutions of which are six radial functions \citep{alterman1959,longman1962,kaula1963}.
Following the conventions of \citet{takeuchi1972}, the scalars associated with radial and tangential displacements are denoted $y_1$ and $y_3$, respectively,
\begin{equation}
\left( u^J_r , u^J_\theta , u^J_\phi \right) = \left( y^J_1 , y^J_3 \, \partial_\theta \, ,  \frac{y^J_3}{\sin\theta} \, \partial_\phi \right) U^J_n \, Y_n(\theta,\phi) \, ,
\label{y1y3def}
\end{equation}
where $Y_n(\theta,\phi)$ is a surface spherical harmonic of degree $n$ (the harmonic order $m$ is implicit).
The scalars associated with the purely radial stress and the radial-tangential shear stress are denoted $y_2$ and $y_4$, respectively:
\begin{equation}
\left( \sigma^J_{rr}, \sigma^J_{r\theta} , \sigma^J_{r\phi} \right) =  \left( y^J_2 \, , y^J_4 \frac{\partial}{\partial \theta} \,  ,  \frac{y^J_4}{\sin\theta} \, \frac{\partial}{\partial \phi} \right) U^J_n \, Y_n(\theta,\phi) \, .
\label{y2y4def}
\end{equation}
The total gravity potential perturbation inside the body is similarly expanded in spherical harmonics with coefficients
\begin{equation}
\Gamma^J_n = y^J_5 \, U^J_n \, ,
\label{y5def}
\end{equation}
while $y^J_6$ is an auxiliary variable defined by $y^J_6=\partial_r y^J_5 - 4\pi G \rho y^J_1 + (n+1)y^J_5/r$.

\subsection{Boundary conditions}
\label{AppendixBoundaryConditions}

The functions $y^J_i$ are continuous within the body except that $y^J_2$ and $y^J_6$ are discontinuous at the shell-ocean boundary if an internal load is applied on that interface.
Though there are six independent solutions in the solid layers, only three are regular at the center of the body, leaving only three degrees of freedom.
For a unit surface load, these regular solutions must be combined in order to satisfy three boundary conditions at the surface \citep{longman1962,saito1974}:
\begin{equation}
\left( y_2^L(R_s) , y_4^L(R_s) , y_6^L(R_s) \right) = \left( - \frac{2n+1}{3}\, \rho_b \, , 0 \, , \frac{2n+1}{R_s} \right) .
\label{bcL}
\end{equation}
For a unit internal load, the surface boundary conditions are homogeneous,
\begin{equation}
\left( y_2^I(R_s) , y_4^I(R_s) , y_6^I(R_s) \right) = \left( 0 \, , 0 \, , 0 \right) ,
\label{bcI}
\end{equation}
while the boundary conditions (or discontinuity conditions) at the shell-ocean boundary are inhomogeneous \citep{grefflefftz1997},
\begin{eqnarray}
y^I_2(R_o^+) &=& y^I_2(R_o^-) + \frac{2n+1}{3} \, \frac{g_o}{g_s} \, \frac{R_s}{R_o} \, \rho_b \, ,
\label{discy2} \\
y^I_6(R_o^+) &=& y^I_6(R_o^-) - \frac{2n+1}{R_o} \, ,
\label{discy6}
\end{eqnarray}
where $R_o^+$ (resp.\ $R_o^-$) denotes the limit $r\rightarrow{}R_o$ within the shell (resp.\ ocean). 

\subsection{Radial and gravitational Love numbers}
\label{AppendixRadialGravLove}

The radial Love number $h^J_j$ (the subscript $j$ denoting that it is evaluated at radius $R_j$) is defined as the transfer function between the unit load potential (divided by $g_s$) and the radial displacement $y^J_1$:
\begin{equation}
h^J_j = g_s \, y_1(R_j) \, .
 \label{LoveNumberh}
\end{equation}
The dependence on the harmonic degree $n$ is implicit.
The normalization factor is $1/g_s$ for both surface and internal loads.
With this normalization, radial Love numbers at different radii are equal to displacements rescaled by a common factor.
This convention follows \citet{hinderer1989} but differs from \citet{grefflefftz1997} where $1/g_o$ is used instead at the shell-ocean boundary.

The gravitational Love number $k^J_j$ is defined by subtracting the direct effect of the load (Eqs.~(\ref{PotExt})-(\ref{PotInt})) from the total gravitational perturbation:
\begin{equation}
k^J_j = \left\{
 \begin{tabular}{ll}
$y_5(R_j) - (R_j/R_J)^n$ & if  $R_j\leq R_J$ \, , \\
$y_5(R_j) - (R_J/R_j)^{n+1}$ & if  $R_j\geq R_J$ \, ,
 \end{tabular}
 \right.
 \label{LoveNumberk}
\end{equation}
where $R_J=R_s$ for a surface load and $R_J=R_o$ for an internal load.

\subsection{Methods of solution if isotropic elasticity}
\label{AppendixMethodsIsotropic}

In this paper, I generally assume that the material is isotropic, in which case there are only two elastic moduli, the Lam\'e parameters $\lambda$ and $\mu$.
If the body is compressible or if the density and rheology vary continuously, the six differential equations must be solved numerically by propagating three independent solutions from the centre to the surface.
For a surface load, the three solutions are combined linearly in order to satisfy the three boundary conditions \citep{longman1963}.
If there is an internal load, a fourth (inhomogeneous) solution must be added to the above linear combination; this solution is obtained by the propagating the internal discontinuities (second terms in RHS of Eqs.~(\ref{discy2})-(\ref{discy2})) from the shell-ocean boundary to the surface.

If the body is incompressible and stratified into homogeneous layers, the differential system can be solved with the propagator matrix method \citep{sabadini2004}.
Analytical solutions are possible if the number of layers is small (in practice, three layers is the maximum).
This method has also the advantage of being stable if the shell is very thin, if the rheology varies exponentially through the shell, or if one is interested in studying the transition to the fluid limit.

Degree one differs from the other harmonic degrees because an additional degree of freedom is associated with rigid translations \citep{farrell1972,saito1974}.
It is thus necessary to add a boundary condition on the gravitational potential which specifies the frame, for example $y_5(R_s)=0$ for the centre-of-mass frame \citep{grefflefftz1997}.

\subsection{Method of solution if local compensation}
\label{AppendixMethodsLocal}

Local isostatic compensation cannot be realized with an isotropic material because the radial-tangential stress components always vanish ($y_4=0$).
The spherical symmetry of the unperturbed model, however, does not require that the elastic properties of the shell be fully isotropic, but only transversely isotropic, in which case there are five independent elastic moduli \citep{backus1967,takeuchi1972,dahlen1999}.

Local compensation is implemented by setting to zero the elastic modulus associated with the radial-tangential strain ($L=0$ in the notation of \citet{takeuchi1972}), whereas the other moduli are related to $\mu$ and $\lambda$ as in the isotropic case.
The third differential equation ($L(ry_3'-y_3+y_1)=ry_4$) is trivially satisfied and drops out of the system.
The fourth differential equation becomes an ordinary equation ($y_4=0$ implies that $y_4'=0$) yielding $y_3$ in terms of $y_1$ and $y_5$.
In the end, the differential system is made of four differential equations for the variables $(y_1,y_2,y_5,y_6)$.
The number of boundary conditions is reduced to two because the equation $y_4=0$ is automatically satisfied.
For a surface load, the differential system is solved numerically by propagating two independent solutions from the center to the surface and combining them linearly in order to satisfy the two surface boundary conditions.
If there is an internal load, a third (inhomogeneous) solution must be added to the above linear combination, similarly to the isotropic case.

\section{Fluid Love numbers}
\label{AppendixFluidCrust}
\renewcommand{\theequation}{B.\arabic{equation}} 
\setcounter{equation}{0}  

\subsection{Fluid layer in the static limit}

In the static limit, fluid layers are treated differently from the solid layers because displacement and stress become indeterminate \citep{saito1974}.
The variables $y_1^J$, $y_2^J$, and $y_5^J$ (with $J=I,L$) are related by the \textit{fluid constraint},
\begin{equation}
y_2^J = \rho_o \left( g_r \, y_1^J - y_5^J \right) \, ,
\end{equation}
where $g_r$ is the gravitational acceleration at radius $r$, while shear stresses vanish (\textit{free-slip} at the ocean boundaries):
\begin{equation}
y^J_4=0 \, .
\label{freeslip}
\end{equation}
There are only two independent variables instead of three, which can be chosen as $y_5^J$ and a new continuous variable denoted $y_7^J$ \citep{saito1974}:
\begin{equation}
y_7^J = y_6^J + (4\pi G/g_r) \, y_2^J \, .
\label{defy7}
\end{equation}

If the ocean is the only fluid layer, the three variables within the elastic core are reduced to one at the core-ocean boundary because of the fluid-constraint and the free-slip condition.
The two fluid variables $y_5^J$ and $y_7^J$ thus depend on one unknown constant.
At the shell-ocean boundary, the indeterminated tangential displacement and radial stress yield two other unknown constants on which the elastic-gravitational equations depend within the shell.
At degree one, $y_7^J=0$ and $y_5^J$ is proportional to $g_r$ in the fluid layer \citep{saito1974}.

\subsection{Fluid limit of load Love numbers}
\label{AppendixFluidLoveNumbers}

In the fluid limit, the shell above the ocean becomes fluid, so that the relevant variables within the shell are now $y_5^J$ and $y_7^J$.
These variables depend on one unknown constant and satisfy one boundary condition at the surface:
\begin{equation}
y_7^J(R_s) = 0 \, .
\end{equation}
This boundary condition is obtained by substituting either Eq.~(\ref{bcL}) or Eq.~(\ref{bcI}) into Eq.~(\ref{defy7}).
The unknown constant is thus equal to zero, so that the fluid variables vanish within the ocean and fluid shell:
\begin{equation}
y_5^J(r)=y_7^J(r)=0 \, .
\end{equation}
Physically, this means that the fluid cannot support surface and internal loads: the loads are gravitationally compensated.
Although $y_6^I$ is discontinuous at the shell-ocean boundary (Eq.~(\ref{discy6})), $y_7^I$ is defined so as to be continuous and vanishes through the fluid shell and the ocean.

The condition that $y_5^J=0$ implies with Eq.~(\ref{LoveNumberk}) that the gravitational fluid Love numbers cancel the direct effect of the load:
\begin{equation}
\left( k_s^{L\circ} , k_o^{L\circ} , k_s^{I\circ} , k_o^{I\circ} \right) = \left( - 1 , - x^n , -x^{n+1} , -1 \right) .
\label{FluidGravLove}
\end{equation}

At the surface, the fluid constraint combined with the condition $y_5^J(r)=0$ yields
\begin{equation}
\rho_s g_s \, y_1^J(R_s) = y_2^J(R_s) \, ,
\end{equation}
from which the radial fluid Love numbers can be computed by application of the surface boundary conditions (Eqs.~(\ref{bcL})-(\ref{bcI})):
\begin{eqnarray}
h_s^{L\circ}&=& - \frac{2n+1}{3\,\xi_{s1}} \, ,
\label{hsL} \\
h_s^{I\circ} &=& 0 \, .
\label{hsI}
\end{eqnarray}
Eq.~(\ref{hsL}) can be found in \cite{wu1982} for a homogeneous model, and with a restriction to high harmonic degrees for an inhomogeneous model; its general validity was stated without proof by \citet{grefflefftz2010}. 

At the shell-ocean boundary, the fluid constraint combined with the condition $y_5^J(r)=0$ yields
\begin{eqnarray}
\rho_s g_s \, y_1^J(R_o^+) = y_2^J(R_o^+) \, ,
\label{ShellOceany1y2plus} \\
\rho_o g_s \, y_1^J(R_o^-) = y_2^J(R_o^-) \, .
\label{ShellOceany1y2minus}
\end{eqnarray}
For a surface load, these equations are only consistent with the continuity of $y_1^J$ and $y_2^J$ if the two variables vanish.
In other words, the shell and ocean are unperturbed below the compensated surface load.
Thus, the radial fluid Love number vanishes at the shell-ocean boundary:
\begin{equation}
h_o^{L\circ} = 0 \, .
\label{hoL}
\end{equation}
For an internal load, Eqs.~(\ref{ShellOceany1y2plus})-(\ref{ShellOceany1y2minus}) combined with the continuity of $y_1^I$ and the discontinuity of $y_2^I$ (Eq.~(\ref{discy2})) yield
\begin{equation}
h_o^{I\circ} = - \frac{2n+1}{3 \left(\xi_{o1}-\xi_{s1}\right)} \, \frac{R_s}{R_o} \, .
\label{hoI}
\end{equation}

\section{Gravity-deformation relations in a 3-layer body}
\label{AppendixRelationLoveIncompressible}
\renewcommand{\theequation}{C.\arabic{equation}} 
\setcounter{equation}{0}  

Consider an incompressible body stratified in layers of homogeneous density (the rheology may be depth-dependent).
In that case, the gravitational perturbation at an arbitrary radius can be computed as a function of the radial deformations of interfaces with a density contrast.
For a 3-layer body (shell/ocean/core), Eqs.~(\ref{PotExt})-(\ref{PotInt}) yield
\begin{eqnarray}
\Gamma_{sn} &=&  g_s \left( \xi_{sn} \, H_{sn} + \Delta \xi_{n} \, x^{n+2} \, H_{on}  + \Delta \xi'_{n} \, y^{n+2} \, H_{cn} \right) ,
\label{GammaHRelation1}
\\
\Gamma_{on} &=& g_s \left(  \xi_{sn} \, x^n \, H_{sn} + \Delta \xi_{n} \, x \, H_{on}  + \Delta \xi'_{n} \, x \, (y/x)^{n+2} \, H_{cn} \right) ,
\label{GammaHRelation2} \\
\Gamma_{cn} &=& g_s \left(  \xi_{sn} \, y^n \, H_{sn} + \Delta \xi_{n} \, x \, (y/x) ^n \, H_{on}  + \Delta \xi'_{n} \, y \, H_{cn} \right) .
\label{GammaHRelation3}
\end{eqnarray}
Using Eqs.~(\ref{ShapeSurf})-(\ref{ShapeCore}) and Eqs.~(\ref{GravSurf})-(\ref{GravCore}), I can write these relations in terms of Love numbers (even better, in terms of deviatoric Love numbers):
\begin{eqnarray}
\hat k_s^J &=&  \xi_{sn} \, \hat  h_s^J + \Delta \xi_{n} \, x^{n+2} \, \hat  h_o^J  + \Delta \xi'_{n} \, y^{n+2} \, \hat h_c^J \, ,
\label{khRelation1} \\
\hat k_o^J &=&  \xi_{sn} \, x^n \, \hat h_s^J +  \Delta \xi_{n} \,x \, \hat h_o^J  + \Delta \xi'_{n} \, x \, (y/x)^{n+2} \, \hat h_c^J \, ,
\label{khRelation2} \\
\hat k_c^J &=& \xi_{sn} \, y^n \, \hat h_s^J + \Delta \xi_{n} \, x \, (y/x) ^n \, \hat h_o^J  +  \Delta \xi'_{n} \, y \,\hat h_c^J \, .
\label{khRelation3}
\end{eqnarray}

If the core is infinitely rigid and $n\geq2$, $\hat h_c^J=0$ so that $(\hat k_s^J,\hat k_o^J)$ do not depend on the properties of the core.
If the core is fluid, the condition that the core surface is an equipotential implies that the shape of the core follows the local geoid:
\begin{equation}
H_{cn} = \frac{\Gamma_{cn}}{g_c}
\,\, \Leftrightarrow \,\, 
\hat k_c^J = \gamma_c \, \hat h_c^J \, ,
\label{kchcFluid}
\end{equation}
where $\gamma_c=g_c/g_s$ (see Eq.~(\ref{gammac})).

If the core is elastic and incompressible, the relation between $k_c^J$ and $h_c^J$ can be derived with the propagator matrix approach.
The six non-singular radial functions $y_{1...6}(r)$ depending on three unknowns are propagated from the center to the core-ocean boundary, where two constraints are applied: the free-slip condition, $y_4(R_c)=0$, and the fluid constraint, $y_2(R_c)=\rho_o(g_oy_1(R_c)-y_5(R_c))$.
Solving these equations for a homogeneous core, I relate the gravitational perturbation $\hat k_c^J$ to the radial deformation $\hat h_c^J$ of the core-ocean boundary:
\begin{equation}
\hat k_c^J = \left( \gamma_c + f_n \,\frac{\bar\mu_{\rm c}}{y \, \Delta\xi'_1} \right) \hat h_c^J \, ,
\label{kchcElastic}
\end{equation}
where $\bar\mu_{\rm c}=\mu_{\rm c}/(\rho_bg_sR_s)$ is the reduced shear modulus of the core and $\Delta\xi'_1$ is the core-ocean density ratio contrast given by Eq.~(\ref{CoreOceanDensContrast}).
The function $f_n$ is defined by
\begin{equation}
f_n = \frac{2\left(n-1\right) \left(2n^2+4n+3\right)}{n\left(2n+1\right)} \, .
\label{fndef}
\end{equation}
At degree one, $f_1=0$ so that Eq.~(\ref{kchcElastic}) reduces to Eq.~(\ref{kchcFluid}): the core is rigidly translated and its surface is equipotential.

Solving Eq.~(\ref{khRelation3}) and Eq.~(\ref{kchcElastic}) for $\hat h_c^J$, I get
\begin{equation}
\hat h_c^J = \frac{\xi_{sn} \, y^n \, \hat  h_s^J +\Delta\xi_n \, x \, (y/x)^n \, \hat h_o^J}
{\gamma_c + f_n \, \bar \mu_{\rm c}/ (y\,\Delta\xi'_1) - y \, \Delta \xi'_n} \, .
\end{equation}
Substituting the result into Eqs.~(\ref{khRelation1})-(\ref{khRelation2}), I write the gravitational perturbations in terms of the radial deformations at the shell boundaries:
\begin{eqnarray}
\hat k_s^J &=& \xi_{sn} \left( 1 + K_{n} \right) \hat  h_s^J + \Delta \xi_{n} \, x^{n+2} \left( 1+  K_{n} \, x^{-2n-1} \right) \hat  h_o^J  \, ,
\label{khRelationGenApp1} \\
\hat k_o^J &=& \left( 1+  K_{n} \, x^{-2n-1} \right) \left( \xi_{sn} \, x^n \, \hat  h_s^J + \Delta \xi_{n} \, x \, \hat  h_o^J  \right) ,
\label{khRelationGenApp2}
\end{eqnarray}
where
\begin{equation}
K_{n} = \frac{\Delta\xi'_n \, y^{2n+2}}{\gamma_c + f_n \, \bar \mu_{\rm c}/ (y\,\Delta\xi'_1) - y \, \Delta \xi'_n} \, .
\label{defAelastic}
\end{equation}
In the main text, these equations are written more compactly as Eq.~(\ref{khRelation}) with the coefficients of Table~\ref{TableCoeff}.
Instead of repeating this equation here, I give the corresponding relations for non-deviatoric Love numbers:
\begin{equation}
\left(
\begin{array}{ll}
k_s^L & k_s^I \\
k_o^L & k_o^I
\end{array}
\right)
=
\left(
\begin{array}{ll}
a & b  \\
c & b \, x^{-n-1}
\end{array}
\right)
\left(
\begin{array}{ll}
h_s^L & h_s^I \\
h_o^L & h_o^I
\end{array}
\right)
+
K_n
\left(
\begin{array}{ll}
1 & x^{-n} \\
x^{-n-1} & x^{-2n-1}
\end{array}
\right) .
\label{khRelationNonDev}
\end{equation}
In the limit $\bar \mu_{\rm s}\rightarrow\infty$, the shell becomes infinitely rigid so that its deformation tends to zero: $h_j^J\rightarrow0$.
Eq.~(\ref{khRelationNonDev}) then implies that $k_s^L\rightarrow K_n$.
In words, $K_n$ measures the gravitational perturbation (due to deformed layers below the shell) in the limit of a non-deformable shell.

In the limit $\bar \mu_{\rm c}\rightarrow\infty$, the core becomes infinitely rigid and $K_{n}\rightarrow0$.
The rigid core constraint, $K_{n}=0$, only make sense if $n\geq2$.
At degree one, $f_1=0$ so that Eq.~(\ref{defAelastic}) yields, whatever the rheology of the core,
\begin{equation}
K_{1} = \frac{\Delta \xi'_1}{\xi_{o1}} \, y^3 = \frac{1 - \xi_{s1} - x^3 \Delta\xi_1}{\xi_{o1}} \, .
\label{Kc1}
\end{equation}
This value does not depend on the shear modulus of the core because the core is rigidly translated.

\section{Love numbers if homogeneous shell}
\label{AppendixLoveNumbersHomogeneousLayers}
\renewcommand{\theequation}{D.\arabic{equation}} 
\setcounter{equation}{0}  

Consider an incompressible 3-layer body with a homogeneous shell (the ocean and the core are not necessarily homogeneous).
This problem is solved analytically with the propagator matrix method (Appendix~\ref{AppendixMethodsIsotropic}).
Since the shell is homogeneous, the shell propagator matrix and its inverse appear each once in the elastic-gravitational equations.
The Love numbers can thus be written as the ratio of two degree-two polynomials in the shear modulus of the shell.

\subsection{Radial Love numbers}

Radial Love numbers are written in the following generic form:
\begin{equation}
h_j^J = \frac{A_j^J + B_j^J \mu_{\rm s} + C_j^J \mu_{\rm s}^2}{D+E \mu_{\rm s}+F \mu_{\rm s}^2} \, ,
\label{hGenE}
\end{equation}
where $j=(s,o)$ and $J=(L,I)$.
The coefficients $(D,E,F)$ are common to all Love numbers (radial, tangential, or gravitational), whatever the type of load (applied at the surface or at the shell-ocean boundary) or the location where they are evaluated (surface or shell-ocean boundary).
All coefficients are polynomials in the interior parameters $(x,y,\xi_{s1},\xi_{o1},\bar\mu_{\rm c})$.

The fluid limit ($h_j^J\rightarrow h_j^{J\circ}$, see Appendix~\ref{AppendixFluidLoveNumbers}) and the limit of an infinitely rigid shell ($h_j^J\rightarrow0$) imply that
\begin{eqnarray}
A^J_j &=& D \, h^{J\circ}_j \, ,
\label{CoeffALS} \\
C_j^J &=& 0 \, .
\label{CoeffCLS}
\end{eqnarray}

Two of the identities required for the $\mu$-invariance of zero-deflection isostasy (Eq.~(\ref{BilinearInvariant})) imply that
\begin{eqnarray}
A_s^L \, B_o^I + A_o^I \, B_s^L &=& (E/D) \, A_s^L \, A_o^I \, ,
\label{IdentityAB} \\
B_s^L \, B_o^I - B_o^L \, B_s^I &=& (F/D) \, A_s^L \, A_o^I \, .
\label{IdentityBB} 
\end{eqnarray}
The identity $\partial_{\mu_{\rm s}}(X_{oo}/X_{ss} ) = 0$ is automatically satisfied because it is proportional to $h_o^L/h_s^I=B_o^L/B_s^I$ which does not depend on $\mu_{\rm s}$.

Deviatoric radial Love numbers read
\begin{equation}
\hat h_j^J = \frac{\hat  A_j^J + \hat  B_j^J \mu_{\rm s} + \hat C_j^J \mu_{\rm s}^2}{D+E \mu_{\rm s}+F \mu_{\rm s}^2} \, ,
\label{hDevGen}
\end{equation}
where
\begin{eqnarray}
\hat A_j^J &=& 0 \, ,
\label{CoeffAdev} \\
\hat  B_j^J &=& B_j^J - E \, ( A_j^J/D ) \, ,
\label{CoeffBdev} \\
\hat C_j^J &=& C_j^J - F \, (A_j^J/D) \, .
\label{CoeffCdev}
\end{eqnarray}
Using Eqs.~(\ref{CoeffALS}) and (\ref{IdentityAB}), I can write the coefficients $\hat  B_j^J$ as
\begin{equation}
\left( \hat B^L_s , \hat B^L_o, \hat B^I_s, \hat B^I_o \right) = \left( - \frac{h_s^{L\circ}}{h_o^{I\circ}} \, B_o^I , B^L_o , B^I_s , - \frac{h_o^{I\circ}}{h_s^{L\circ}} \, B_s^L \right) .
 \label{coeffBdev}
\end{equation}

Partial derivatives of radial Love numbers (deviatoric or not, see Eq.~(\ref{LovePartial})) with respect to $\mu_{\rm s}$ and evaluated at $\mu_{\rm s}=0$ are equal to
\begin{equation}
\dot h^J_j = \hat  B_j^J/D \, .
 \label{hdotGen}
\end{equation}

\subsection{Gravitational Love numbers}

Gravitational Love numbers are written in the same generic form as the radial Love numbers, except that the coefficients are denoted by a `prime' superscript:
\begin{equation}
k_j^J = \frac{A'^J_j + B'^J_j \mu_{\rm s} + C'^J_j \mu_{\rm s}^2}{D+E \mu_{\rm s}+F \mu_{\rm s}^2} \, .
\label{kGen}
\end{equation}
The fluid limit (Eq.~(\ref{FluidGravLove})) and the limit of an infinitely rigid shell (see after Eq.~(\ref{khRelationNonDev})) imply that
\begin{eqnarray}
A'^J_j &=&D \, k^{J\circ}_j \, ,
\\
\left( C'^L_s , C'^L_o , C'^I_s , C'^I_o \right) &=& \left( 1 , x^{-n-1} ,x^{-n} , x^{-2n-1} \right) K_n \, F \, .
\end{eqnarray}

Deviatoric gravitational Love numbers read
\begin{equation}
\hat k_j^J = \frac{\hat  A'^J_j + \hat  B'^J_j \mu_{\rm s} + \hat C'^J_j \mu_{\rm s}^2}{D+E \mu_{\rm s}+F \mu_{\rm s}^2} \, ,
\label{kDevGen}
\end{equation}
where $\hat  A'^J_j =0$.
The coefficients $(\hat  B'^J_j,\hat C'^J_j)$ are related to $(B'^J_j,C'^J_j)$ by equations similar to Eqs.~(\ref{CoeffBdev})-(\ref{CoeffCdev}) and to $(\hat  B^J_j,\hat C^J_j)$ by Eq.~(\ref{khRelation}).

The Saito-Molodensky relation (Eq.~(\ref{SMrelation})) gives the following (equivalent) constraints:
\begin{eqnarray}
B^I_s - B'^I_s &=& x \left( \gamma_o \, B^L_o - B'^L_o \right) \, ,
\\
\hat B^I_s - \hat B'^I_s &=& x \left( \gamma_o \, \hat B^L_o - \hat B'^L_o \right) \, .
\label{SMrelationBcoeff}
\end{eqnarray}

\subsection{Rigid core, fluid core, and thin shell limits}

If the core is elastic, the analytical expressions for the Love number coefficients are quite long.
More manageable formulas are obtained by relating the elastic core model to simpler models in which the core is infinitely rigid, fluid-like, or point-like.
Since the core is homogeneous and extends to the centre of the body, the propagator matrix for the core is used only once and its inverse is not needed.
As a result, the numerator and denominator of the Love number coefficients depend linearly on the shear modulus of the core.
Denoting generically the coefficients ($B_j^J$ and so forth) of the elastic core model by the symbol $Z_e$, I can thus write
\begin{equation}
Z_e = Z_f + p \, \bar\mu_{\rm c} \, Z_r \, ,
\end{equation}
where $Z_f$ and $Z_r$ are associated with the limits of a fluid core ($\bar\mu_{\rm c}\rightarrow0$) and of a rigid core ($\bar\mu_{\rm c}\rightarrow\infty$), respectively.
The factor of proportionality $p$ is common to all coefficients, so that it simplifies when taking the rigid core limit of a Love number.
Moreover, the fluid core model should reduce to the rigid core model in the point-core limit ($y\rightarrow0$) because the gravity field outside a rigid core is indistinguishable from the one produced by a point-like core.
Thus, I can write
\begin{equation}
Z_f = q \, Z_r + Z_{f0} \, ,
\end{equation}
where $q$ is a factor of proportionality common to all coefficients.
$Z_{f0}$ is the remainder function which vanishes in the point-core limit.
Solving explicitly the elastic core model, I get
\begin{eqnarray}
p &=& 2\left(n-1\right) \left(2n^2+4n+3\right) x^{2n+1} \, y^4 \, ,
\\
q &=& 2 n\left(n-1\right) x^{2n+1} (1-\xi_{s1}-x^3\Delta\xi_1)^2 \, .
\end{eqnarray}
In this way, the lengthy coefficients $Z_e$ can be expressed in terms of the shorter coefficients $Z_r$ and $Z_{f0}$.

In the thin shell limit, the limiting values of Love numbers (Eqs.~(\ref{hThinShell})-(\ref{kThinShell})) imply that the generic coefficients tend to
\begin{equation}
\lim_{\varepsilon\rightarrow0} \frac{( D , F , B^J_j , B'^J_j )}{E} = \left( 0 , 0 , - \frac{1}{\xi_{on}} , -1 \right) .
\end{equation}
The identity (\ref{IdentityBB}) then entails that the following ratios of $B_j^J$ coefficients are equal up to order $\varepsilon$ included:
\begin{equation}
B_s^L/B_o^L = B_s^I/B_o^I +{\cal O}(\varepsilon^2) \, .
\label{IdentityBBthinshell}
\end{equation}
Combined with Eq.~(\ref{ShapeRatioZDIGen}), this result shows that elastic isostasy in a thin incompressible homogeneous shell does not depend on the choice of boundary conditions (represented by the isostatic family parameter).

\section{Integral identities and variation of Love numbers}
\label{AppendixIntegralIdentities}
\renewcommand{\theequation}{E.\arabic{equation}} 
\setcounter{equation}{0}  

\subsection{Integral of motion}

Let $y^J_i$ and $y^K_i$ ($i=1...6$) be two solutions of the equations of motion, with $(J,K$) denoting the type of load (and the the boundary conditions).
Consider the following bilinear form:
\begin{equation}
Y^{JK} = r^2 \left( y^J_1 \, y^K_2 - \delta_n \, y^J_3 \, y^K_4 + \frac{y^J_5 \, y^K_6}{4\pi{}G} \right) ,
\end{equation}
where $\delta_n=-n(n+1)$.
In the solid layers (shell and core), the following quantities are integrals of motion \citep{takeuchi1972,okubo1983,tobie2005}:
\begin{eqnarray}
\left[ Y^{JK} \right]_{0}^{R_c}
&=& \int_{0}^{R_c} {\cal L}^{JK}_s \, dr \, ,
\label{Ycore} \\
\left[ Y^{JK} \right]_{R_o}^{R_s}
&=& \int_{R_o}^{R_s} {\cal L}^{JK}_s \, dr \, ,
\label{Ycrust}
\end{eqnarray}
where
\begin{equation}
{\cal L}^{JK}_s = \mu \, {\cal H}^{JK}_\mu + \kappa \, {\cal H}^{JK}_\kappa + \rho \, {\cal H}^{JK}_\rho + {\cal H}^{JK}_0  \, ,
\label{defLsolid}
\end{equation}
in which $(\mu,\kappa,\rho)$ are the depth-dependent values of the shear modulus, bulk modulus, and density.
Each term of this sum represents a different contribution to the angular average of the energy density.
In particular, $\mu{\cal H}^{JK}_\mu/2$ measures the shear strain energy density (see Eq.~(\ref{defEJKmu})).
For the sake of completeness, the other functions are given by
\begin{eqnarray}
{\cal H}^{JK}_\kappa &=& \Big( r \partial_r y^J_1 + 2 y^J_1 + \delta_n y^J_3 \Big) \Big( r \partial_r y^{K}_1 + 2 y^{K}_1 + \delta_n y^{K}_3 \Big) ,
\\
{\cal H}^{JK}_\rho &=& 
(n+1) r \left( y^J_1 \, y^K_5 + y^J_5 \, y^K_1 \right) + r \delta_n \left( y^J_3 \, y^K_5 + y^J_5 \, y^K_3 \right)
\nonumber \\
&& - \, 4 g_r r \, y^J_1 \, y^K_1 - g_r r \, \delta_n \left( y^J_1 \, y^K_3 + y^J_1 \, y^K_3 \right) ,
\\
{\cal H}^{JK}_0 &=& \frac{1}{4\pi{}G} \, r^2 \, y^J_6 \, y^K_6 \, .
\end{eqnarray}
In the fluid layer (i.e.\ the ocean), the function $y_3$ (tangential displacement potential) is undetermined in the static limit.
Eq.~(\ref{Ycrust}) must then be replaced by
\begin{equation}
\left[ Y^{JK} \right]_{R_c}^{R_o} = \int_{R_c}^{R_o} {\cal L}^{JK}_f \, dr +  \left[ \frac{r^2}{\rho_o g_r} \, y^J_2 \, y^K_2 \right]_{R_c}^{R_o} ,
\label{Yocean}
\end{equation}
where $y^J_4=y^K_4=0$ in the LHS (no shear stress in a fluid layer), while the integrand in the RHS is given by
\begin{equation}
{\cal L}^{JK}_f = 2 (n-1) \, \frac{\rho r}{g} \, y^J_5 \, y^K_5 + \frac{r^2}{4\pi{}G} \, y^J_7 \, y^K_7 \, ,
\label{defLocean}
\end{equation}
in which $y^J_7=y^J_6+(4\pi{}G/g_r)y^J_2$.
Adding Eqs.~(\ref{Ycore})-(\ref{Ycrust}) and Eq.~(\ref{Yocean}) yields Eq.~(12) of \citet{okubo1983} modified to allow for discontinuities at the shell-ocean boundary::
\begin{equation}
\left[ Y^{JK} \right]_{0}^{R_c} + \left[ Y^{JK} \right]_{R_c}^{R^-_o} + \left[ Y^{JK} \right]_{R^+_o}^{R_s}
=  \int_{0}^{R_s} {\cal L}^{JK} \, dr +  \left[ \frac{r^2}{\rho_o g_r} \, y^J_2 \, y^K_2 \right]_{R_c}^{R_o} \, ,
\label{Ytotal}
\end{equation}
where ${\cal L}^{JK}={\cal L}^{JK}_s$ in the solid layers and ${\cal L}^{JK}={\cal L}^{JK}_f$ in the fluid layer.

\subsection{Saito-Molodensky relation}
\label{AppendixSaitoMolodensky}

The integral of motion implies a relation between two sets of solutions of the equations of motion.
Since the RHS of Eq.~(\ref{Ytotal}) is symmetric under the exchange of $J$ and $K$,
\begin{equation}
\left[ Y^{JK} -Y^{KJ} \right]_{0}^{R_c} + \left[ Y^{JK} -Y^{KJ} \right]_{R_c}^{R_o} + \left[ Y^{JK} -Y^{KJ} \right]_{R_o}^{R_s} = 0 \, .
\end{equation}
Applying the boundary and discontinuity conditions for the surface load ($J=L$) and the internal load ($K=I$), I obtain a relation between the surface load and internal load Love numbers:
\begin{equation}
k_s^I - h_s^I = x \, \Big( k_o^L - \gamma_o \, h_o^L \Big) \, ,
\label{SMappendix}
\end{equation}
where $\gamma_o=g_o/g_s$.
This is a new example of the Saito-Molodensky relations already known between tidal, surface load, and shear Love numbers \citep{molodensky1977,saito1978}.

\subsection{Variation of Love numbers}
\label{AppendixVariationLove}

Applying variational calculus on the integral of motion (Eq.~(\ref{Ytotal})), \citet{okubo1983} determine how variations of $(\mu,\kappa,\rho)$ affect the tidal, surface load and shear Love numbers.
I will only consider here variations with respect to the elastic parameters (the density does not change).
The equations of \citet{okubo1983} must be slightly modified to allow for discontinuities at the shell-ocean boundary.
Define
\begin{equation}
Z^{JK} = r^2 \left( \Big( y^J_1 \, \delta y^K_2 - y^J_2 \, \delta y^K_1 \Big) - \delta_n \Big( y^J_3 \, \delta y^K_4 - y^J_4 \, \delta y^K_3 \Big) + \frac{y^J_5 \, \delta y^K_6 - y^J_6 \, \delta y^K_5}{4\pi{}G} \right) ,
\end{equation}
where $\delta_n=-n(n+1)$.
Then Eq.~(16) of \citet{okubo1983} becomes
\begin{equation}
\left[ Z^{JK} \right]_{0}^{R_c} + \left[ Z^{JK} \right]_{R_c}^{R^-_o} + \left[ Z^{JK} \right]_{R^+_o}^{R_s}
= \int_{R_o}^{R_s} \left( {\cal H}^{JK}_\mu \, \delta \mu + {\cal H}^{JK}_\kappa \, \delta \kappa \right) dr \, .
\end{equation}
Applying the boundary and discontinuity conditions for the surface load ($J=L,I$) and the internal load ($K=L,I$), I obtain
\begin{eqnarray}
\delta h_s^L - \delta k_s^L &=& \chi^{-1} \int_d \left( {\cal H}^{LL}_\mu \, \delta\mu + {\cal H}^{LL}_\kappa \, \delta\kappa \right) dr   \, , 
\label{LoveVar1} \\
\delta h_s^I - \delta  k_s^I &=& \chi^{-1} \int_d \left( {\cal H}^{LI}_\mu \,  \delta\mu + {\cal H}^{LI}_\kappa \,  \delta\kappa \right) dr \, ,
\\
x \, \Big( \gamma_o \, \delta h_o^L - \delta k_o^L \Big) &=& \chi^{-1}  \int_d \left( {\cal H}^{IL}_\mu \,  \delta\mu + {\cal H}^{IL}_\kappa \,  \delta\kappa \right) dr \, ,
 \\
x \, \Big( \gamma_o \,  \delta h_o^I -  \delta k_o^I \Big) &=& \chi^{-1}    \int_d \left( {\cal H}^{II}_\mu \,  \delta\mu +{\cal H}^{II}_\kappa \,  \delta\kappa \right) dr \, ,
\label{LoveVar4}
\end{eqnarray}
where $\chi=(2n+1)R_s/(4 \pi G)$.
As ${\cal H}^{JK}_\mu$ and ${\cal H}^{JK}_\kappa$ are symmetric under the exchange of $J$ and $K$, the LHS of the second and third equations must be identical, which is guaranteed by the Saito-Molodensky relation (Eq.~(\ref{SMappendix})).
In the main text, these identities are expressed in terms of partial derivatives of Love numbers (Eqs.~(\ref{LoveDeriv1})-(\ref{LoveDeriv2})).

\section{Zero deflection isostasy}
\label{AppendixZDI}
\renewcommand{\theequation}{F.\arabic{equation}} 
\setcounter{equation}{0}  

\subsection{ZDI family parameter}
\label{AppendixZDIalpha}

In Section~\ref{ZeroDeflectionIsostasy}, zero deflection isostasy (ZDI) is defined as a one-parameter isostatic family by imposing that the ratio between the deflections of the shell boundaries is equal to a constant $\alpha$.
The range of $\alpha$ is unnecessarily large because some values of $\alpha$ lead to manifestly non-isostatic models.
For example, the shape ratio and compensation factor diverge if the surface shape vanishes, which occurs if $\alpha$ is equal to
\begin{equation}
\alpha_{sing} = x \, \frac{\Delta\xi_1}{\xi_{s1}} \, \frac{ \dot h_s^I }{ \dot h_s^L } \, .
\label{alphaSing}
\end{equation}
This value (found by equating Eqs.~(\ref{zetaSingular0}) and (\ref{zetaZDI0})) is close to $-1$ if $n=2$ and tends to zero as $n$ increases.
Since the shape ratio is $\mu$-invariant (Eqs.~(\ref{SnZDI}) and (\ref{BilinearInvariant})), the singular value $\alpha_{sing}$ given by Eq.~(\ref{alphaSing}) is valid at finite $\mu$.
Non-isostatic models can be avoided if negative values of $\alpha$ are excluded, but that constraint is too restrictive when establishing a one-to-one correspondence between
zero deflection isostasy and minimum stress isostasy (Section~\ref{MinimumStressFluidLimit}).
Under the assumption of a homogeneous shell, the constraint on $\alpha$ resulting from this correspondence (Eqs.~(\ref{beta0})-(\ref{betainf})) excludes the range
\begin{equation}
 \alpha_0 < \alpha <  \alpha_\infty  \, ,
\label{ExcludedRange}
\end{equation}
where $\alpha_0 = - d/c$ and $\alpha_\infty= - b/(1-a)$ with the coefficients $(a,b,c,d)$ of Table~\ref{TableCoeff}.
The excluded range lies on the negative $\alpha$-axis and its extension increases with the harmonic degree: as  $n\rightarrow\infty$, $\alpha_0$ and $\alpha_\infty$ tend to
$(-2/3)(\gamma_o/\xi_{s1})(n\,x^{-n})$ and $(-3/2)\Delta\xi_1(x^{n+2}/n)$, respectively (Fig.~\ref{FigAlpha01}).
In practice, the excluded range can be assimilated to the whole negative axis because the value of any ratio changes very little between $\alpha=-\infty$ and $\alpha=\alpha_0$, on the one hand, and between $\alpha=\alpha_\infty$ and $\alpha=0$, on the other.

\subsection{Incompressible body with homogeneous shell}
\label{ZDI3layers}

Consider an incompressible body with a homogeneous shell.
The deviatoric Love numbers take the generic form given in Appendix~\ref{AppendixLoveNumbersHomogeneousLayers}.
Assuming $\mu$-invariance, I compute the shape ratio in the fluid limit, as a function of the generic Love number coefficients $B_j^J$, by substituting Eq.~(\ref{zetaZDI0}) and Eq.~(\ref{hdotGen}) into Eq.~(\ref{ShapeRatio0}):
\begin{equation}
S_n^{\rm ZDI} = - \frac{\xi_{s1}}{x\Delta\xi_1} \, \frac{B_s^L + \alpha \, B_o^L}{B_s^I + \alpha \, B_o^I} \, .
\label{ShapeRatioZDIGen}
\end{equation}
For a shell of finite strength, substituting Eqs.~(\ref{Xjs})-(\ref{Xjo}) into Eq.~(\ref{SnZDI}) leads to the same result, but the conditions for $\mu$-invariance must be satisfied in any case.
The coefficients $B_j^J$ can be computed analytically with the propagator matrix method (Appendix~\ref{AppendixMethodsIsotropic}) for an incompressible body with three homogeneous layers.
The compensation factor and topographic ratio are either obtained from the shape ratio with Eqs.~(\ref{TnSnRelation})-(\ref{FnSnRelation}) or computed from Love numbers with Eqs.~(\ref{TopoRatio0})-(\ref{CompensFactor0}).
Full results are given in the complementary software.

The ZDI topographic ratio has the remarkable property of being independent of the internal structure below the shell.
For an incompressible body with three homogeneous layers, this means being independent of the size and shear modulus of the core.
In terms of the generic Love number coefficients of Appendix~\ref{AppendixLoveNumbersHomogeneousLayers}, the topographic ratio (Eq.~(\ref{TopoRatio0})) reads
\begin{equation}
T_n^{\rm ZDI} = \frac{1}{\gamma_o} \, \frac{ \left( \gamma_o \hat B_o^L - \hat B'^L_o \right) + \zeta_n^{ \circ \rm ZDI} \left( \gamma_o \hat B_o^I - \hat B'^I_o \right)}{ \left( \gamma_s \hat B_s^L - \hat B'^L_s \right) + \zeta_n^{ \circ \rm ZDI} \left( \gamma_s \hat B_s^I - \hat B'^I_s \right)} \, .
\label{TopoRatioGen}
\end{equation}
Although the combinations $\gamma_j \hat B^J_j - \hat B'^J_j$ depend on the core parameters, their ratios do not:
\begin{eqnarray}
\frac{\gamma_o \hat B^L_o - \hat B'^L_o}{\gamma_s \hat B^I_s - \hat B'^I_s} &=& \frac{1}{x} \, ,
\\
\frac{\gamma_o \hat B^I_o - \hat B'^I_o}{\gamma_s \hat B^I_s - \hat B'^I_s} &=& - \frac{1}{x^2} \, \frac{\xi_{s1}}{\Delta\xi_1} \, \frac{Q_n(x)}{P_n(x)} \, ,
\\
\frac{\gamma_s \hat B^L_s - \hat B'^L_s}{\gamma_s \hat B^I_s - \hat B'^I_s} &=& - \frac{1}{x} \, \frac{\Delta\xi_1}{\xi_{s1}} \, \frac{Q'_n(x)}{P_n(x)} \, ,
\end{eqnarray}
where the functions $(P_n(x),Q_n(x),Q'_n(x))$ are polynomials in $x$ with coefficients defined in Table~\ref{TablePoly}.
The first relation results from the Saito-Molodensky relation (see Eq.~(\ref{SMrelationBcoeff})).
The two other ones must be computed with explicit formulas for Love numbers (see complementary software).

\begin{table}[ht]\centering
\ra{1.3}
\scriptsize
\caption[Non-zero coefficients of polynomials appearing in the ZDI geometrical factor]{\small
Non-zero coefficients of polynomials appearing in the ZDI geometrical factor (Eq.~(\ref{GeometricalFactorT})).}
\begin{tabular}{@{}llrr@{}}
\hline
\multicolumn{4}{l}{Polynomial  $P_n(x) = \sum_j p_j \, x^{j}$ } \\
Coefficient & \hspace{2mm} Arbitrary $n$ & $n=1$ & \hspace{1mm} $n=2$ \\
$p_{n-1}$ & $\,\,\,\,\, n^2(n+2)^2(2n-1)$ & 9 &  $3\times 64$\\
$p_{n+1}$ & $-(n^2-1)^2(2n+3)$ & 0 & $-3\times 21$ \\
$p_{3n}$ & $-p_{n+1}$ & 0 & $3\times 21$\\
$p_{3n+2}$ & $-p_{n-1}$ &  $-9$ & $-3\times 64$ \\
 \hline
\multicolumn{4}{l}{ Polynomials $Q_n(x) = \sum_j q_j \, x^j$ \hspace{1mm} and  \hspace{1mm} $Q'_n(x) = - \sum_j q_j \, x^{4n+3-j} $ } \\
Coefficient & \hspace{2mm} Arbitrary $n$ & $n=1$ & \hspace{1mm} $n=2$ \\
$q_0$ & $\,\,\,\,\, n (n+2) (2n^2+1)$ & 9 & $3\times 24$ \\
$q_{2n-1}$ & $\,\,\,\,\, n (n^2-1) (n+2) (2n+1)$ & 0 & $3\times 40$ \\
$q_{2n+3}$ & $- (2n+1) (n^4+2n^3-n^2-2n+3)$ & $-9$ & $- \, 3\times 45$ \\
$q_{4n+2}$ & $- (n^2-1) (2n^2+4n+3)$ & 0 & $- \,3\times 19$ \\
\hline
\end{tabular}
\label{TablePoly}
\end{table}%

The topographic ratio can thus be written as
\begin{equation}
T_n^{\rm ZDI} = - \frac{\xi_{s1}}{\Delta\xi_1} \, \frac{1}{\gamma_o} \, {\cal T}_n(x) \, ,
\label{TopoRatioZDI}
\end{equation}
where ${\cal T}_n(x)$ is a geometrical factor (that is, independent of density and rheology) defined by
\begin{equation}
{\cal T}_n(x) = \frac{\alpha \, P_n(x) + Q_n(x)}{\alpha \, Q'_n(x) + x^2 \, P_n(x)} \, .
\label{GeometricalFactorT}
\end{equation}
If the conjecture about the topographic ratio being independent of the internal structure below the shell is indeed correct, then Eqs.~(\ref{TopoRatioZDI})-(\ref{GeometricalFactorT}) are valid for bodies with stratified core and stratified ocean as long as the shell is homogeneous and incompressible.

\subsection{Dependence on parameters $\alpha$ and $n$}

\begin{figure}
\centering
    \includegraphics[width=0.4\textwidth]{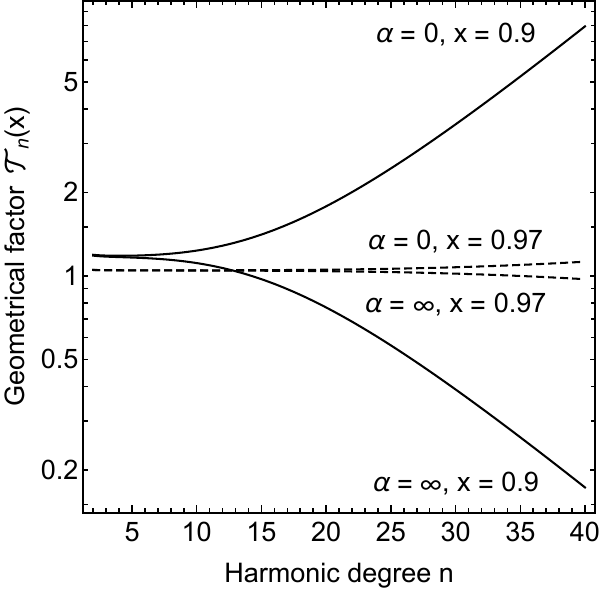}
   \caption[Geometrical factor for zero deflection isostasy as a function of harmonic degree]{
   Geometrical factor for zero deflection isostasy (Eq.~(\ref{GeometricalFactorT})), as a function of harmonic degree.
   The body is incompressible with three homogeneous layers.
   The shell is either thick ($x=0.9$) or thin ($x=0.97$).
   The geometrical factor can be interpreted as the bottom-to-surface pressure ratio of topographic loads.
 }
   \label{FigGeometricalFactor}
\end{figure}

The geometrical factor is the ratio of the radial pressures associated with the bottom and surface topographic loads: ${\cal T}_n(x)=-p_o/p_s$ where $p_s=\rho_sg_s H'_{sn}$ and $p_o=(\rho_o-\rho_s)g_o H'_{on}$ with $H'_{jn}=H_{jn}-\Gamma_{jn}/g_j$ being the topography at interface $j$.
Fig.~\ref{FigGeometricalFactor} shows the geometrical factor as a function of harmonic degree for either a thick shell or a thin shell; the isostatic family parameter $\alpha$ is set to zero or infinity.
The effect of $\alpha$ increases with harmonic degree and shell thickness.

At small harmonic degree, the geometrical factor depends little on $\alpha$ and gets close to the degree-one value which is independent of $\alpha$ (since $P_1=Q_1=Q_1'/x^2$):
\begin{equation}
{\cal T}_1(x)=1/x^2 \, .
\label{Tau1}
\end{equation}
Thus, Eq.~(\ref{TopoRatioZDI}) predicts the correct degree-one topographic ratio (Eq.~(\ref{T1})), although it was actually derived for $n\geq2$.
Such degree-one extensions do not exist for the shape and compensation factor which are frame-dependent at degree one (Section~\ref{DegreeOneCompensation}).

\begin{figure}
\centering
    \includegraphics[width=0.83\textwidth]{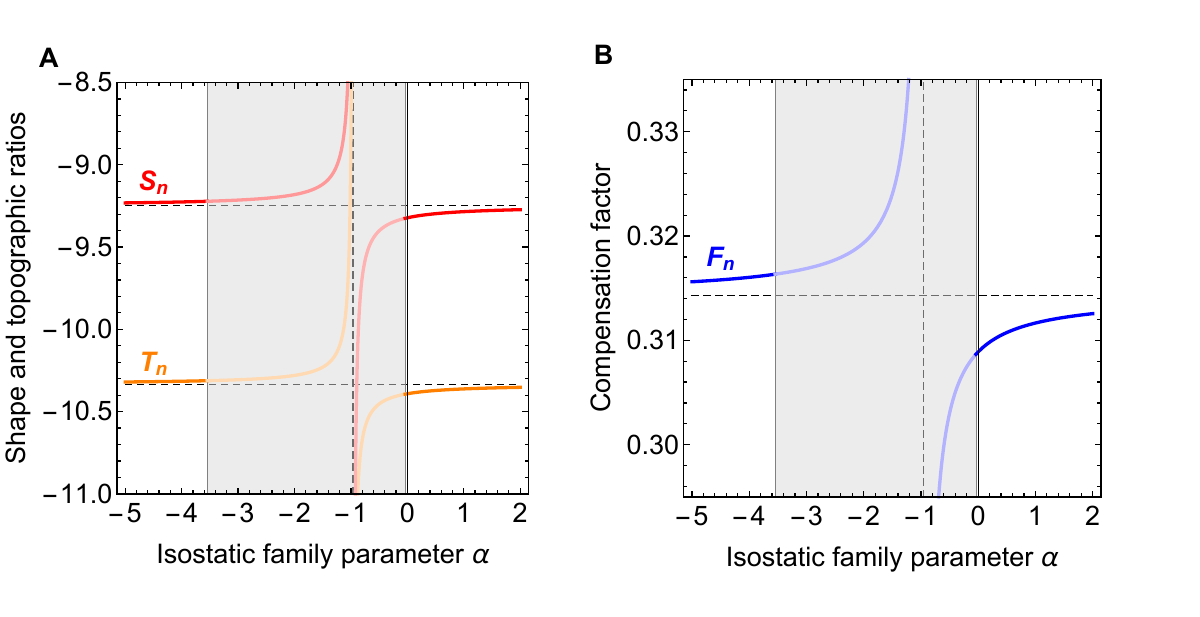}
   \caption[Zero deflection isostasy: shape ratio, topographic ratio, and compensation factor]{
   Zero deflection isostasy: (A) Shape and topographic ratios at degree 2 as a function of the isostatic family parameter $\alpha$;
   (B) compensation factor at degree 2 as a function of $\alpha$.
   The interior model is the one for Enceladus used in Fig.~\ref{FigComparisonfn}.
   The shaded zone covers the excluded range $\alpha_0<\alpha<\alpha_\infty$, with $\alpha_0=-3.55$ and $\alpha_\infty=-0.039$ (see Eq.~(\ref{ExcludedRange})).
   Horizontal dashed lines are asymptotes to $\alpha\rightarrow\pm\infty$.
   Vertical dashed lines are the asymptotes to diverging $S_n$ and $F_n$ (Eq.~(\ref{alphaSing})).
   The shape ratio and compensation factors diverge at $\alpha=\alpha_{sing}=-0.961$ while the topographic ratio diverges at $\alpha=-0.963$.
 }
   \label{FigureZDIdegree2}
\end{figure}

The shape ratio and compensation factor can be expressed in terms of the polynomials $(P_n,Q_n,Q'_n)$ by transforming back the topographic ratio with the $S_n-T_n$ and $F_n-T_n$ relations (Appendix~\ref{AppendixIsoRatiosIncompressible}).
Unless the core is infinitely rigid, the shape ratio and compensation factor depend on the properties of the core.
Fig.~\ref{FigureZDIdegree2} shows the degree-two shape ratio, topographic ratio, and compensation factor of Enceladus in zero deflection isostasy as a function of the isostatic family parameter.
If the range $\alpha_0<\alpha<\alpha_\infty$ is excluded (Eq.~(\ref{ExcludedRange})), the various ratios are minimum at $\alpha=\alpha_\infty$ and maximum at $\alpha=\alpha_0$.
It makes little difference to consider that the minimum occurs at $\alpha=0$ and the maximum at $\alpha=-\infty$ (e.g.\ error of $0.6\%$ in Fig.~\ref{FigureZDIdegree2}B).
At degree 2, the differences between the minimum and maximum values are about 1\% for the shape and topographic ratio and $2.5\%$ for the compensation factor.
For an Europa-like model, these differences are smaller by an order of magnitude.

Consider now the the thin shell limit of the same model (incompressible body with homogeneous shell).
Up to first order in $\varepsilon=1-x$, the ZDI topographic ratio is independent of the isostatic family parameter $\alpha$:
\begin{equation}
T_n^{\rm ZDI}
\cong - \frac{\xi_{s1}}{\Delta\xi_1} \, \frac{1}{\gamma_o} \left( 1 + \varepsilon \, \frac{3n \left( n+1 \right)}{2n^2+2n-1} \right) + {\cal O}(\varepsilon^2) \, .
\label{TopoRatioZDIthinshell}
\end{equation}
The shape ratio and compensation factor have thus a unique thin shell limit (that is, independent of $\alpha$), since they are related to the topographic ratio by Eqs.~(\ref{TnSnRelation})-(\ref{TnFnRelation}).
The unicity of the thin shell limit can be deduced from $\mu$-invariance without computing Love numbers (see Eq.~(\ref{IdentityBBthinshell})).
Furthermore, the ZDI topographic ratio in the thin shell limit is the same as the topographic ratio found in Dahlen's thin shell isostasy (Eq.~(\ref{TopoRatioTSI})).
If the core is rigid, $F_n$ and $S_n$ are thus given in the thin shell limit by Eqs.~(\ref{CompensFactorTSI})-(\ref{ShapeRatioTSI}), otherwise they should be computed from the topographic ratio with Eqs.~(\ref{TnSnRelation})-(\ref{TnFnRelation}).

\section{MSI-ZDI duality without fluid limit}
\label{AppendixFluidCrustLoadingRatio}
\renewcommand{\theequation}{G.\arabic{equation}} 
\setcounter{equation}{0}  

For an incompressible body with a homogeneous shell, ZDI $\mu$-invariance implies MSI $\mu$-invariance.
I will prove this statement by establishing the MSI-ZDI duality without using the fluid limit, but assuming instead ZDI $\mu$-invariance.
The MSI shape ratio is given by
\begin{equation}
S^{\rm MSI}_n = \frac{ E_{os,p} + \beta \, E_{oo,p} }{ E_{ss,p} + \beta \, E_{so,p} } \, .
\label{SnMINGbis}
\end{equation}
This formula is equivalent to Eq.~(\ref{SnMING}) because $E_{os,p} = E_{so,p}$, but we will see that there is a benefit to keep these coefficients formally distinct when proving $\mu$-invariance.
The essence of the proof consists in expressing stress-energy coefficients in terms of partial derivatives of Love numbers, eliminating gravitational Love numbers with gravity-deformation relations, and finally factoring out derivatives with the conditions for ZDI $\mu$-invariance.
The resulting MSI shape ratio is dual to the ZDI shape ratio.

For the purpose of the proof, the stress/energy coefficients are written as
\begin{equation}
E_{jk,p} = \frac{1}{N^2 } \left( \hat h_j^I \left( \hat h_k^I \, E^{LL}_p - \hat h_k^L \, E^{LI}_p \right) + \hat h_j^L \left( \hat h_k^L \,  E^{II}_p - \hat h_k^I \, E^{IL}_p \right) \right) .
\label{EjkpBis}
\end{equation}
This formula is equivalent to Eq.~(\ref{Ejkp}) because $E^{IL}_p=E^{LI}_p$.
Using Eqs.~(\ref{LoveDeriv1})-(\ref{LoveDeriv2}), I write the quantities within the interior brackets as
\begin{eqnarray}
\hat h_k^I \, E^{LL}_p - \hat h_k^L \, E^{LI}_p &=& \bar \chi \left( \hat h_k^I \, \partial_\mu \Big( h_s^L - k_s^L \Big) - \hat h_k^L \, \partial_\mu \Big( h_s^I - k_s^I \Big) \right) ,
\label{expr1} \\
\hat h_k^L \,  E^{II}_p - \hat h_k^I \, E^{IL}_p &=& \bar \chi \, x \left( \hat h_k^L \, \partial_\mu \Big( \gamma_o \, h_o^I - k_o^I \Big) - \hat h_k^I \, \partial_\mu \Big( \gamma_o \, h_o^L - k_o^L \Big) \right) \, ,
\label{expr2}
\end{eqnarray}
where common constants are gathered into
\begin{equation}
\bar\chi = \frac{p}{2} \, (\mu_{\rm s})^p \, (g_s R_s)^2 \, \frac{ (2n+1)R_s}{4 \pi G} \, .
\end{equation}
Eliminating the gravitational Love numbers with Eq.~(\ref{khRelation}), I get
\begin{eqnarray}
\hat h_k^I \, E^{LL}_p - \hat h_k^L \, E^{LI}_p &=& \bar \chi \left( \left( a-1 \right) Y_{ks} + b \, Y_{ko} \right) ,
\\
\hat h_k^L \,  E^{II}_p - \hat h_k^I \, E^{IL}_p &=& \bar \chi \, x \left( -c \, Y_{ks} + d \, Y_{ko} \right) \, ,
\end{eqnarray}
where
\begin{equation}
Y_{jk} = \hat h_j^L \, \partial_\mu h_k^I - \hat h_j^I \, \partial_\mu h_k^L \, .
\end{equation}
Noting that $X_{oo}=-h_o^L/(x\Delta\xi_n)$ and $X_{ss}=h_s^I/\xi_{sn}$ (see Table~\ref{TableLove}), one can see that
the conditions for ZDI $\mu$-invariance (Eq.~(\ref{BilinearInvariant})) are equivalent to
\begin{eqnarray}
\partial_{\mu} \Big( \hat h_s^I - \hat h_o^I \, h_s^L /\hat h_o^L \big) &=& 0 \, ,
\label{ZDIinvariance1} \\
\partial_{\mu} \left( \hat h_o^L - \hat h_s^L\, h_o^I /\hat h_s^I \right) &=& 0 \, ,
\label{ZDIinvariance2} \\
\partial_{\mu} \left( \Big( h_o^{I\circ} \, \hat h_s^L - h_s^{L\circ} \, \hat h_o^I \Big)/\hat h_s^I \right) &=& 0 \, .
\label{ZDIinvariance3}
\end{eqnarray}
Recall that $h_o^L=\hat h_o^L$ and $h_s^I=\hat h_s^I$ while $h_s^L=h_s^{L\circ}+\hat h_s^L$ and $h_o^I=h_o^{I\circ}+\hat h_o^I$.
Expanding Eqs.~(\ref{ZDIinvariance1})-(\ref{ZDIinvariance3}) and rearranging terms, I obtain three relations between the four different $Y_{jk}$:
\begin{eqnarray}
Y_{os}/h_s^L &=& Y_{oo}/h_o^L \, ,
\label{Yrel1} \\
Y_{so}/h_o^I &=& Y_{ss}/h_s^I \, ,
\label{Yrel2} \\
Y_{oo}/h_o^L &=& - (c x/b) \, Y_{ss}/h_s^I \, .
\label{Yrel3}
\end{eqnarray}
For the last equation, I relied on the fact that $h^{I\circ}_o/h^{L\circ}_s=cx/b$.
With these relations, I can express Eqs.~(\ref{expr1})-(\ref{expr2}) in terms of only one $Y_{jk}$, for example $Y_{ss}$, and substitute the results into Eq.~(\ref{EjkpBis}).
Using in addition the Saito-Molodensky relation (Eq.~(\ref{SMrelationRadial})), I write the stress/energy coefficients as
\begin{eqnarray}
E_{js,p} &=& \Big( - c \, X_{js} + d \, X_{jo} \Big) \left(x \, \frac{\bar \chi}{N^2} \, \frac{Y_{ss}}{h^I_s} \right) ,
\label{Ejsp} \\
E_{jo,p} &=& \frac{c}{b} \, \Big( (a-1) \, X_{js} + b \, X_{jo} \Big) \left(x \, \frac{\bar \chi}{N^2} \, \frac{Y_{ss}}{h^I_s} \right) ,
\label{Ejop}
\end{eqnarray}
where $X_{jk}$ is the bilinear of Love numbers defined by Eq.~(\ref{Xjkdef}).
Substituting Eqs.~(\ref{Ejsp})-(\ref{Ejop}) into (Eq.~(\ref{SnMINGbis}), I simplify the derivatives of Love numbers out of the shape ratio:
\begin{equation}
S^{\rm MSI}_n = \frac{ -c \left( \beta \left(1-a\right) +b \right) X_{os} + b \left( \beta \, c + d\right) X_{oo} }{ -c \left(\beta\left(1-a\right) +b \right) X_{ss} + b \left( \beta \, c + d\right) X_{so} } \, .
\label{SnMINGter}
\end{equation}
A quick comparison with the ZDI shape ratio (Eq.~(\ref{SnZDI})) shows that both formulas depend on the same combinations of Love numbers $X_{jk}$.
Therefore the conditions for ZDI $\mu$-invariance also imply MSI $\mu$-invariance, at least for an incompressible body with a homogeneous shell.
The additional benefit mentioned at the start of the section is that the ZDI and MSI shape ratios are actually identical if the isostatic family parameters $\alpha$ and $\beta$ are related by
\begin{equation}
\alpha = - \frac{b}{c} \, \frac{\beta \, c + d}{ \beta \left(1-a\right) +b }   \, .
\end{equation}
In the main text, I derive this duality in a simpler way by taking the fluid limit under the assumption of $\mu$-invariance.

Although not required for the computation of the shape ratio, it is interesting to know the dependence on $\mu$ of the stress/energy coefficients.
It is found by using the generic form of Love numbers appropriate to a homogeneous shell.
Using the identities (\ref{IdentityAB})-(\ref{IdentityBB}) and (\ref{coeffBdev}), one can check that
\begin{eqnarray}
N &=&  \frac{h_s^{L\circ} \, h_o^{I\circ} \, F \mu_s^2}{D+E \mu_{\rm s}+F \mu_{\rm s}^2} \, ,
\label{N3layers} \\
X_{js} &=& - h_s^{L\circ} \, \frac{\hat B_j^I \, \mu_{\rm s}}{D+E \mu_{\rm s}+F \mu_{\rm s}^2} \, ,
\label{Xjs} \\
X_{jo} &=& h_o^{I\circ} \, \frac{\hat B_j^L \, \mu_{\rm s}}{D+E \mu_{\rm s}+F \mu_{\rm s}^2} \, ,
\label{Xjo} \\
\frac{ Y_{ss} }{h_s^I} &=& h_s^{L\circ} \, \frac{F \mu_{\rm s}}{D+E \mu_{\rm s}+F \mu_{\rm s}^2} \, .
\end{eqnarray}
The stress/energy coefficients (Eqs.~(\ref{Ejsp})-(\ref{Ejop})) become
\begin{eqnarray}
E_{js,p} &=& \frac{\bar \chi}{\mu_{\rm s}^2} \, \frac{b \, \hat B_j^I + d \, x \, \hat B_j^L}{h_s^{L\circ} \, h_o^{I\circ} \, F} \, ,
 \\
E_{jo,p} &=& \frac{\bar \chi}{\mu_{\rm s}^2} \, \frac{(1-a) \, \hat B_j^I + c \, x \, \hat B_j^L}{h_s^{L\circ} \, h_o^{I\circ} \, F} \, .
\end{eqnarray}
Since $\bar\chi\sim\mu_{\rm s}^p$, the stress/energy coefficients are proportional to $(\mu_{\rm s})^{p-2}$.
Observe that the substitution of $X_{jk}$ (Eqs.~(\ref{Xjs})-(\ref{Xjo})) into the general ZDI shape ratio (Eq.~(\ref{SnZDI})), taking into account Eq.~(\ref{coeffBdev}), confirms the formula derived in the fluid limit (Eq.~(\ref{ShapeRatioZDIGen})).

\section{Isostatic ratios if three incompressible homogeneous layers}
\label{AppendixIsoRatiosIncompressible}
\renewcommand{\theequation}{H.\arabic{equation}} 
\setcounter{equation}{0}  

 Linear fractional transformations (LFTs) are defined by
\begin{equation}
f(z) = \frac{a'z+b'}{c'z+d'}
\,\, \leftrightarrow \,\,
f(z) = LFT \left[
\left(
\begin{array}{ll}
a' & b'  \\
c'& d'
\end{array}
\right)
\right]
(z) \, .
\label{LFT}
\end{equation}
In the matrix representation, the composition of two LFTs is given by the product of the associated matrices.
Isostatic ratios are LFTs in three different ways: of the loading ratio, of their isostatic family parameter, and of other isostatic ratios via the isostatic relations.

Consider an incompressible body with three homogeneous layers.
The ZDI topographic ratio, given by Eq.~(\ref{TopoRatioZDI}), is a LFT of $\alpha$.
The shape ratio and compensation factor are related to this formula by the $S_n-T_n$ and $F_n-T_n$ relations (Eqs.~(\ref{TnSnRelation}) and (\ref{TnFnRelation})).
Since these relations, as well as Eq.~(\ref{TopoRatioZDI}), are LFTs, they can be composed through the product of their associated matrices.
The MSI isostatic ratios are obtained by considering $\alpha_{\rm MSI}$ as a LFT of $\beta$ (Eq.~(\ref{alphaMSI})).
In the following, $(a,b,c,d)$ are the parameters defined in Table \ref{TableCoeff} in terms of the interior structure and of the harmonic degree $n$.
The dependence of $P_n$, $Q_n$, and $Q'_n$ on $x$ is implicit.

The ZDI shape ratio results from the composition of $S_n=LFT(\gamma_oT_n)$ with $T_n=T_n^{\rm ZDI}=LFT(\alpha)$:
\begin{eqnarray}
S_n^{\rm ZDI} &=&
LFT \left[
\left(
\begin{array}{cc}
1-a & c  \\
b & d
\end{array}
\right)
\left(
\begin{array}{cc}
-\xi_{s1} \, P_n & -\xi_{s1} \, Q_n \\
\Delta\xi_1 \, Q'_n & \Delta\xi_1 \, x^2 P_n
\end{array}
\right)
\right]
(\alpha)
\nonumber \\
&=& \frac{ \alpha \left( - \left(1-a\right) \xi_{s1} \, P_n  + c \, \Delta\xi_1 \, Q'_n \right)  - \left(1-a\right) \xi_{s1} \, Q_n + c \, \Delta\xi_1 \, x^2 P_n }{  \alpha \left( - b \, \xi_{s1} \, P_n + d \, \Delta\xi_1 \, Q'_n \right)  - b \, \xi_{s1} \, Q_n + d \, \Delta \xi_1\, x^2 P_n } \, .
\label{ShapeRatioZDI}
\end{eqnarray}

The ZDI compensation factor results from the composition of $\xi_{sn}F_n=LFT(\gamma_oT_n)$ with $T_n=T_n^{\rm ZDI}=LFT(\alpha)$:
\begin{eqnarray}
\xi_{sn} F_n^{\rm ZDI} &=&
LFT \left[
\left(
\begin{array}{cc}
b & a \, d + b \, c \\
b & d
\end{array}
\right)
\left(
\begin{array}{cc}
-\xi_{s1} \, P_n & -\xi_{s1} \, Q_n \\
\Delta\xi_1 \, Q'_n & \Delta\xi_1 \, x^2 P_n
\end{array}
\right)
\right]
(\alpha)
\nonumber \\
&=& \frac{ \alpha \left( - b \, \xi_{s1} \, P_n + \left(a d + b c \right) \Delta\xi_1 \, Q'_n  \right)  - b \, \xi_{s1} \, Q_n + \left(a d + b c \right) \Delta\xi_1 \, x^2 P_n }{ \alpha \left( - b \, \xi_{s1} \, P_n + d \, \Delta\xi_1 \, Q'_n \right) - b \, \xi_{s1} \, Q_n + d \, \Delta \xi_1\, x^2 P_n } \, .
\label{CompensFactorZDI}
\end{eqnarray}

The MSI topographic ratio results from the composition of $T_n^{\rm ZDI}=LFT(\alpha)$ with $\alpha=LFT(\beta)$:
\begin{eqnarray}
\gamma_o \, T_n^{\rm MSI}
&=&
LFT \left[
\left(
\begin{array}{cc}
-\xi_{s1} \, P_n & -\xi_{s1} \, Q_n \\
\Delta\xi_1 \, Q'_n & \Delta\xi_1 \, x^2 P_n
\end{array}
\right)
\left(
\begin{array}{cc}
- b \, c & - b \, d  \\
\left(1-a\right) c & b \, c
\end{array}
\right)
\right]
(\beta)
\nonumber \\
&=& - \frac{\xi_{s1}}{\Delta\xi_1} \,
\frac{ \beta \, c \left( - b \, P_n + \left(1-a\right) Q_n \right) + b \left( - d \, P_n  + c \, Q_n \right) }{ \beta \, c \left( - b \, Q'_n + \left(1-a\right) x^2 \, P_n \right) + b \left( c \, x^2 \, P_n - d \, Q'_n \right) } \, .
\label{TopoRatioMSI}
\end{eqnarray}

The MSI shape ratio results from the composition of $S_n=LFT(\gamma_oT_n)$ with $T_n=T_n^{\rm ZDI}=LFT(\alpha)$ and $\alpha=LFT(\beta)$, which is a bit too long to expand here:
\begin{equation}
S_n^{\rm MSI} =
LFT \left[
\left(
\begin{array}{cc}
1-a & c  \\
b & d
\end{array}
\right)
\left(
\begin{array}{cc}
-\xi_{s1} \, P_n & -\xi_{s1} \, Q_n \\
\Delta\xi_1 \, Q'_n & \Delta\xi_1 \, x^2 P_n
\end{array}
\right)
\left(
\begin{array}{cc}
- b \, c & - b \, d  \\
\left(1-a\right) c & b \, c
\end{array}
\right)
\right]
(\beta) \, .
\label{SnMSILFT}
\end{equation}

The MSI compensation factor results from the composition of $\xi_{sn}F_n=LFT(\gamma_oT_n)$ with $T_n=T_n^{\rm ZDI}=LFT(\alpha)$ and $\alpha=LFT(\beta)$:
\begin{equation}
\xi_{sn} F_n^{\rm MSI} =
LFT \left[
\left(
\begin{array}{cc}
b & a \, d + b \, c \\
b & d
\end{array}
\right)
\left(
\begin{array}{cc}
-\xi_{s1} \, P_n & -\xi_{s1} \, Q_n \\
\Delta\xi_1 \, Q'_n & \Delta\xi_1 \, x^2 P_n
\end{array}
\right)
\left(
\begin{array}{cc}
- b \, c & - b \, d  \\
\left(1-a\right) c & b \, c
\end{array}
\right)
\right]
(\beta) \, .
\label{FnMSILFT}
\end{equation}

\end{appendices}

\bibliographystyle{agufull04}

\end{document}